\documentclass[11pt]{article}

\title{The Relative Consistency of the Axiom of Choice\\
   Mechanized Using Isabelle/ZF}
\author{Lawrence C. Paulson\\
Computer Laboratory, University of Cambridge\\
         15 JJ Thomson Avenue\\
         Cambridge CB3 0FD, England}

\usepackage{isabelle,isabellesym}
\usepackage{amsmath,amssymb}
\usepackage{basic,proof,url}
\usepackage{times,newtxmath}
\usepackage{graphicx}

\newlength{\hsbw}
\newcommand{\MLSpacing}{13pt}

\newenvironment{session}{\begin{flushleft}
 \begin{tabular}{@{}|c@{}@{}}
 \begin{minipage}[b]{\hsbw}
 \vspace*{3pt}
\begingroup\small\baselineskip\MLSpacing}{\endgroup\end{minipage}\\
 \end{tabular}
 \end{flushleft}}
\newenvironment{defsession}{\par\noindent\begin{session}\begin{isabelle}\footnotesize}%
     {\end{isabelle}\end{session}}

\newcommand{\BL}{\ensuremath{\mathbf{L}}}
\newcommand{\BM}{\ensuremath{\mathbf{M}}}

\newcommand{\BON}{\ensuremath{\mathbf{ON}}}
\newcommand{\BV}{\ensuremath{\mathbf{V}}}
\newcommand{\BC}{\ensuremath{\mathbf{C}}}

\newcommand{\ZF}{\textrm{ZF}}

\newcommand{\ZFL}{\textrm{ZFL}}

\newcommand{\abbrevs}{\quad\text{abbreviates}\quad}
\newcommand{\quot}[1]{\ulcorner#1\urcorner}
\newcommand{\pow}{{\cal P}}
\newcommand{\dpow}{{\cal D}}

\usepackage[pdftex,colorlinks=true,linkcolor=blue,citecolor=blue,%
filecolor=blue,pagecolor=blue,urlcolor=blue]{hyperref}

\begin{document}


\date{}
\maketitle
\begin{abstract}
The proof of the relative consistency of the axiom of choice has been
mechanized using Isabelle/ZF\@.  The proof builds upon a previous
mechanization of the reflection theorem~\cite{paulson-reflection}. 
The heavy reliance on metatheory in the original proof makes the
formalization unusually long, and not entirely satisfactory: two parts
of the proof do not fit together.  It seems impossible to solve these
problems without formalizing the metatheory.  However, the present
development follows a standard textbook, Kunen's \emph{Set
Theory}~\cite{kunen80}, and could support the  formalization of further
material from that book.  It also serves as an example of what to
expect when deep mathematics is formalized.
\end{abstract}

\newpage
\tableofcontents

\newpage
\section{Introduction} \label{sec:introduction}

In 1940, G\"{o}del~\cite{goedel40} published his famous monograph
proving that the axiom of choice
(AC) and the generalized continuum hypothesis (GCH) are consistent
with respect to the other axioms of set theory. This theorem
addresses the first of Hilbert's celebrated list of mathematical
problems. I have attempted to reproduce this work in Isabelle/ZF\@. 

When so much mathematics has already been checked mechanically, what is the point of checking any more?  Obviously, the theorem's
significance makes it a challenge, as does its size and complexity,
but the real challenge comes from its reliance on metamathematics.  As
I have previously noted~\cite{paulson-reflection}, some theorems seem
difficult to formalize even in their statements, let alone in their
proofs.  G\"{o}del's work is not a single formal theorem.  It
consists of several different theorems which, taken collectively, can
be seen as expressing the relative consistency  of the axiom of
choice.  At the end of Chapter~VII,  G\"{o}del remarks that given a
contradiction from the axioms of set theory augmented with AC, a
contradiction in basic set theory ``could actually be constructed'' 
\cite[p.\ts87]{goedel40}.  This claim is crucial: logicians prefer
consistency proofs to be constructive.

G\"odel's idea~\cite{goedel38,goedel39} is to define a very lean model, 
called $\BL$, of set theory.  $\BL$ contains just the
sets that must exist because they can be defined by formulae.  Then,
prove that $\BL$ satisfies the ZF axioms and the additional axiom
``every  set belongs to~$\BL$,'' which is abbreviated
$\BV=\BL$.  We now know that $\BV=\BL$ is consistent with ZF, and can
assume this axiom.  (The conjunction of ZF and
$\BV=\BL$ is abbreviated ZFL\@.)  We conclude by proving that AC and GCH are
theorems of ZFL and therefore are also consistent with ZF\@.

A complication in G{\"o}del's proof is its use of classes.  Intuitively
speaking, a \emph{class} is a collection of sets that is defined by
comprehension,
$\{x\mid\phi(x)\}$.  Every set $A$ is trivially a class, namely
$\{x\mid x\in A\}$, but a \emph{proper class} is too big to be a set. 
Formal set theories restrict the use of classes in order to eliminate
the danger of paradoxes.  Modern set theorists use Zermelo-Fraenkel
(ZF) set theory, where classes exist only in the metalanguage.  That
is, the class
$\{x\mid\phi(x)\}$ is just an alternative notation for the formula
$\phi(x)$, and $a\in
\{x\mid\phi(x)\}$ is just an alternative notation for $\phi(a)$.  The
universal class, $\BV$, corresponds to the formula \isa{True}.  An
``equation'' like 
$\BV = \union_{\alpha\in\BON}\, V_\alpha$ stands for
$\forall x.\,\exists \alpha.\, \BON(\alpha)\land x\in V_\alpha$. 
(Here, $\BON$ denotes the class of ordinal numbers.)
G{\"o}del's monograph~\cite{goedel40}
uses von Neumann-Bernays-G{\"o}del (NBG) set theory, which allows
quantification over classes but restricts their use in other
ways.  With either axiom system, classes immensely complicate the
reasoning, making it highly syntactic.

Why did G{\"o}del use classes?  Working entirely with sets, he could have
used essentially the same techniques to prove that if
$M$ is a model of ZF then there exists a model $L(M)$ of ZFC\@. 
(ZFC refers to the ZF axioms plus AC\@.)
Therefore, if ZFC has no models, then neither does ZF\@. But with
this approach, he can no longer claim that if he had a contradiction in
ZFC then a contradiction in ZF ``could actually be constructed.''%
\footnote{Consistency proofs using
sets in ZF can be constructive provided we take care to keep track of
the number of instances of the separation axiom that are required. We can
prove in ZF (using the reflection theorem) that every finite conjunction
of the ZF axioms has a model. Therefore, the absence of a model yields
a contradiction in ZF\@. Proofs using \emph{forcing} are often formalized using this
technique. I am indebted to Kunen for this observation.}
For the sake of this remark, which is not part of any theorem
statement, G{\"o}del chose a more difficult route.  Classes
create more difficulties for formal proof checkers (which have to face
foundational issues squarely) than they do for mathematicians writing
in English.

The proof uses metatheoretic reasoning extensively.  G{\"o}del writes
\cite[p.\ts34]{goedel40},
\begin{quote}
However, the only purpose of these general metamathematical
considerations is to show how the proofs for theorems of a certain
kind can be accomplished by a general method. And, since applications
to only a finite number of instances are necessary $\ldots$, the general metamathematical
considerations could be left out entirely, if one took the trouble to
carry out the proofs separately for any instance.
\end{quote}
I decided to take the trouble, not using metatheory but relying instead
on a mechanical theorem prover.

This paper describes the Isabelle/ZF proofs.  It
indicates the underlying mathematical ideas and sometimes discusses
practical issues such as proof length or machine resources used.  It
necessarily omits much material that would be too long or too
repetitious.  The paper concerns how existing mathematics is
formalized; it contains no original mathematics.

\paragraph*{Overview.}  The paper begins by outlining G\"odel's
relative consistency proof (\S\ref{sec:outline}).  After a brief
overview of Isabelle/ZF, the paper describes the strategy guiding the
formalization (\S\ref{sec:isabelle}) and presents some elementary
absoluteness proofs (\S\ref{sec:basics}).  It then discusses
relativization issues involving well-founded recursion
(\S\ref{sec:wfrec}).  Turning away from absoluteness, the paper
proceeds to describe the formalization of the constructible universe
and the proof that $\BL$ satisfies the ZF axioms
(\S\ref{sec:defining-l}); then, it describes how the reflection
theorem is used to prove that $\BL$ satisfies the separation axiom
(\S\ref{sec:comprehension}).  Absoluteness again takes centre stage
as the paper presents the relativization of two essential datatypes
(\S\ref{sec:datatypes}) and finally presents the absoluteness
of~$\BL$ itself (\S\ref{sec:absoluteness-l}).  Finally, the paper
presents the Isabelle proof that AC holds in~$\BL$
(\S\ref{sec:ac-in-l}), and offers some conclusions
(\S\ref{sec:conclusions}).

\section{Proof Outline} \label{sec:outline}

\begin{figure}\centering
\includegraphics[scale=0.3]{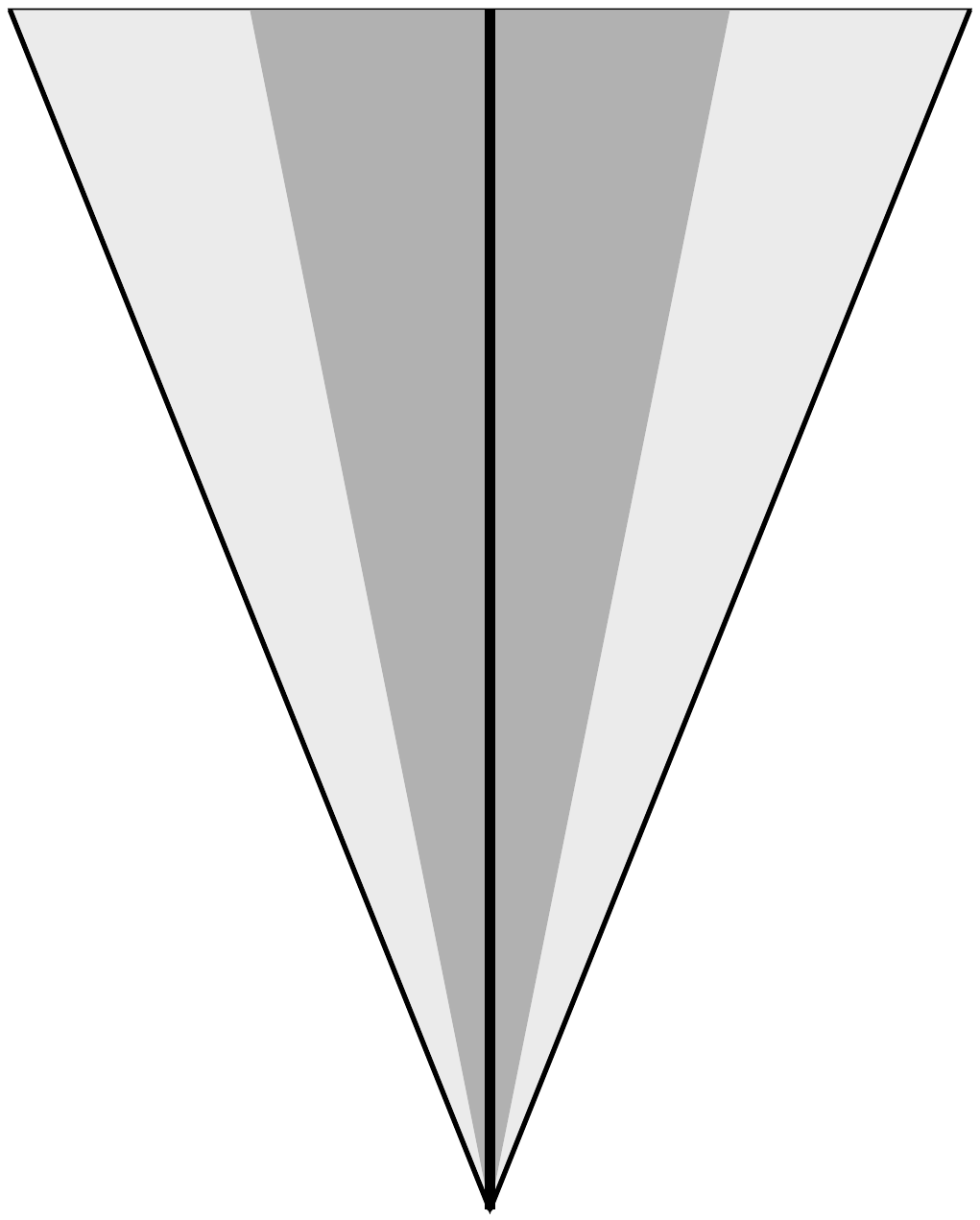}
\caption{The Constructible Universe, $\BL$}
\label{fig:L-diagram}
\end{figure}

Recall that G\"odel's idea is to define a lean model of set theory,
the class $\BL$ of the \emph{constructible sets}.  
Figure~\ref{fig:L-diagram} shows $\BL$ (shaded) as a subclass of the
universe, $\BV$.  The vertical line represents the class $\BON$ of the
ordinals.

G\"odel's proof involves four main tasks:
\begin{enumerate}
\item defining the class $\BL$ within ZF
\item proving that $\BL$ satisfies the ZF axioms
\item proving that $\BL$ satisfies $\BV=\BL$
\item proving that $\BV=\BL$ implies the axiom of choice
\end{enumerate}
As we shall see, Isabelle is well-suited for completing the first and last 
parts. Both the definition of~$\BL$ and the proof of the axiom of choice 
are straightforward exercises in mechanized set theory. The second task
cannot quite be completed: all of the ZF axioms can be verified apart from
\emph{separation}, which is an axiom scheme, and each instance requires
its own proof. As for the third task, the Isabelle proof that $\BL$ 
satisfies 
$\BV=\BL$ is much longer than I would like because the metamathematical
techniques that abbreviate textbook proofs are not available.

Once we have completed the first three tasks, we should be able to conclude
that if ZF is consistent, then so is ZFL\@. (And from the fourth task, 
if ZF is consistent, then so is ZFC\@.) This inference requires reasoning
in the metatheory, which is not possible using Isabelle/ZF, so the
machine formalization omits it.  Standard 
treatments also gloss over this step, regarding it as obvious. 
Section~\ref{sec:proof-theory} below expands on this issue.

\subsection{The Problem With Class Models} 

Because $\BL$ is a proper
class, we cannot adopt the usual notion of satisfaction.  To
formalize the standard Tarski definition of truth
\cite[p.\ts60]{mendelson} requires first defining, in set theory, a
set~$F$ to represent the syntax of first-order formulae.  $F$ is easily
defined, either using G{\"o}del-numbering or as a recursive data
structure.  If $M$ is a set, $p\in F$ represents a formula with $k$
free variables, and $m_1$, \ldots,~$m_k\in M$ then
$M\models p(m_1,\ldots,m_k)$ can be defined by recursion on the
structure of~$p$.  If
$\BM$ is a proper class, then the obvious definition of $\BM\models
p(\vec m)$ cannot be formalized in set theory; the environments that
hold the bindings of free variables would have to belong to a function
space whose range was all of~$\BM$.  Tarski's theorem on 
non-definability of truth~\cite[p.\ts41]{kunen80} asserts that
 no formula
$\chi(p)$ expresses $\BV\models p$.  If for each formula~$\phi$ we
write
$\quot{\phi}$ for the corresponding element of~$F$, then
$\psi\bimp\lnot\chi(\quot{\psi})$ is a theorem  for some
sentence~$\psi$.  Satisfaction cannot be defined, at least if
$\BM=\BV$.

\subsection{Relativization}

G{\"o}del instead expressed satisfaction for class models 
syntactically.  This approach abandons
the set $F$ of formula representatives in favour of real formulae.  Set theory uses a first-order language with no constant
symbols, no function symbols and no relation symbols other than $\in$
and~$=$.  Variables are the only terms.  

G{\"o}del's key concept is
\emph{relativization}.%
\footnote{See G{\"o}del \cite[p.\ts76]{goedel40} or for a modern
treatment Kunen
\cite[p.\ts112]{kunen80}.}
If
$\BM$ is a class and $\phi$ is a formula, define $\phi^\BM$
recursively as follows:
\begin{align*}
  (x=y)^\BM & \abbrevs x=y\\
  (x \in y)^\BM & \abbrevs x\in y\\
  (\phi\conj\psi)^\BM & \abbrevs \phi^\BM \conj\psi^\BM\\
  (\lnot\phi)^\BM & \abbrevs \lnot(\phi^\BM)\\
  (\exists x.\, \phi)^\BM & \abbrevs \exists x.\,x\in\BM\conj\phi^\BM
\end{align*}
Dually $(\forall x.\, \phi)^\BM$  abbreviates 
$\forall x.\,x\in\BM\imp\phi^\BM$, if  universal quantifiers are
defined as usual.  (When working in ZF, we should
write
$\BM(x)$ instead of $x\in\BM$ above.)
Relativization bounds all
quantifiers in $\phi$ by~\BM.  It is intuitively clear that
$\phi^\BM$ expresses that $\phi$ is true in~$\BM$.  But while the
satisfaction relation ($\models$) can be defined within set theory,
relativization can only be defined in the metalanguage: it combines
two arguments, $\phi$ and $\BM$, which lie outside ZF\@.  

\subsection{The Formal Treatment of Terms}

Despite the lack of terms in their formal language, set theorists use
elaborate notational conventions.  In other
branches of mathematics, an expression like
$f(x)g(y)-h(x,y)$ means what it says: functions $f$, $g$
and $h$ are applied and the results combined by multiplication and
subtraction.  But in set theory, each expression~$E(x)$ abbreviates a
formula $\phi(x,y)$, which reduces the meaning of $y=E(x)$ 
to a combination of $\in$ and~$=$. For example, we can express the
meaning of
$Y=A\un B$ by the predicate  $\isa{union}(A,B,Y)$, defined by 
\[ \forall z\,.\; z\in Y \bimp z\in A\disj z\in B.\]
We can similarly define $\isa{inter}(A,B,Y)$ to express $Y=A\Int B$. 
Combining these predicates gives meaning to more complex terms; for
example, 
$Y=(A\un B)\Int C$ abbreviates 
\[ \exists X.\, \isa{union}(A,B,X)\conj \isa{inter}(X,C,Y). \]

Variable binding notation, ubiquitous in set theory, causes
complications.  In $\union_{x\in A}\,B(x)$, what is $B$? 
Syntactically, $B(x)$ is a term with parameter~$x$, so we can take it
as an abbreviation for some formula $\phi(x,y)$.  But then $\union$
becomes an operation on formulae rather than one on sets.  An equally
legitimate alternative~\cite[p.\ts34]{halmos60} is to regard $B$ as a
function in set theory --- formally, the set of pairs
$\{\pair{x,B(x)}\mid x\in A\}$.  

Set theorists generally say little about these notational conventions
and act as if terms were meaningful in themselves.  But
relativization forces us to make the translation from terms to
formulae explicit.  In the
Isabelle formalization, I have defined relational equivalents of
dozens of term formers.  I have included a class argument in each
one to perform relativization at the same time; we can
express the relativized term $((A\un B)\Int C)^\BM$ as  
\[ \exists X\in\BM.\; \isa{union}(\BM,A,B,X)\conj
\isa{inter}(\BM,X,C,Y) \]
The hardest tasks were (1) to define relational equivalents of the
complicated expressions generated by Isabelle/ZF for recursively
defined sets and functions and (2) to cope with the sheer bulk of the
definitions.

\subsection{G{\"o}del's Claim Viewed Proof-Theoretically}
\label{sec:proof-theory}

The purpose of relativization is to express claims of the form
``$\phi$ is true in $\BM$.''  To prove that $\BL$
satisfies the ZF axioms and $\BV=\BL$, we must prove $\phi^\BL$ for
each ZF axiom~$\phi$, and we must prove $(\BV=\BL)^\BL$. 
Now we can consider G{\"o}del's claim that from a contradiction in
ZFL a contradiction in ZF ``could actually be constructed.''  His
claim is proof-theoretic.  A contradiction in ZFL is a proof, $\Pi$, of
$\bot$ from finitely many ZF axioms and $\BV=\BL$:
\[ \deduce[\strut\Pi]{\bot}{\phi_1 & \ldots & \phi_n & \BV=\BL}
\]
Once we have proved
that $\BL$ satisfies the axioms of ZFL, we have the $n+1$ proofs
\[ \ZF\turn \phi^\BL_1 \quad \ldots \quad \ZF\turn \phi^\BL_n \quad
(\BV=\BL)^\BL.\]
Verifying G{\"o}del's claim reduces to showing that we can
always construct a proof~$\Pi^\BL$ of $\bot^\BL$ from the
relativized premises:
\[ \deduce[\strut\Pi^\BL]{\bot^\BL}{\phi^\BL_1 & \ldots & \phi^\BL_n &
(\BV=\BL)^\BL}
\]
For then we get a proof of $\ZF\turn
\bot^\BL$, which is just
$\ZF\turn\bot$.   

So how we obtain $\Pi^\BL$ from~$\Pi$?  To be
concrete, suppose we are working with a natural deduction
formalization of first-order logic.  By the normal
form theorem~\cite{prawitz71}, 
since the conclusion of the proof is
atomic, we can assume that $\Pi$ applies only elimination rules.  We
must modify
$\Pi$ so that it accepts relativized versions of its premises and
delivers a relativized version of its conclusion.  
The only hard
cases involve quantifiers.  Where
$\Pi$ applies the existential elimination rule to $\exists
x.\,\phi(x)$, it delivers the formula~$\phi(x)$ to the rest of the
proof.  (Assume that $x$ has already been renamed, if necessary.)  At
the corresponding position,
$\Pi^\BL$ should apply the existential and conjunction elimination
rules to
$\exists x.\,x\in\BL\conj\phi(x)$, delivering the
formulae~$x\in\BL$ and $\phi(x)$ to the rest of the
proof.  

Universal quantifiers require a bit more work.  First,
recall that the language of set theory has no terms other than
variables.  Where
$\Pi$ applies the universal elimination rule to $\forall
x.\,\phi(x)$, it delivers the formula~$\phi(y)$ to the rest of the
proof, where $y$ is a variable.  At the corresponding position,
$\Pi^\BL$ should apply the existential and conjunction elimination
rules to
$\forall x.\,x\in\BL\imp\phi(x)$.  But before it can deliver the
formula $\phi(y)$, it requires a proof of~$y\in\BL$.  We will 
indeed have $y\in\BL$ if the variable $y$ is obtained by a
previous existential elimination, but what if $y$ was
chosen arbitrarily?  We can handle such cases by inserting at this
point an application of the empty set axiom, which will yield a new
variable (say~$z$) and the assumption~$z\in\BL$.  Intuitively, we are
replacing all free variables in~$\Pi$ by~0.  

The sketchy argument above cannot be called a rigorous proof of
G{\"o}del's claim.  But it is more detailed than standard expositions
of G{\"o}del's proof.  Kunen relegates the relevant lemma to an
appendix, and for the proof he merely remarks ``Similar to the easy
direction of the G{\"o}del Completeness Theorem''
\cite[p.\ts141]{kunen80}.  To G{\"o}del, it was all presumably
trivial.  I have not formalized the argument in Isabelle/ZF because
that would require formalizing the metatheory. 

\subsection{Defining the Class $\BL$}

The equation $\BV = \union_{\alpha\in\BON}\,
V_\alpha$ expresses the universe of sets as the union
of the cumulative hierarchy
$\{V_\alpha\}_{\alpha\in\BON}$, which is recursively defined by
$V_0 = 0$, 
$V_{\alpha+1} = \pow(V_\alpha)$ and
$V_\alpha  = \union_{\xi<\alpha}\,V_\xi$ when $\alpha$ is a limit
ordinal.  We obtain $\BL$ by a similar construction, replacing the
powerset operator~$\pow$ by the \emph{definable powerset}
operator,~$\dpow$\@.  Essentially, $\dpow(A)$ yields the set of all
subsets of~$A$ that can be defined by a formula taking parameters
over~$A$.  If we define the set~$F$
of formulae and the satisfaction relation $A\models p$ as outlined
above, then we can make the definition
\begin{multline*}
 \dpow(A) = 
   \{X\in\pow(A)\mid \\
     \exists p\in F.\; \exists a_1 \ldots a_k\in A.\;
     X = \{x\in A\mid A\models p(x,a_1,\ldots,a_l)\}\}.
\end{multline*}
(The ellipsis can be eliminated in favour of lists
over~$A$.)  Finally, we define the \emph{constructible universe}:
$\BL = \union_{\alpha\in\BON}\,
L_\alpha$, where
$L_0 = 0$, 
$L_{\alpha+1} = \dpow(L_\alpha)$ and
$L_\alpha  = \union_{\xi<\alpha}\,L_\xi$ when $\alpha$ is limit.  

Kunen proves that $\BL$ satisfies the ZF axioms, remarking
\cite[p.\ts170]{kunen80} ``only the Comprehension Axiom required any
work.''  His remark applies to the Isabelle/ZF proofs.  $\BL$ inherits
most of the necessary properties from~$\BV$.  Even the axiom scheme
of replacement can be proved as the theorem
\isa{replacement(L,P)}; the proof is independent of the
formula~\isa{P}\@.  However, the proof of comprehension for the
formula $\phi$ requires an instance of the reflection theorem
for~$\phi$, which requires recursion over the structure
of~$\phi$.  Each instance of comprehension therefore has a different
proof from the ZF axioms.  At the metalevel, of course, all these
proofs are instances of a single algorithm.  For Isabelle/ZF, this
means that each instance of comprehension must be proved separately,
although the proof scripts are nearly identical.  

\subsection{Absoluteness: Proving $(\BV=\BL)^\BL$}

Proving that $\BL$ satisfies $\BV=\BL$ is a key part of the proof, and
despite first appearances, it is far from trivial.  It amounts to
saying that the  construction of
$\BL$ is idempotent: $\BL^\BL=\BL$. 
The underlying concept is called \emph{absoluteness}, which expresses
that a given operator or formula behaves the same in a class
model~$\BM$ as it does in~$\BV$, the universe.  A class $\BM$ is
\emph{transitive} if $x\in\BM$ implies
$x\subseteq\BM$, and we shall only be concerned with transitive
models below.

Most constructions are absolute.  The empty set
can only be a set having no elements, and $A\subseteq B$ can only mean
that every element of $A$ belongs to~$B$.  If $A$ and $B$ are sets then
their union can only be the set containing precisely the elements of
those sets.  Many complicated notions are also absolute: domains and
ranges of relations, bijections, well-orderings, order-isomorphisms,
ordinals.  With some effort, we can show the absoluteness of
recursively defined data structures and functions.

Powersets, except in trivial cases, are not absolute.  For example,
$\pow(\omega)$ might contain subsets of the natural numbers that
cannot be shown to exist.  The function space $A\to B$ is not
absolute because of the obvious connection between $\pow(A)$ and
$A\to\{0,1\}$.  More subtly,
cardinality is not absolute: if
$M$ is a countable model of set theory, and
$\alpha$ is an uncountable cardinal according to
$M$, then obviously $\alpha$ must be really be countable, with the
bijections between $\alpha$ and $\omega$ lying outside $M$.  This
situation is called Skolem's paradox \cite[p.\ts141]{kunen80}.

Metamathematical arguments are an efficient means of proving
absoluteness.  For example, any concept that is provably equivalent
(in ZF) to a formula involving only bounded quantifiers is
absolute~\cite[p.\ts119]{kunen80}.  This is the class of
$\Delta_0^{\ZF}$ formulae.  The larger class of $\Delta_1^{\ZF}$
formulae can also be shown to be absolute.  Unfortunately, all such
arguments are beyond our reach unless we formalize the metatheory. 

\subsection{The Consequences of $\BV=\BL$}

Once we have proved that $\BL$ is absolute, we obtain
$\ZF\turn(\BV=\BL)^\BL$.  We can then investigate the consequences of
assuming  $\BV=\BL$.  To prove the
axiom of choice, it suffices to prove that every set can be
well-ordered.  The key step, given a well-ordering of $A$, is to
construct a well-ordering of $\dpow(A)$.  It comes from the
lexicographic ordering on tuples $\langle p,a_1,\ldots,a_k\rangle$ for
$p\in F$ and $a_1$, \ldots,~$a_k\in A$.
So if $L_\alpha$ is well-ordered, so is $L_{\alpha+1}$.  By transfinite
induction, each level of the construction of $\BL$ is well-ordered. 

The axiom $\BV=\BL$ is very strong.  G\"{o}del proved that it implies
the generalized continuum hypothesis.  Jensen later proved that it
implies the combinatorial principle known as $\lozenge$, and it has many
additional consequences.  
But it is important to note that such proofs are
entirely separate from that of $\ZF\turn(\BV=\BL)^\BL$.  We prove
$\ZFL\turn\textrm{AC}$, $\ZFL\turn\textrm{GCH}$ and
$\ZFL\turn\lozenge$, but we do not prove $\ZF\turn\textrm{AC}^\BL$,
$\ZF\turn\textrm{GCH}^\BL$ and
$\ZF\turn\lozenge^\BL$.  Those results, if we want them, are most
easily obtained in the metatheory, using the general fact that if
$\phi\turn\psi$ then $\phi^\BL\turn\psi^\BL$.

\section{Introduction to the Isabelle/ZF Formalization}
\label{sec:isabelle}

\href{http://isabelle.in.tum.de/}{Isabelle}~\cite{isa-tutorial,paulson-isa-book}
is an interactive theorem prover that supports a  variety of
logics, including set theory and higher-order logic. Isabelle
provides automatic tools for simplification and  logical reasoning.
They can be combined with single-step inferences  using a traditional
tactical style or as structured proof texts.  The
\href{http://www.proofgeneral.org/}{Proof
  General} user interface provides an
effective interactive  environment. Isabelle has been applied to a
huge number of verification  tasks, including the semantics of the
Java language~\cite{strecker-formal} and the correctness  of
cryptographic protocols~\cite{paulson-security}. Most of these proofs
use Isabelle/HOL, the version of Isabelle for higher-order logic.
Isabelle/HOL's polymorphic type system is ideal for modelling problems
in computer science.

Isabelle also supports Zermelo-Fraenkel set theory. Formalized 
material includes the traditional concepts of functions, 
ordinals, order types and cardinals.  Isabelle/ZF also accepts
definitions of recursive functions and data structures; in this it
resembles other computational logics, with the important difference of
being typeless.  Some problems do call for a typeless logic. 
Isabelle/ZF is also  good for investigating foundational issues, and,
of course, for  formalizing proofs in axiomatic set theory. 

Previous published work on Isabelle/ZF describes its basic 
development~\cite{paulson-set-I} and its treatment of recursive
functions~\cite{paulson-set-I} and inductive
definitions~\cite{paulson-fixedpt-milner}.  Another paper
describes proofs drawn from set theory textbooks~\cite{paulson-gr}.
Particularly noteworthy are the proofs of equivalence between various 
formulations of the axiom of choice.  Those proofs, formalized by
Gr\c{a}bczewski, are highly technical, demonstrating that advanced set theory
proofs can be replicated in Isabelle/ZF given enough time and 
effort.  That is precisely why we should investigate G\"{o}del's proof
of the relative consistency of AC: much of the reasoning takes place
outside set theory. 

The previous section has presented many reasons why we should 
formalize G\"{o}del's proof directly in the metatheory.  That
strategy does not require a set theory prover.  We could use any
system that lets us define the first-order
formulae, the set theory axioms, and the set of
theorems  derivable from any given axioms.  We would enjoy a number of
advantages.  
\begin{itemize}
\item Relativization could easily be defined by recursion on the
structure of formulae.  

\item Metatheorems about absoluteness --- for example, that all
$\Delta_0^{\ZF}$ formulae are absolute --- could be proved and used to
obtain simple proofs of many absoluteness results.  

\item The constructiveness of the consistency result could be
stated and proved. 
\end{itemize}
However, the metatheoretical strategy also presents difficulties. 
We would have to work in the pure language of set theory, which 
reduces all concepts to membership and equality, and is unreadable;
an alternative would be to formalize the familiar term language. 
We would constantly be reasoning about an explicitly formalized 
inference system for ZF rather than using our prover's built-in 
reasoning tools. I believe this strategy would involve as much 
work as the strategy I adopted, although the work would be distributed
differently. 

The choice resembles the standard one we face when we model a 
formal language: shall we adopt a deep or a shallow embedding? 
A shallow embedding maps phrases in the language to corresponding 
phrases in the prover's logic.  It works well for 
reasoning about specific examples, but does not allow
metareasoning (proofs  about the language).  A deep embedding involves
formalizing the language's syntax and semantics in the prover's logic. 
The extra mechanism allows metareasoning but complicates reasoning
about specific examples.  Compared with a shallow embedding, the
strengths and weaknesses are exchanged.

I have chosen to formalize G\"{o}del's theorem in set theory,
minimizing any excursions into the metatheory.  This strategy still
requires defining relational equivalents for each element of set
theory's term language, while limiting my exposure to
unreadable relational formulae.  After  all, the critical proofs
involve showing that various concepts are absolute, which means
that they do not vary from one model of set theory to another.
Each absoluteness proof justifies replacing some primitive of
the relational language by its counterpart in the term
language.  Thus Isabelle's simplifier can transform relational
formulae into ones using terms, exploiting the existing
formalization of set theory.  

This plan worked well for basic concepts such as union, intersection,
relation, function, domain, range, image, inverse image and even
ordinal. The absoluteness proofs for well-orderings, recursive functions
and recursive data types were harder:
\begin{itemize}
\item If a concept is defined in terms of non-absolute primitives, such
as powerset, it must be proved equivalent to a suitable alternative
definition.

\item Much of the theory of well-founded recursion must be formalized
from scratch in the relational language.

\item Higher-order functions complicate the relational language.

\item Recursive functions generate complicated
fixedpoint definitions that must be converted into relational form
manually.
\end{itemize}

\section{Relativization and Absoluteness: Basics} \label{sec:basics}

The first step is to define the relational language, introducing
predicates for all the basic concepts of set theory. 
Each predicate takes a class as an argument so that it can express
relativization.  This relational language will later allow appeals to
the reflection theorem.  Space permits only a few of the
predicates to appear below. 
Note that the class quantifications $\forall
x\in\BM$ and $\exists x\in\BM$ are written
\isa{\isasymforall x[M]} and \isa{\isasymexists x[M]}\@ in Isabelle.
For example, \isa{\isasymforall x[M].\ P(x)} is
definitionally equivalent to \isa{\isasymforall x.\ M(x) \isasymlongrightarrow P(x)}.

\subsection{From the Empty Set to Functions}
\label{sec:union_predicate}

We begin with definitions of trivial concepts such as the empty set and
the subset relation.  A set~\isa{z} is empty if it has no elements:
\begin{defsession}
"empty(M,z)\ \ \ \ ==\ \isasymforall x[M].\ x\ \isasymnotin \
z"\isanewline "subset(M,A,B)\ ==\ \isasymforall x[M].\ x\ \isasymin\ A\
\isasymlongrightarrow\ x\ \isasymin \ B"
\end{defsession}
All Isabelle definitions in this paper are indicated by a vertical
line, as shown.

A set~\isa{z} is the unordered pair of~\isa{a} and~\isa{b} if
it contains those two sets and no others.  The Kuratowski definition
of ordered pairs $\pair{a,b} = \{\{a,a\},\{a,b\}\}$ is then expressed
using the predicate \isa{upair}:
\begin{defsession}
"upair(M,a,b,z)\ ==\ a\ \isasymin \ z\ \&\ b\ \isasymin \ z\ \&\ (\isasymforall x[M].\ x\isasymin z\ \isasymlongrightarrow\ x=a\ |\ x=b)"\isanewline
"pair(M,a,b,z)\ \ ==\ \isasymexists x[M].\
upair(M,a,a,x)\ \&\isanewline
\ \ \ \ \ \ \ \ \ \ \ \ \ \ \ \ \ \ \ (\isasymexists y[M].\
upair(M,a,b,y)\ \&\ upair(M,x,y,z))"
\end{defsession}
A set~\isa{z} is the union of~\isa{a} and~\isa{b} if
it contains their elements and no others.  The general union
$\union(A)$, also written as
$\union\{X\mid X\in A\}$, has an analogous definition.
\begin{defsession}
"union(M,a,b,z)\ \ ==\ \isasymforall x[M].\ x\ \isasymin \ z\ \isasymlongleftrightarrow\ x\
\isasymin \ a\ |\ x\ \isasymin \ b"\isanewline
"big\_union(M,A,z)\ ==\
\isasymforall x[M].\ x\ \isasymin \ z\ \isasymlongleftrightarrow\ (\isasymexists y[M].\
y\isasymin A\ \&\ x\ \isasymin \ y)"
\end{defsession}
A set~\isa{z} is the domain of the relation~\isa{r} if
it consists of each element \isa{x} such that
\isa{\isasymlangle x,y\isasymrangle\ \isasymin\ r} for some~\isa{y}. 
\begin{defsession}
"is\_domain(M,r,z)\
==\ \isasymforall x[M].\ x\ \isasymin \ z\ \isasymlongleftrightarrow\isanewline
\ \ \ \ \ \ \ \ \ \ \ \ \ \ \ \ \ \ \ \ \ \ (\isasymexists w[M].\
w\isasymin r\ \&\ (\isasymexists y[M].\ pair(M,x,y,w)))"
\end{defsession}

\subsection{Relativizing the Ordinals}

Now we can define relational versions of ordinals and related
concepts.  The formalization is straightforward.

An ordinal is a transitive set of
transitive sets. 
\begin{defsession}
"transitive\_set(M,a)\ ==\ \isasymforall x[M].\ x\isasymin a\ \isasymlongrightarrow\ subset(M,x,a)"\isanewline
"ordinal(M,a)\ ==\isanewline
\ \ \ \ \ transitive\_set(M,a)\ \&\ (\isasymforall
x[M].\ x\isasymin a\ \isasymlongrightarrow\ transitive\_set(M,x))"
\end{defsession}
A limit ordinal is a non-empty, successor-closed
ordinal. 
\begin{defsession}
"limit\_ordinal(M,a)\ ==\isanewline
\ \ \ \ ordinal(M,a)\ \&\ \isachartilde \ empty(M,a)\ \&\isanewline
\ \ \ \ (\isasymforall x[M].\ x\isasymin a\ \isasymlongrightarrow\ (\isasymexists y[M].\
y\isasymin a\ \&\ successor(M,x,y)))"
\end{defsession}
A successor ordinal is any ordinal that is neither empty nor
limit. 
\begin{defsession}
"successor\_ordinal(M,a)\ ==\isanewline
\ \ \ \ ordinal(M,a)\ \&\ \isachartilde \ empty(M,a)\ \&\ \isachartilde
\ limit\_ordinal(M,a)"
\end{defsession}
The set of natural numbers, $\omega$, is a limit ordinal that contains
no limit ordinals. 
\begin{defsession}
"omega(M,a)\ ==\isanewline
\ \ \ \ limit\_ordinal(M,a)\ \&\ (\isasymforall x[M].\
x\isasymin a\ \isasymlongrightarrow\ \isachartilde \ limit\_ordinal(M,x))"
\end{defsession}

\subsection{Defining the Zermelo-Fraenkel Axioms} 
\label{sec:defining-ZF}

Formally defining the ZF
axioms relative to a class~\isa{M}\@ lets us express
that \isa{M} satisfies those axioms.  Each axiom is relativized so
that all quantified variables range over~\isa{M}\@.

We begin with extensionality:
\begin{defsession}
"extensionality(M)\ ==\isanewline
\ \ \ \ \isasymforall x[M].\ \isasymforall y[M].\ (\isasymforall z[M].\
z\ \isasymin \ x\ \isasymlongleftrightarrow\ z\ \isasymin \ y)\ \isasymlongrightarrow\ x=y"
\end{defsession}
The separation axiom is also known as comprehension:
\begin{defsession}
"separation(M,P)\ ==\isanewline
\ \ \ \ \isasymforall z[M].\ \isasymexists y[M].\ \isasymforall x[M].\
x\ \isasymin \ y\ \isasymlongleftrightarrow\ x\ \isasymin \ z\ \&\ P(x)"
\end{defsession}
This only yields a valid instance of separation if the formula~\isa{P}
obeys certain syntactic restrictions.  All quantifiers in
\isa{P} must be relativized to \isa{M}, and the free variables in
\isa{P} must range over elements of~\isa{M}\@.  These restrictions
prevent us from assuming separation as a scheme by leaving \isa{P} as
a free variable.  We must separately note every instance of separation
that we need.  If it meets the syntactic restrictions,
then later we shall be able to prove that $\BL$ satisfies it. 

That looks bad when we recall that the native separation axiom in
Isabelle/ZF, and the theorems using it, are schematic in \isa{P}\@. 
But if we formalize Bernays-G\"odel set theory as a new Isabelle logic
(creating the system Isabelle/BG) then the same problem occurs
elsewhere.  The analogue of separation in BG set theory is the General
Existence Theorem, which is a metatheorem: proving each instance
requires a separate construction.  To compensate, at least BG has no
axiom schemes.

The axioms of unordered pairs, unions and powersets all state that
\isa{M} is closed under the given operation:
\begin{defsession}
"upair\_ax(M)\ ==\ \isasymforall x[M].\ \isasymforall y[M].\ \isasymexists z[M].\ upair(M,x,y,z)"\isanewline
"Union\_ax(M)\ ==\ \isasymforall x[M].\ \isasymexists z[M].\
big\_union(M,x,z)"\isanewline "power\_ax(M)\ ==\ \isasymforall x[M].\
\isasymexists z[M].\ powerset(M,x,z)"
\end{defsession}

The foundation axiom states that every non-empty set has a
$\in$-minimal element:
\begin{defsession}
"foundation\_ax(M)\ ==\isanewline
\ \ \isasymforall x[M].\ (\isasymexists y[M].\ y\isasymin x)\ \isasymlongrightarrow\ (\isasymexists y[M].\ y\isasymin x\ \&\ \isachartilde (\isasymexists z[M].\ z\isasymin x\ \&\ z\isasymin y))"
\end{defsession}

Call a formula \emph{univalent} over a set if it describes a class
function on that set.   The replacement axiom holds for univalent formulae:
\begin{defsession}
"univalent(M,A,P)\ ==\isanewline
\ \ \ \ \isasymforall x[M].\ x\isasymin A\ \isasymlongrightarrow\ (\isasymforall y[M].\
\isasymforall z[M].\ P(x,y)\ \&\ P(x,z)\ \isasymlongrightarrow\ y=z)"\isanewline
"replacement(M,P)\ ==\isanewline
\ \ \ \ \isasymforall A[M].\ univalent(M,A,P)\ \isasymlongrightarrow\isanewline
\ \ \ \ \ \ \ \ \ \ \ \ (\isasymexists Y[M].\ \isasymforall b[M].\
(\isasymexists x[M].\ x\isasymin A\ \&\ P(x,b))\ \isasymlongrightarrow\ b\ \isasymin \
Y)"
\end{defsession}
Intuitively, if $\cal F$ is a class function and and $A$ is a set,
then replacement says that ${\cal F}``A$ (the image of $A$ under~$\cal
F$) is a set.  However, the axiom formalized above is weaker: it
merely asserts (relative to the class~$\BM$) that ${\cal F}``A
\subseteq B$ for some set~$B$.  To get the set
we really want, namely ${\cal F}``A$, we must apply the axiom of
separation to~$B$.  The weak form of replacement can be proved
schematically for~$\BL$.  The strong form cannot be proved
schematically because of its reliance on separation.
\begin{defsession}
"strong\_replacement(M,P)\ ==\isanewline
\ \ \isasymforall A[M].\ univalent(M,A,P)\ \isasymlongrightarrow\isanewline
\ \ (\isasymexists Y[M].\ \isasymforall b[M].\ b\ \isasymin \ Y\ \isasymlongleftrightarrow\ (\isasymexists x[M].\ x\isasymin A\ \&\
P(x,b)))"
\end{defsession}

\subsection{Introducing a Transitive Class Model}

The absoluteness proofs are carried out with respect to an arbitrary
class model $\BM$, although they are only needed for $\BL$. 
Generalizing the proofs over other models has two advantages: it
separates the absoluteness proofs from reasoning about $\BL$ and it
allows the proofs to be used with other class models.  

Isabelle's locale mechanism \cite{kammueller-locales}  makes the
generalization possible.  A locale packages the many properties
required of~$\BM$, creating a context in which they are implicitly
available.  A proof within a locale may refer to those properties and to
other theorems proved in the same locale.  A locale can
extend an older one, creating a context that includes
everything available in the ancestor locales.  
\begin{defsession}
 \isacommand{locale}\ M\_trivial\ =\isanewline
\ \ \isakeyword{fixes}\ M\isanewline
\ \ \isakeyword{assumes}\ transM:\ \ \ \ \ \ \ "\isasymlbrakk y\isasymin
x;\ M(x)\isasymrbrakk \ \isasymLongrightarrow\ M(y)"\isanewline
\ \ \ \ \ \isakeyword{and}\ upair\_ax:\ \ \ \ \ "upair\_ax(M)"\isanewline
\ \ \ \ \ \isakeyword{and}\ Union\_ax:\ \ \ \ \ "Union\_ax(M)"\isanewline
\ \ \ \ \ \isakeyword{and}\ power\_ax:\ \ \ \ \ "power\_ax(M)"\isanewline
\ \ \ \ \ \isakeyword{and}\ replacement:\
\ts"replacement(M,P)"\isanewline
\ \ \ \ \ \isakeyword{and}\ M\_nat\ [iff]:\ \ "M(nat)"
\end{defsession}
The class $\BM$ is assumed to be transitive (\isa{transM}) and to
satisfy some relativized ZF axioms, such as unordered pairing
(\isa{upair\_ax}) and replacement.  It contains the set of natural
numbers, \isa{nat} (which is also the ordinal~$\omega$).  This
locale does not assume any instances of separation.

\subsection{Easy Absoluteness Proofs}
\label{sec:M_trivial}

Here is a canonical example of an absoluteness result.  The phrase
\isa{\isakeyword{in}\ M\_trivial} includes the lemma in the locale. 
\begin{isabelle}\footnotesize
\isacommand{lemma}\ (\isakeyword{in}\ M\_trivial)\ empty\_abs\ [simp]:\isanewline
\ \ \ \ \ "M(z)\ \isasymLongrightarrow\ empty(M,z)\ \isasymlongleftrightarrow\
z=0"\isanewline
\isacommand{apply}\ (simp\ add:\ empty\_def)\isanewline
\isacommand{apply}\ (blast\ intro:\ transM)\isanewline
\isacommand{done}
\end{isabelle}
The proof refers to the definition of empty set (\isa{empty\_def}) and
to the transitivity of~\isa{M} (the locale assumption
\isa{transM}); it uses \isa{blast}, an automatic prover.  The attribute
\isa{[simp]} declares
\isa{empty\_abs} as a simplification rule: the simplifier will replace
any occurrence of \isa{empty(M,z)} by \isa{z=0} provided it can prove
\isa{M(z)}. From now on, usually just the statements of theorems will
be shown, not header lines and proofs.

Here are some similar absoluteness results, also proved in locale
\isa{M\_trivial} and declared to the simplifier.  Most have trivial
proofs like the one shown above.
\begin{isabelle}\footnotesize
"M(A)\ \isasymLongrightarrow\ subset(M,A,B)\ \isasymlongleftrightarrow\ A\ \isasymsubseteq \ B"\isanewline
"M(z)\ \isasymLongrightarrow\ upair(M,a,b,z)\ \isasymlongleftrightarrow\ z=\isacharbraceleft
a,b\isacharbraceright "\isanewline
"M(z)\ \isasymLongrightarrow\ pair(M,a,b,z)\ \isasymlongleftrightarrow\ z=\isasymlangle a,b\isasymrangle"\isanewline
"\isasymlbrakk M(r);\ M(A);\ M(z)\isasymrbrakk \ \isasymLongrightarrow\ image(M,r,A,z)\ \isasymlongleftrightarrow\ z\ =\
r``A"\isanewline 
"\isasymlbrakk M(A);\ M(B);\ M(z)\isasymrbrakk \
\isasymLongrightarrow\ cartprod(M,A,B,z)\ \isasymlongleftrightarrow\ z\ =\ A\isasymtimes
B"\isanewline "\isasymlbrakk M(a);\ M(b);\ M(z)\isasymrbrakk \
\isasymLongrightarrow\ union(M,a,b,z)\ \isasymlongleftrightarrow\ z\ =\ a\ \isasymunion\
b"\isanewline "\isasymlbrakk M(A);\ M(z)\isasymrbrakk \
\isasymLongrightarrow\ big\_union(M,A,z)\ \isasymlongleftrightarrow\ z\ =\ Union(A)"
\end{isabelle}
These theorems express absoluteness because the class~\isa{M}
disappears from the right-hand side: the meaning of subset, image,
etc., is the same as its meaning in~$\BV$\@.  Each theorem also
expresses the correctness of an element of the relational language,
for example that
\isa{big\_union} captures the meaning of \isa{Union}.

Absoluteness results involving ordinals are also easily proved:
\begin{isabelle}\footnotesize
"M(a)\ \isasymLongrightarrow\ ordinal(M,a)\ \isasymlongleftrightarrow\ Ord(a)"\isanewline
"M(a)\ \isasymLongrightarrow\ limit\_ordinal(M,a)\ \isasymlongleftrightarrow\ Limit(a)"\isanewline
"M(a)\ \isasymLongrightarrow\ successor\_ordinal(M,a)\ \isasymlongleftrightarrow\isanewline
\ \ \ \ \ \ \ \ \ \ Ord(a)\ \&\
(\isasymexists b[M].\ a\ =\ succ(b))"
\end{isabelle}
Thus we see that the simplifier can rewrite relational formulae into
term notation, provided we are able to prove that they refer to
elements of~\isa{M}\@.  For this purpose, there are many results
showing that \isa{M} is closed under the usual set-theoretic
constructions.  In particular, we can use the separation axiom for a
specific formula~\isa{P}\@:
\begin{isabelle}\footnotesize
"M(A)\ \isasymLongrightarrow\ M(Union(A))"\isanewline
"\isasymlbrakk M(A);\ M(B)\isasymrbrakk \ \isasymLongrightarrow\ M(A\ \isasymunion\ B)"\isanewline
"\isasymlbrakk separation(M,P);\ M(A)\isasymrbrakk \ \isasymLongrightarrow\ M(\{x\isasymin A.\ P(x)\})"
\end{isabelle}
Also useful are logical equivalences to simplify assertions
involving~\isa{M}\@:
\begin{isabelle}\footnotesize
"M(\isacharbraceleft a,b\isacharbraceright )\ \isasymlongleftrightarrow\ M(a)\ \&\
M(b)"\isanewline
"M(\isasymlangle a,b\isasymrangle)\ \isasymlongleftrightarrow\ M(a)\ \&\ M(b)"
\end{isabelle}

\subsection{Absoluteness Proofs Assuming Instances of Separation}
\label{sec:M_basic}

All the theorems shown above are proved without recourse to the axiom
of separation.  Obviously many set-theoretic operators are defined
using separation --- possibly in the guise of strong replacement --- so
we now extend locale \isa{M\_trivial} accordingly.  
\begin{defsession}
 \isacommand{locale}\ M\_basic\ =\ M\_trivial\ +\isanewline
\isakeyword{assumes}\ Inter\_separation:\isanewline
\ \ \ \ \ "M(A)\ \isasymLongrightarrow\ separation(M,\ \isasymlambda x.\ \isasymforall y[M].\ y\isasymin A\ \isasymlongrightarrow\ x\isasymin y)"\isanewline
\ \ \isakeyword{and}\ Diff\_separation:\isanewline
\ \ \ \ \ "M(B)\ \isasymLongrightarrow\ separation(M,\ \isasymlambda x.\ x\ \isasymnotin \ B)"\isanewline
\ \ \isakeyword{and}\ cartprod\_separation:\isanewline
\ \ \ \ \ "\isasymlbrakk M(A);\ M(B)\isasymrbrakk \isanewline
\ \ \ \ \ \ \isasymLongrightarrow\ separation(M,\ \isasymlambda z.\ \isasymexists x[M].\ x\isasymin
A\ \&\isanewline
\ \ \ \ \ \ \ \ \ \ \ \ \ \ \ \ \ \ \ \ \ \ \ \ \ \ \ \ (\isasymexists
y[M].\ y\isasymin B\ \&\ pair(M,x,y,z)))"\isanewline
\ \ \isakeyword{and}\ image\_separation:\isanewline
\ \ \ \ \ "\isasymlbrakk M(A);\ M(r)\isasymrbrakk \isanewline
\ \ \ \ \ \ \isasymLongrightarrow\ separation(M,\ \isasymlambda y.\ \isasymexists p[M].\ p\isasymin r\ \&\isanewline
\ \ \ \ \ \ \ \ \ \ \ \ \ \ \ \ \ \ \ \ \ \ \ \ \ \ \ \
(\isasymexists x[M].\ x\isasymin A\ \&\ pair(M,x,y,p)))"\isanewline
\ \ \isakeyword{and}\ converse\_separation:\isanewline
\ \ \ \ \ "M(r)\ \isasymLongrightarrow\ separation(M,\ \isasymlambda z.\ \isasymexists
p[M].\ p\isasymin r\ \&\isanewline
\ \ \ \ \ \ \ \ \ \ \ \ \ \ \ \ \ \ \ \ \ \ \ \ \ \ \ \ \ \ \ \ \
(\isasymexists x[M].\ \isasymexists y[M].\ pair(M,x,y,p)\ \&\isanewline
\ \ \ \ \ \ \ \ \ \ \ \ \ \ \ \ \ \ \ \ \ \ \ \ \ \ \ \ \ \ \
\ \ pair(M,y,x,z)))"
\end{defsession}
Only a few of the 11 instances of separation appear above.  Omitted
are the more complicated ones, for example concerning well-founded
recursion. 

By \isa{Inter\_separation} it follows that \isa{M} is
closed under intersections.  
\begin{isabelle}\footnotesize
 \isacommand{lemma}\ (\isakeyword{in}\ M\_basic)\
Inter\_closed:\isanewline
\ \ \ \ \ "M(A)\ \isasymLongrightarrow\ M(Inter(A))"
\end{isabelle}
From the lemma declaration, you can see that the proof takes place in
locale \isa{M\_basic}.  All results proved in locale
\isa{M\_trivial} remain available. 

By \isa{cartprod\_separation} it follows that the class \isa{M} is
closed under Cartesian products.  The proof is complicated
because the powerset operator (which is not absolute) occurs in the
definition.  A trivial corollary is that \isa{M} is
closed under disjoint sums.
\begin{isabelle}\footnotesize
"\isasymlbrakk M(A);\ M(B)\isasymrbrakk \ \isasymLongrightarrow\ M(A\isasymtimes B)"\isanewline
"\isasymlbrakk M(A);\ M(B)\isasymrbrakk \ \isasymLongrightarrow\ M(A+B)"
\end{isabelle}
I devoted some effort to minimizing the number of instances of
separation required.  For example, the inverse image operator is
expressed in terms of the image and converse operators.  Then the
domain and range operators can be expressed in terms of inverse image
and image.  We obtain five closure theorems from the two
assumptions \isa{image\_separation} and
\isa{converse\_separation}:
\begin{isabelle}\footnotesize
"\isasymlbrakk M(A);\ M(r)\isasymrbrakk \ \isasymLongrightarrow\ M(r``A)"\isanewline
"\isasymlbrakk M(A);\ M(r)\isasymrbrakk \ \isasymLongrightarrow\ M(r-``A)"\isanewline
"M(r)\ \isasymLongrightarrow\ M(converse(r))"\isanewline
"M(r)\ \isasymLongrightarrow\ M(domain(r))"\isanewline
"M(r)\ \isasymLongrightarrow\ M(range(r))"
\end{isabelle}
These five operators are also absolute.  Here is the result for
\isa{domain}: 
\begin{isabelle}\footnotesize
"\isasymlbrakk M(r);\ M(z)\isasymrbrakk \ \isasymLongrightarrow\ is\_domain(M,r,z)\ \isasymlongleftrightarrow\ z\ =\ domain(r)"
\end{isabelle}

Although we assume that \isa{M} satisfies the powerset axiom, we
cannot hope to prove \isa{M(A) \isasymLongrightarrow M(Pow(A))}.  The powerset of
\isa{A} relative to~\isa{M} is smaller than the true powerset,
containing only those subsets of~\isa{A} that belong to~\isa{M}\@. 
Similarly, we cannot show that \isa{M} contains all functions from
\isa{A} to~\isa{B}\@.  However, it holds for a finite case, essentially
the set
$B^n$ of
$n$-tuples:
\begin{isabelle}\footnotesize
"\isasymlbrakk n\isasymin nat;\ M(B)\isasymrbrakk \ \isasymLongrightarrow\ M(n->B)"
\end{isabelle}
This lemma will be needed later to prove the absoluteness of
transitive closure.

\subsection{Some Remarks About Functions}

In set theory, a function is a single-valued relation and thus is a
set of ordered pairs.  Operators such as powerset and union, which
apply to all sets, are not functions.  (Strictly speaking, there are no
operators in the formal language of set theory, since the only terms
are variables.)  Isabelle/ZF distinguishes functions from operators
syntactically.  
\begin{itemize}
\item The application of the function~\isa{f} to the
argument~\isa{x} is written~\isa{f`x}.  On
the other hand, application of an operator to its operand is written
using parentheses, as in \isa{Pow(X)}, or using infix notation.
\item Function abstraction over a set~\isa{A} is indicated by
\isa{\isasymlambda x\isasymin A}, and yields a set of pairs.  For
instance, 
\isa{\isasymlambda x\isasymin A.\ts x} denotes the identity function
on~\isa{A}\@.  Operators are essentially abstractions over the
universe, as in \isa{\isasymlambda x.\ts Pow(Pow(x))}.  Abstraction
can also express predicates; for instance, 
\isa{\isasymlambda x.\ts P(x) \& Q(x)} is the conjunction
of the two predicates  \isa{P} and~\isa{Q}\@.
\end{itemize}

Kunen \cite[p.\ts14]{kunen80} defines function application in the
usual way: 
$f`x$ is ``the unique~$y$ such that $\pair{x,y}\in f$.''  Isabelle/ZF
originally adopted a formal version of this definition, using a
description operator \cite[\S7.5]{paulson-set-I}.  The relational
version of the operator, namely \isa{fun\_apply(M,f,x,y)}, 
held if the pair
\isa{\isasymlangle x,y\isasymrangle} belongs to\isa{f} for that unique~\isa{y}.

My original definitions of function application, in its
infix and relational forms, both followed Kunen's definition. 
However, the absoluteness theorem relating them was
conditional on the function application's being well-defined.  That
made it harder to simplify 
\isa{fun\_apply(M,f,x,y)} to \isa{f`x = y} and often forced
proofs to include what was essentially type information.

Redefining function application by $f`x=\union(f``\{a\})$ solved these
problems by eliminating the definite description.  The new definition
looks peculiar, but it agrees with the old one when the latter is
defined.  Its relational version is straightforward:
\begin{isabelle}\footnotesize
"fun\_apply(M,f,x,y)\ ==\isanewline
\ \ \ \ (\isasymexists xs[M].\ \isasymexists fxs[M].\isanewline
\ \ \ \ \ upair(M,x,x,xs)\ \&\ image(M,f,xs,fxs)\ \&\
big\_union(M,fxs,y))"
\end{isabelle}
Thus it follows that \isa{M} is closed under function application,
which is also absolute:
\begin{isabelle}\footnotesize
"\isasymlbrakk M(f);\ M(a)\isasymrbrakk \ \isasymLongrightarrow\ M(f`a)"\isanewline
"\isasymlbrakk M(f);\ M(x);\ M(y)\isasymrbrakk \ \isasymLongrightarrow\ fun\_apply(M,f,x,y)\ \isasymlongleftrightarrow\ f`x\ =\
y"
\end{isabelle}

\section{Well-Founded Recursion}
\label{sec:wfrec}

The hardest absoluteness proofs concern recursion.  Well-founded
recursion is the most general form of recursive function definition. 
The proof that well-founded relations are absolute consists of several
steps.  Well-orderings, which are well-founded linear orderings, are
somewhat easier to prove absolute.

\subsection{Absoluteness of Well-orderings}

The concept of well-ordering is the first we encounter whose absoluteness
proof is hard.  One direction is easy:
if relation $r$ well-orders $A$, then it also well-orders $A$ relative
to~$\BM$.  For if every nonempty subset of~$A$ has an $r$-minimal element,
then trivially so does every nonempty subset of~$A$ that belongs to~$\BM$;
this is Lemma IV 3.14 in Kunen~\cite[p.\ts123]{kunen80}. 
For proving the converse direction, Kunen (Theorem IV~5.4, page 127)
reasons that ``every well-ordering is isomorphic to an
ordinal.''  We can obtain this result by showing that order types exist
in~$\BM$ and are absolute.  The proof requires some instances of
separation and replacement for~$\BM$.  

The theory defines various properties of relations, relative to a
class~\isa{M}\@.  Transitivity, linearity, and other simple properties
have the obvious definitions and are easily demonstrated to be
absolute.  The definition of well-founded refers to the existence
of~\isa{r}-minimal elements, as discussed above.  
\begin{defsession}
"wellfounded\_on(M,A,r)\ ==\isanewline
\ \ \ \ \isasymforall x[M].\ x\isasymnoteq0\
\isasymlongrightarrow\ x\ \isasymsubseteq\ A\isanewline
\ \ \ \ \ \ \ \ \ \ \ \ \ \ \ \ \isasymlongrightarrow\ (\isasymexists y[M].\ y\isasymin
x\
\&\ \isachartilde (\isasymexists z[M].\ z\isasymin x\ \&\ \isasymlangle z,y\isasymrangle\ \isasymin \
r))"
\end{defsession}
A well-ordering is a well-founded relation that is also linear and
transitive.
\begin{defsession}
"wellordered(M,A,r)\ ==\isanewline
\ \ \ \ transitive\_rel(M,A,r)\ \&\ linear\_rel(M,A,r)\ \&\isanewline
\ \ \ \ wellfounded\_on(M,A,r)"\
\end{defsession}
Kunen's lemma IV 3.14 takes the following form:
\begin{isabelle}\footnotesize
"well\_ord(A,r)\ \isasymLongrightarrow\ wellordered(M,A,r)"
\end{isabelle}

The definition of order types is standard; see Theorem I 7.6
of Kunen~\cite[p.\ts17]{kunen80}.  We use replacement to construct a
function that maps elements of~\isa{A} to ordinals, proving that its
domain is the whole of~\isa{A} and that each element of its range is
an ordinal.  Its range is the desired order type.  But the construction
must be done relative to~\isa{M}\@.  In particular, when we need
well-founded induction on~\isa{r}, we must apply a relativized
induction rule:
\begin{isabelle}\footnotesize
"\isasymlbrakk a\isasymin A;\ \ wellfounded\_on(M,A,r);\ \ M(A);\isanewline
\ \ separation(M,\ \isasymlambda x.\ x\isasymin A\
\isasymlongrightarrow\
\isachartilde P(x));\isanewline
\ \ \isasymforall x\isasymin A.\ M(x)\ \&\ (\isasymforall
y\isasymin A.\ \isasymlangle y,x\isasymrangle\ \isasymin \ r\ \isasymlongrightarrow\ P(y))\ \isasymlongrightarrow\ P(x)\isasymrbrakk \isanewline \isasymLongrightarrow\ P(a)"
\end{isabelle}
One premise is an instance of the separation axiom involving the
negation of the induction formula.  Each time we apply induction, we
must assume another instance of separation.

After about 250 lines of proof script, we arrive at Kunen's Theorem IV~5.4. 
The notion of well-ordering is absolute:
\begin{isabelle}\footnotesize
"\isasymlbrakk M(A);\ M(r)\isasymrbrakk \ \isasymLongrightarrow\ wellordered(M,A,r)\ \isasymlongleftrightarrow\
well\_ord(A,r)"
\end{isabelle}
Order types are absolute.  That is, if \isa{f} is an order-isomorphism
from between \isa{(A,r)} and some ordinal~\isa{i}, then
\isa{i} is the order type of~\isa{(A,r)}.
\begin{isabelle}\footnotesize
"\isasymlbrakk wellordered(M,A,r);\ f\ \isasymin \ ord\_iso(A,\ r,\ i,\
Memrel(i));\isanewline
\ \ M(A);\ M(r);\ M(f);\ M(i);\ Ord(i)\isasymrbrakk \
\isasymLongrightarrow\ i\ =\ ordertype(A,r)"
\end{isabelle}
These results are not required in the sequel, but I found their proofs
useful preparation for tackling the more general problem of well-founded
recursion.

\subsection{Functions Defined by Well-founded Recursion Are Absolute}
\label{sec:wfrec-abs}

It is essential to show that functions can be defined by well-founded
recursion in~\isa{M} and that such functions are absolute. 
This is Kunen's theorem IV~5.6, page 129.

Let $r$ be a well-founded relation.  If $f$ is recursively defined
over~$r$ then $f(a)$ is derived from~$a$ and from various $f(y)$ where
$y$ ranges over the set of $r$-predecessors of~$a$.  This set is just
$r^{-1}``\{a\}$, the inverse image of $\{a\}$ under~$r$, more
explicitly $\{y\mid\pair{y,a}\in r\}$.  Writing the body of~$f$ as $H(x,g)$,
with free variables $x$ and~$g$, we get the recursion equation:
\begin{equation}
 f(a)  =  H(a, f\restriction(r^{-1}``\{a\})) \label{eqn:recfun}
\end{equation}
Note that $f\restriction(r^{-1}``\{a\})$ denotes the function obtained by
restricting~$f$ to $r$-predecessors of~$a$.

If $r$ and~$H$ are given, then the existence of a suitable function~$f$ follows by
well-founded induction over~$r$, as I have described in previous
work~\cite{paulson-set-II}.
I have had to repeat some of these proofs relative to~\isa{M}\@.  The theorems
may assume only the relativized assumption
\isa{wellfounded(M,r)}, which for the moment is weaker than
\isa{wf(r)}.  About 200
lines of proof script are necessary, but fortunately much of this
material is based on earlier proofs.  We reach a key result concerning
the existence of recursive functions:
\begin{isabelle}\footnotesize
"\isasymlbrakk wellfounded(M,r);\ \ trans(r);\isanewline
\ \ separation(M,\ \isasymlambda x.\ \isachartilde \ (\isasymexists f[M].\ is\_recfun(r,x,H,f)));\isanewline
\ \ strong\_replacement(M,\ \isasymlambda x\ z.\isanewline
\ \ \ \ \ \isasymexists y[M].\ \isasymexists g[M].\
z=\isasymlangle x,y\isasymrangle\
\&\ is\_recfun(r,x,H,g)\ \&\ y\ =\ H(x,g));\isanewline
\ \ M(r);\ \ M(a);\isanewline
\ \ \isasymforall x[M].\ \isasymforall g[M].\ function(g)\ \isasymlongrightarrow\ M(H(x,g))\isasymrbrakk \isanewline
\isasymLongrightarrow\ \isasymexists f[M].\ is\_recfun(r,a,H,f)"
\end{isabelle}
The predicate \isa{is\_recfun(r,a,H,f)} expresses that \isa{f} satisfies the
recursion equation~(\ref{eqn:recfun}) for the given relation~\isa{r} and
body~\isa{H} for all \isa{r}-predecessors of~\isa{a}.
So the theorem states that if \isa{r} is well-founded
and transitive then there exists \isa{f} in \isa{M}
satisfying the recursion equation below~\isa{a}.  
Obviously \isa{r} and \isa{a} must belong to the class
\isa{M}, which moreover must be closed under~\isa{H}\@.  Two
additional premises list instances of separation and replacement, which depend
upon~\isa{r} and~\isa{H}\@.  Before we can assume such instances, we must express
them relative to~\isa{M}\@.  That in turn requires a relativized version of
\isa{is\_recfun}:
\begin{defsession}
"M\_is\_recfun(M,MH,r,a,f)\ ==\isanewline
\ \ \isasymforall z[M].\ z\ \isasymin \ f\ \isasymlongleftrightarrow\isanewline
\ \ \ \ \ \ \ \ \ (\isasymexists x[M].\ \isasymexists y[M].\ \isasymexists xa[M].\ \isasymexists sx[M].\ \isasymexists r\_sx[M].\ \isasymexists f\_r\_sx[M].\isanewline
\ \ \ \ \ \ \ \ \ \ \ \ pair(M,x,y,z)\ \&\ pair(M,x,a,xa)\ \&\ upair(M,x,x,sx)\ \&\isanewline
\ \ \ \ \ \ \ \ \ \ \ \ pre\_image(M,r,sx,r\_sx)\ \&\ restriction(M,f,r\_sx,f\_r\_sx)\ \&\isanewline
\ \ \ \ \ \ \ \ \ \ \ \ xa\ \isasymin \ r\ \&\ MH(x,\ f\_r\_sx,\ y))"
\end{defsession}
This definition is the translation of equation~(\ref{eqn:recfun}) into
relational language.  (Observe how quickly
this language becomes unreadable.)  In particular, the binary operator
\isa{H} becomes the ternary relation~\isa{MH}\@.  The argument \isa{H}
makes \isa{is\_recfun} a higher-order function, which complicates
subsequent work.  We cannot relativize \isa{is\_recfun} once and for all,
but if \isa{MH} is expressed in relational language, then so is
\isa{M\_is\_recfun}.

The predicate \isa{relation2} expresses that \isa{is\_f} is the
relational form of \isa{f} over~\isa{M}:
\begin{defsession}
"relation2(M,is\_f,f)\ ==\isanewline
\ \ \ \ \ \isasymforall x[M].\ \isasymforall y[M].\
\isasymforall z[M].\ is\_f(x,y,z)\ \isasymlongleftrightarrow\ z\ =\ f(x,y)"
\end{defsession}

The predicate \isa{is\_wfrec} expresses that \isa{z} is computed from
\isa{a} and \isa{MH} by well-founded recursion over~\isa{r}.  The body of
the definition expresses the existence of a function~\isa{f} satisfying
equation~(\ref{eqn:recfun}), with \isa{z = H(a,f)}.
\begin{defsession}
"is\_wfrec(M,MH,r,a,z)\ ==\isanewline
\ \ \ \isasymexists f[M].\ M\_is\_recfun(M,MH,r,a,f)\ \&\ MH(a,f,z)"
\end{defsession}

We now reach two lemmas, stating that \isa{M\_is\_recfun} and
\isa{is\_wfrec} behave as intended.  The first result is absoluteness of 
\isa{is\_recfun}.  Among the premises are that \isa{M} is closed under
\isa{H} and that \isa{MH} is the
relational form of \isa{H}: 
\begin{isabelle}\footnotesize
"\isasymlbrakk \isasymforall x[M].\ \isasymforall g[M].\ function(g)\ \isasymlongrightarrow\
M(H(x,g));\ \ M(r);\ M(a);\ M(f);\isanewline
\ \ relation2(M,MH,H)\isasymrbrakk \isanewline
\ \isasymLongrightarrow\ M\_is\_recfun(M,MH,r,a,f)\ \isasymlongleftrightarrow\ is\_recfun(r,a,H,f)"
\end{isabelle}
Under identical premises, we get the corollary
\begin{isabelle}\footnotesize
"is\_wfrec(M,MH,r,a,z)\ \isasymlongleftrightarrow\ (\isasymexists g[M].\ is\_recfun(r,a,H,g)\ \&\
z=H(a,g))"
\end{isabelle}

\subsection{Making Well-founded Recursion Available}
\label{sec:wfrec_abs}

Mathematically speaking, we have already proved the absoluteness of
well-founded recursion.  Pragmatically speaking, unfortunately, more
work must be done to package the results so that they can be used in
formal proofs.  In particular, we need a theorem relating the
predicate \isa{is\_wfrec} defined above with the function
\isa{wfrec} provided by Isabelle/ZF \cite[\S3.1]{paulson-set-II};
\isa{wfrec(r,a,H)} denotes the result of \isa{f(a)}, where \isa{f} is
the function with body~\isa{H} defined by recursion over~\isa{r}.

The development of well-founded recursion assumes \isa{r} to be
transitive.  To apply well-founded recursion to other relations
requires a theory of transitive closure.  Isabelle/ZF defines the 
transitive closure of a relation
inductively~\cite[\S2.5]{paulson-set-II}.  Inductive definitions are
abstract and elegant, but they do not lend themselves to absoluteness
proofs because they use the powerset operator.  We must find an
alternative definition, and an obvious one is based on the intuition
\[ x\prec^*y \iff x=s_0\prec s_1\prec\cdots\prec s_n=y. \]
The sequence $s_0, s_1,\ldots, s_n$ can be modelled as a finite
function: as noted in Section \ref{sec:M_basic}, finite functions
are absolute.  From $x\prec^*y$ it is trivial to define the transitive
closure, $x\prec^+y$.  In the definition below,
\isa{f} is the sequence and
\isa{A} is intended to represent the field of~\isa{r}:
\begin{defsession}
"rtrancl\_alt(A,r)\ ==\isanewline
\ \ \ \isacharbraceleft p\ \isasymin \ A*A.\ \isasymexists n\isasymin nat.\ \isasymexists f\ \isasymin \ succ(n)\ ->\ A.\isanewline
\ \ \ \ \ \ \ \ \ \ \ \ \ (\isasymexists x\ y.\ p\ =\ \isasymlangle x,y\isasymrangle\ \&\ \ f`0\ =\ x\ \&\ f`n\ =\ y)\ \&\isanewline
\ \ \ \ \ \ \ \ \ \ \ \ \ \ \ \ \ \ \ (\isasymforall i\isasymin n.\ \isasymlangle f`i,\ f`succ(i)\isasymrangle\ \isasymin \ r)\isacharbraceright
"
\end{defsession}
It is easy to prove that this definition coincides with
Isabelle/ZF's inductive one:
\begin{isabelle}\footnotesize
"rtrancl\_alt(field(r),r)\ =\ r\isacharcircum *"
\end{isabelle}
Since every concept used in the new definition is absolute, we merely
have to relativize this definition to~\isa{M}, defining 
\isa{rtran\_closure\_mem(M,A,r,p)} to hold when \isa{p} is
an element of
\isa{rtrancl\_alt(A,r)}.  I omit the definition because the relational
language is unreadable.  We cannot even use the constant~\isa{0} but
must introduce a variable
\isa{zero} and constrain it by \isa{empty(M,zero)}.

The next two predicates relativize the reflexive-transitive and
transitive closure of a relation:
\begin{defsession}
"rtran\_closure(M,r,s)\ ==\isanewline
\ \ \ \ \isasymforall A[M].\ is\_field(M,r,A)\ \isasymlongrightarrow\isanewline
\ \ \ \ \ (\isasymforall p[M].\ p\ \isasymin \ s\ \isasymlongleftrightarrow\ rtran\_closure\_mem(M,A,r,p))"\isanewline
"tran\_closure(M,r,t)\ ==\isanewline
\ \ \ \ \ \isasymexists s[M].\ rtran\_closure(M,r,s)\ \&\
composition(M,r,s,t)"
\end{defsession}
Once we assume an instance of separation involving
\isa{rtran\_closure\_mem}, closure and absoluteness results follow
directly:
\begin{isabelle}\footnotesize
"M(r)\ \isasymLongrightarrow\ M(rtrancl(r))"\isanewline
"\isasymlbrakk M(r);\ M(z)\isasymrbrakk \ \isasymLongrightarrow\ rtran\_closure(M,r,z)\ \isasymlongleftrightarrow\ z\ =\
rtrancl(r)"\isanewline
"M(r)\ \isasymLongrightarrow\ M(trancl(r))"\isanewline
"\isasymlbrakk M(r);\ M(z)\isasymrbrakk \ \isasymLongrightarrow\ tran\_closure(M,r,z)\ \isasymlongleftrightarrow\ z\ =\
trancl(r)"
\end{isabelle}

If a relation is well-founded then so is its transitive closure.  The
following lemma use useful because at this point we do not know that
\isa{wellfounded(M,r)} is equivalent to \isa{wf(M,r)}.
\begin{isabelle}\footnotesize
"\isasymlbrakk wellfounded(M,r);\ M(r)\isasymrbrakk \ \isasymLongrightarrow\ wellfounded(M,r\isacharcircum
+)"
\end{isabelle}
After about 130 lines of proof script, we arrive at some important
theorems.  One asserts absoluteness, relating the predicate
\isa{is\_wfrec} with the operator \isa{wfrec}:
\begin{isabelle}\footnotesize
"\isasymlbrakk wf(r);\ \ trans(r);\ \ relation(r);\ \ M(r);\ \ M(a);\ \ M(z);\isanewline
\ \ wfrec\_replacement(M,MH,r);\ \ relation2(M,MH,H);\isanewline
\ \ \isasymforall x[M].\ \isasymforall g[M].\ function(g)\
\isasymlongrightarrow\ M(H(x,g))\isasymrbrakk \isanewline
\ \isasymLongrightarrow\ is\_wfrec(M,MH,r,a,z)\ \isasymlongleftrightarrow\ z=wfrec(r,a,H)"
\end{isabelle}
Another states that the class \isa{M} is closed under well-founded
recursion:
\begin{isabelle}\footnotesize
"\isasymlbrakk wf(r);\ trans(r);\ relation(r);\ M(r);\ M(a);\isanewline
\ \ wfrec\_replacement(M,MH,r);\ \ relation2(M,MH,H);\isanewline
\ \ \isasymforall x[M].\ \isasymforall g[M].\ function(g)\
\isasymlongrightarrow\ M(H(x,g))\isasymrbrakk \isanewline
\ \isasymLongrightarrow\ M(wfrec(r,a,H))"
\end{isabelle}
The theorems fortunately require identical instances of replacement.
Both theorems assume \isa{trans(r)}; omitted are more general theorems
that relax the assumption of transitivity.

Both theorems use the predicate \isa{wfrec\_replacement} to
express a necessary instance of replacement.  Its arguments are the
ternary predicate \isa{MH}, which represents the body of the recursive
function, and the well-founded relation~\isa{r}.
\begin{defsession}
"wfrec\_replacement(M,MH,r)\ ==\isanewline
\ \ \ \ \ strong\_replacement(M,\isanewline
\ \ \ \ \ \ \ \ \ \ \isasymlambda x\ z.\ \isasymexists y[M].\ pair(M,x,y,z)\ \&\
is\_wfrec(M,MH,r,x,y))"
\end{defsession}

\section{Defining First-Order Formulae and the Class $\BL$}
\label{sec:defining-l}

We pause from proving absoluteness results in order to consider our
main objective, namely the class $\BL$ and its properties.  The most
logical order of presentation might have been to develop $\BL$ first
and then to prove that constructibility is absolute.  The order of
presentation adopted here better represents how I actually carried out
the proofs.  Kunen similarly presents general absoluteness results
before he introduces~$\BL$.

\subsection{Internalized First-Order Formulae}

The idea of $\BL$ is to introduce, at each stage, the sets that can be
defined from existing ones by a first-order formula with parameters. 
Neither G\"odel~\cite{goedel40} nor Kunen actually use first-order formulae, preferring
more abstract constructions that achieve the goal more easily. 
However, Isabelle/ZF's recursive datatype package automates the task of
defining the set of first-order formulae and the satisfaction relation
on them.  G\"odel's earlier proof~\cite{goedel39} also uses
first-order formulae.

The obvious representation of first-order formulae is de
Bruijn's~\cite{debruijn72}, where there are no variable names. 
Instead, each variable reference is a non-negative integer, where zero
refers to the innermost quantifier and larger numbers refer to
enclosing quantifiers.  If the integer is greater than the number of
enclosing quantifiers, than it is a free variable.  This
representation eliminates the danger of name confusion.  It is
particularly useful for formulae with parameters, since their order
is determined numerically rather than by name.  
\begin{defsession}
\isacommand{datatype}
\ \ "formula"\ =\ Member\ ("x\ \isasymin\ nat",\ "y\ \isasymin\ nat")\isanewline
\ \ \ \ \ \ \ \ |\ Equal\ \ ("x\ \isasymin\ nat",\ "y\ \isasymin\ nat")\isanewline
\ \ \ \ \ \ \ \ |\ Nand\ ("p\ \isasymin\ formula",\ "q\ \isasymin\ formula")\isanewline
\ \ \ \ \ \ \ \ |\ Forall\ ("p\ \isasymin\ formula")
\end{defsession}
Having only four cases simplifies the relativization of functions on
formulae.  All propositional connectives are expressed in terms of
\isa{Nand}. 
\begin{defsession}
\ \ \ \ "Neg(p)\ ==\ Nand(p,p)"\isanewline
\ \ \ \ "And(p,q)\ ==\ Neg(Nand(p,q))"\isanewline
\ \ \ \ "Or(p,q)\ ==\ Nand(Neg(p),Neg(q))"\isanewline
\ \ \ \ "Implies(p,q)\ ==\ Nand(p,Neg(q))"\isanewline
\ \ \ \ "Iff(p,q)\ ==\ And(Implies(p,q),\ Implies(q,p))"\isanewline
\ \ \ \ "Exists(p)\ ==\ Neg(Forall(Neg(p)))"
\end{defsession}

\subsection{The Satisfaction Relation}
\label{sec:sats}

Satisfaction is a primitive recursive function on formulae.  Thanks to
the nameless representation, interpretations are simply lists rather
than functions from variable names to values.  The familiar
list function
\isa{nth}, defined below, looks up variables in
interpretations: 
\begin{defsession}
"nth(0, Cons(a, l)) = a"\isanewline
"nth(succ(n), Cons(a,l)) = nth(n,l)"\isanewline
"nth(n, Nil) = 0"
\end{defsession}
The second of these equations is subject to the condition \isa{n\
\isasymin
\ nat}.  Note that element zero is the head of the list.
Another useful function is \isa{bool\_of\_o}, which converts a
truth value to an integer:
\begin{defsession}
"bool\_of\_o(P) == (if P then 1 else 0)"
\end{defsession}
This conversion is necessary because Isabelle/ZF is based on
first-order logic.  Formulae are not values, so we encode them using
integers.  We thus define a recursive predicate as a
recursive integer-valued function.  We are now able to define the
function
\isa{satisfies}, which takes a set (the domain of discourse), a formula
and an interpretation (written \isa{env} for environment).  It returns
1 or 0, depending upon whether or not the formula evaluates to true or
false:
\begin{defsession}
"satisfies(A,Member(x,y))\ =\isanewline
\ \ \ \ (\isasymlambda env\ \isasymin \ list(A).\ bool\_of\_o\ (nth(x,env)\ \isasymin \ nth(y,env)))"\isanewline
"satisfies(A,Equal(x,y))\ =\isanewline
\ \ \ \ (\isasymlambda env\ \isasymin \ list(A).\ bool\_of\_o\ (nth(x,env)\ =\ nth(y,env)))"\isanewline
"satisfies(A,Nand(p,q))\ =\isanewline
\ \ \ \ (\isasymlambda env\ \isasymin \ list(A).\ not\ ((satisfies(A,p)`env)\ and\isanewline
\ \ \ \ \ \ \ \ \ \ \ \ \ \ \ \ \ \ \ \
\ \ \ \ \ \ (satisfies(A,q)`env)))"\isanewline
"satisfies(A,Forall(p))\ =\isanewline
\ \ \ \ (\isasymlambda env\ \isasymin \ list(A).\ bool\_of\_o\isanewline
\ \ \ \ \ \ \ \ \ \ \ \ \ \ \ \ \ \ \ \ \ \ (\isasymforall
x\isasymin A.\ satisfies(A,p)`(Cons(x,env))\ =\ 1))"
\end{defsession}

The abstraction and explicit function applications involving
environments are necessary because the environments can vary in the
recursive calls.   The last line of
\isa{satisfies} deserves attention.  The universal
formula \isa{Forall(p)} evaluates to 1 just if \isa{p} evaluates to 1
in every environment obtainable from \isa{env} by adding an
element of~\isa{A}\@.  Such environments have the form
\isa{Cons(x,env)} for \isa{x\isasymin A}.

The satisfaction predicate, \isa{sats}, is a macro that
refers to the function \isa{satisfies}. 
\begin{defsession}
\isacommand{translations}\ "sats(A,p,env)"\ ==\ "satisfies(A,p)`env\
=\ 1"
\end{defsession}

The satisfaction predicate enjoys a number of properties that
relate the internalized formulae to real formulae.  All the
equivalences are subject to the typing condition
\isa{env\ \isasymin \ list(A)}.  For example, the membership and
equality relations behave as they should:
\begin{isabelle}\footnotesize
"sats(A,\ Member(x,y),\ env)\ \isasymlongleftrightarrow\ nth(x,env)\ \isasymin
\ nth(y,env)"\isanewline 
"sats(A,\ Equal(x,y),\ env)\ \isasymlongleftrightarrow\ nth(x,env)\
=\ nth(y,env)"
\end{isabelle}

The propositional connectives also work:
\begin{isabelle}\footnotesize
"sats(A,\ Neg(p),\ env)\ \isasymlongleftrightarrow\
\isachartilde \ sats(A,p,env)"\isanewline 
"(sats(A,\ And(p,q),\ env))\
\isasymlongleftrightarrow\ sats(A,p,env)\ \&\ sats(A,q,env)"\isanewline 
"(sats(A,\ Or(p,q),\
env))\ \isasymlongleftrightarrow\ sats(A,p,env)\ |\ sats(A,q,env)" 
\end{isabelle}

Quantifiers work too.  Notice how the environment is extended:
\begin{isabelle}\footnotesize
"sats(A,\ Exists(p),\
env)\ \isasymlongleftrightarrow\ (\isasymexists x\isasymin A.\ sats(A,\ p,\
Cons(x,env)))"
\end{isabelle}

\subsection{The Arity of a Formula}

The arity of a formula is, intuitively, its set of free variables.  In
\isa{sats(A,p,env)}, if the arity of \isa{p} does not exceed the
length of \isa{env}, then the environment supplies values to all of
\isa{p}'s free variables.

Take each de Bruijn reference, adjusted for the depth of quantifier
nesting at that point; the
arity is the maximum of the resulting values.
The recursive definition of function \isa{arity} is simpler than
this description.
\begin{defsession}
"arity(Member(x,y))\ =\ succ(x)\ \isasymunion \ succ(y)"\isanewline
"arity(Equal(x,y))\ =\ succ(x)\ \isasymunion \ succ(y)"\isanewline
"arity(Nand(p,q))\ =\ arity(p)\ \isasymunion \ arity(q)"\isanewline
"arity(Forall(p))\ =\ Arith.pred(arity(p))"
\end{defsession}
Note that $m\un n=\max\{m,n\}$ in set theory and that
\isa{Arith.pred} denotes the predecessor function.  
Trivial corollaries of this definition tell us how to compute the
arities of other connectives:
\begin{isabelle}\footnotesize
"arity(Neg(p))\ =\ arity(p)"\isanewline
"arity(And(p,q))\ =\ arity(p)\ \isasymunion \ arity(q)"
\end{isabelle}

The following result is more interesting.
Extra items in the environment (exceeding the arity) are ignored. 
Here \isa{@} is the list ``append'' operator, so \isa{env\
@ \ extra} is \isa{env} with additional items added.
\begin{isabelle}\footnotesize
"\isasymlbrakk arity(p)\ \isasymle \ length(env);\ p\ \isasymin \
formula;\isanewline
\ \ env\ \isasymin \ list(A); extra\
\isasymin \ list(A)\isasymrbrakk \isanewline
\ \isasymLongrightarrow\ sats(A,\ p,\ env@extra)\ \isasymlongleftrightarrow\ sats(A,\ p,\
env)"
\end{isabelle}

\subsection{Renaming (Renumbering) Free Variables}
\label{sec:renaming}

If $A$ is a set, then the subset
\[ \{x\in A\mid\phi(x,a_1,\ldots,a_m)\} \]
is determined by the choice of $\phi$ and of the
parameters $a_1$, \ldots, $a_m$, which are elements of~$A$.  These are
the definable subsets of~$A$.  

Now, consider the problem of showing
that the definable sets are closed under intersection.  Suppose
another subset of~$A$ is defined by a formula $\psi$ and parameters 
$a_{m+1}$, \ldots, $a_{m+n}$:
\[ \{x\in A\mid\psi(x,a_{m+1},\ldots,a_{m+n})\} \]
Then, their intersection can presumably be defined by
\[ \{x\in A\mid\phi(x,a_1,\ldots,a_m)\conj
               \psi(x,a_{m+1},\ldots,a_{m+n})\} 
\]
Our aim is to regard the conjunction $\phi\conj\psi$ as having the
free variables $x$, $a_1$, \ldots, $a_n$.  The occurrences
of $x$ in both formulae must be identified, while the parameter lists
of the two formulae must be kept disjoint.  To achieve our aim may
require renaming one of the formula's free variables.

The de Bruijn representation refers to variables by number rather
than by name.  The variables shown as $x$ above always have the de
Bruijn index zero, so they will be identified automatically.  We
keep the parameter lists disjoint by renumbering
the free variables in one of the
formulae.  Since $x$ must be left alone, we only renumber the variables
having an index greater than zero.

Renumbering functions are often necessary with the de Bruijn
approach, though normally they rename variables during substitution. 
When efficiency matters, the renumbering functions take an argument
specifying what number should be added to the variables.  Here, the
definitions are for reasoning about rather than for execution, so
renaming for us means adding one; repeating this allows renaming by
larger integers.
In the following definitions, \isa{nq} refers to the number of
quantifiers enclosing the current point.  Any de Bruijn index
smaller than \isa{nq} must not be renamed.

\subsubsection{The Renaming Function}

First, we need a one-line function that renames a de Bruijn
variable:
\begin{defsession}
"incr\_var(x,nq)\ ==\ if\ x<nq\ then\ x\ else\
succ(x)"
\end{defsession}
Now we can define the main renaming function.  As with
\isa{satisfies} above, abstraction and explicit function
applications  are necessary: the argument \isa{nq} (``nesting of
quantifiers'') varies in the recursive calls.  In the \isa{Member} and
\isa{Equal} case, the variables are simply renamed.  The \isa{Nand} case
recursively renames the subformulae using the same nesting depth, while
the
\isa{Forall} case renames its subformula using an increased nesting
depth.
\begin{defsession}
"incr\_bv(Member(x,y))\ =\isanewline
\ \ \ \ (\isasymlambda nq\ \isasymin \ nat.\ Member\ (incr\_var(x,nq),\ incr\_var(y,nq)))"\isanewline
\isanewline
"incr\_bv(Equal(x,y))\ =\isanewline
\ \ \ \ (\isasymlambda nq\ \isasymin \ nat.\ Equal\ (incr\_var(x,nq),\ incr\_var(y,nq)))"\isanewline
\isanewline
"incr\_bv(Nand(p,q))\ =\isanewline
\ \ \ \ (\isasymlambda nq\ \isasymin \ nat.\ Nand\ (incr\_bv(p)`nq,\ incr\_bv(q)`nq))"\isanewline
\isanewline
"incr\_bv(Forall(p))\ =\isanewline
\ \ \ \ (\isasymlambda nq\ \isasymin \ nat.\ Forall\ (incr\_bv(p)\ `\
succ(nq)))"
\end{defsession}

Recall the example at the start of this section, concerning a set
defined by the conjunction $\phi\conj\psi$.  If we are to conjoin the
formulae $\phi$ and $\psi$ and combine their sets of parameters, then
we need to ensure that some of the parameters are only visible to
$\phi$ and the rest are only visible to $\psi$.  The following lemma
makes this possible:
\begin{isabelle}\footnotesize
"\isasymlbrakk p\ \isasymin \ formula;\ bvs\ \isasymin \
list(A);\ env\ \isasymin \ list(A);\ x\ \isasymin \ A\isasymrbrakk
\isanewline
\ \isasymLongrightarrow\ sats(A,\ incr\_bv(p)\ `\ length(bvs),\ bvs\ @\
Cons(x,env))\ \isasymlongleftrightarrow\isanewline
\ \ \ \ \ sats(A,\ p,\ bvs@env)"
\end{isabelle}
For the intuition, suppose that \isa{bvs} is the list
$[x_0,\ldots,x_{m-1}]$ (and therefore has length \isa{m}).  Then
the conclusion essentially says
\begin{isabelle}\footnotesize
\ \ \ \ \ sats(A,\ incr\_bv(p)\ `\ m,\
\([x_0,\ldots,x_{m-1},x,x_m,\ldots,x_n]\))\ \isasymlongleftrightarrow\isanewline
\ \ \ \ \ sats(A,\ p,\ \([x_0,\ldots,x_{m-1},x_m,\ldots,x_n]\))"
\end{isabelle}
and thus the renaming allows an additional value to be put into the
environment at position~\isa{m}.  The renamed formula will ignore
the new value.  By repeated renaming, we can construct a formula that
will ignore a section of the parameter list that is intended for
another formula.

The next result describes the obvious relationship between
\isa{arity} and renaming.  Renaming increases a formula's arity by
one, unless the variable being renamed does not exist, when renaming
has no effect.
\begin{isabelle}\footnotesize
"\isasymlbrakk p\ \isasymin \ formula;\ n\ \isasymin \ nat\isasymrbrakk
\isanewline
\ \isasymLongrightarrow\ arity\ (incr\_bv(p)\ `\ n)\ =\isanewline
\ \ \ \ \ (if\ n\ <\ arity(p)\ then\ succ(arity(p))\
else\ arity(p))"
\end{isabelle}
Considering how trivial the notion of arity is, many proofs about it
(including this one) are complicated by innumerable case splits. 
Getting the simplifier to prove most of them automatically requires
some ingenuity.  Many other tiresome proofs about arities are
omitted.  

\subsubsection{Renaming all but the first bound variable}

One more thing is needed before we can define sets using 
conjunctions.  As discussed at the beginning of
Sect.\ts\ref{sec:renaming}, when a formula $\phi$ defines a set, the
variable with de Bruijn index zero gives the extension
of that set, while the remaining free variables serve as parameters.  
Therefore, our basic renaming operator must only rename variables
having a de Bruijn index of one or more:
\begin{defsession}
"incr\_bv1(p)\ ==\ incr\_bv(p)`1"
\end{defsession}

Finally we reach a lemma justifying our intended use of renaming.
\begin{isabelle}\footnotesize
"\isasymlbrakk p\ \isasymin \ formula;\ bvs\ \isasymin \
list(A);\ x\
\isasymin \ A;\ env\ \isasymin \ list(A);\isanewline
\ \ length(bvs)\ =\
n\isasymrbrakk \isanewline
\ \isasymLongrightarrow\ sats(A,\ iterates(incr\_bv1,\ n,\ p),\ Cons(x,\ bvs@env))\ \isasymlongleftrightarrow\isanewline
\ \ \ \ \ sats(A,\ p,\ Cons(x,env))"
\end{isabelle}
If the environment has an initial segment \isa{bvs} of length
\isa{n} and if we apply the \isa{incr\_bv1} \isa{n} times, then the
modified formula ignores the \isa{bvs} part.  But the renamed and
original formulae agree on the first element of the environment,
shown above as~\isa{x}.

\subsection{The Definable Powerset Operation}
\label{sec:DPow}

The definable powerset operator is called \isa{DPow}:
\begin{defsession}
"DPow(A)\ ==\ \isacharbraceleft X\ \isasymin \ Pow(A).\isanewline
\ \ \ \ \ \ \ \ \ \ \ \ \ \isasymexists env\ \isasymin \ list(A).\ \isasymexists p\ \isasymin \ formula.\isanewline
\ \ \ \ \ \ \ \ \ \ \ \ \ \ \ arity(p)\ \isasymle \ succ(length(env))\ \&\isanewline
\ \ \ \ \ \ \ \ \ \ \ \ \ \ \ X\ =\ \isacharbraceleft x\isasymin A.\ sats(A,\ p,\ Cons(x,env))\isacharbraceright \isacharbraceright
"
\end{defsession}
A set \isa{X} belongs to \isa{DPow(A)} provided there is an
environment~\isa{env} (a list of values drawn from~\isa{A}) and a
formula~\isa{p}.  The constraint \isa{arity(p)\ \isasymle \
succ(length(env))} indicates that the environment should interpret all
but one of
\isa{p}'s free variables.  The variable whose de Bruijn index is zero
determines the extension of~\isa{X} via the satisfaction relation:
\isa{sats(A,\ p,\ Cons(x,env))}.  You may want to compare this with the
informal discussion in the previous section, or with Definition VI 1.1
of Kunen~\cite[p.\ts165]{kunen80}.

Some consequences of this definition are easy to prove.  The empty set
is defined by the predicate $\lambda x. x\neq x$, and singleton sets by
$\lambda x.\, x=a$.
\begin{isabelle}\footnotesize
"0\ \isasymin \ DPow(A)"\isanewline
"a\ \isasymin \ A\ \isasymLongrightarrow\
\isacharbraceleft a\isacharbraceright \ \isasymin \ DPow(A)"
\end{isabelle}

The complement of a set~\isa{X} is defined by negating the formula
used to define~\isa{X}\@.  Intersection is done by conjoining the
defining formulae, using the renaming techniques developed in the
previous section.  Union is then trivial by de Morgan's laws. 
\begin{isabelle}\footnotesize
"X\ \isasymin \ DPow(A)\ \isasymLongrightarrow\ (A-X)\ \isasymin \ DPow(A)"\isanewline
"\isasymlbrakk X\ \isasymin \ DPow(A);\ Y\ \isasymin \
DPow(A)\isasymrbrakk \ \isasymLongrightarrow\ X\ Int\ Y\ \isasymin \ DPow(A)"\isanewline
"\isasymlbrakk X\ \isasymin \ DPow(A);\ Y\ \isasymin \
DPow(A)\isasymrbrakk \ \isasymLongrightarrow\ X\ Un\ Y\ \isasymin \ DPow(A)"
\end{isabelle}
And thus \isa{DPow} coincides with \isa{Pow} (the real
powerset operator) for finite sets:
\begin{isabelle}\footnotesize
"Finite(A)\ \isasymLongrightarrow\ DPow(A)\ =\ Pow(A)"
\end{isabelle}

\subsection{Proving that the Ordinals are Definable}
\label{sec:subset_fm}

In order to show that \isa{DPow} is closed under other operations,
we must be able to code their defining formulae as elements of the set
\isa{formula}.  The treatment of the subset relation is typical. 
We begin by encoding the formula $\forall z.\,z\in x\imp z\in y$. 
Below, \isa{x} and \isa{y} are de Bruijn indices, which are
incremented to \isa{succ(x)} and \isa{succ(x)} because the quantifier
introduces a new variable binding.
\begin{defsession}
"subset\_fm(x,y)\ ==\isanewline
\ \ \ Forall(Implies(Member(0,succ(x)),\
Member(0,succ(y))))"
\end{defsession}
The arguments are just de Bruijn indices because internalized
formulae have no terms other than variables.  It is trivial to prove
that
\isa{subset\_fm} maps a pair of de Bruijn indices to a formula:
\begin{isabelle}\footnotesize
"\isasymlbrakk x\ \isasymin \
nat;\ y\ \isasymin \ nat\isasymrbrakk \ \isasymLongrightarrow\ subset\_fm(x,y)\
\isasymin \ formula"
\end{isabelle}
The arity of the formula is the maximum of those of its
operands:
\begin{isabelle}\footnotesize
"\isasymlbrakk x\ \isasymin \ nat;\ y\ \isasymin \ nat\isasymrbrakk \ \isasymLongrightarrow\ arity(subset\_fm(x,y))\ =\ succ(x)\ \isasymunion \
succ(y)"
\end{isabelle}

The following equivalence involves absoluteness, since it
relates
\isa{subset\_fm} to the real subset relation, \isa{\isasymsubseteq}.
To reach this conclusion requires the
additional assumption \isa{Transset(A)}, saying that
\isa{A} is a transitive set.  The premise \isa{x\ <\
length(env)} puts a bound on~\isa{x} (which is a de Bruijn index),
ensuring that \isa{nth(x,env)} belongs to~\isa{A}\@.
\begin{isabelle}\footnotesize
"\isasymlbrakk x\ <\ length(env);\ y\ \isasymin \ nat;\ env\ \isasymin \ list(A);\ Transset(A)\isasymrbrakk \isanewline
\ \isasymLongrightarrow\ sats(A,\ subset\_fm(x,y),\ env)\ \isasymlongleftrightarrow\ nth(x,env)\
\isasymsubseteq \ nth(y,env)"
\end{isabelle}

We must repeat this exercise (details omitted) for the concepts of
transitive set and ordinal.  This lets us prove that ordinals are
definable, leading to a result involving ordinals and \isa{DPow}.
\begin{isabelle}\footnotesize
"Transset(A)\ \isasymLongrightarrow\ \isacharbraceleft x\ \isasymin \ A.\
Ord(x)\isacharbraceright \ \isasymin \ DPow(A)"
\end{isabelle}
This lemma ultimately leads to a proof that $\BL$ contains all
the ordinals.

\subsection{Defining $\BL$, The Constructible Universe}

The constant \isa{Lset} formalizes the family of sets
$\{L_\alpha\}_{\alpha\in\BON}$.  Its definition in Isabelle/ZF uses a
standard operator for transfinite recursion.  We also define 
$\BL = \union_{\alpha\in\BON}\, L_\alpha$:
\begin{defsession}
"Lset(i)\ ==\ transrec(i,\ \%x\ f.\ \isasymUnion y\isasymin x.\
DPow(f`y))"\isanewline
"L(x)\ ==\ \isasymexists i.\ Ord(i)\ \&\ x\ \isasymin \ Lset(i)"
\end{defsession}
Some effort is required before we can transform the cryptic
definition of \isa{Lset} into the usual recursion equations.  First, we
prove Kunen's~\cite[p.\ts167]{kunen80} lemma VI 1.6, which states the
transitivity and monotonicity of the $L_\alpha$:
\begin{isabelle}\footnotesize
"Transset(A)\ \isasymLongrightarrow\ Transset(DPow(A))"\isanewline
"Transset(Lset(i))"\isanewline
"i\isasymle j\ \isasymlongrightarrow\ Lset(i)\ \isasymsubseteq\ Lset(j)"
\end{isabelle}
Then we reach the 0, successor and limit equations for the $L_\alpha$:
\begin{isabelle}\footnotesize
"Lset(0)\ =\ 0"\isanewline
"Lset(succ(i))\ =\ DPow(Lset(i))"\isanewline
"Limit(i)\ \isasymLongrightarrow\ Lset(i)\ =\ (\isasymUnion y\isasymin i.\ Lset(y))"
\end{isabelle}
The basic properties of $\BL$, as presented in Kunen's IV \S1, are not
hard to prove.  For example, $\BL$ contains the ordinals:
\begin{isabelle}\footnotesize
"Ord(i)\ \isasymLongrightarrow\ i\ \isasymin \ Lset(succ(i))"\isanewline
"Ord(i)\ \isasymLongrightarrow\ L(i)"
\end{isabelle}

\subsection{Eliminating the Arity Function}
\label{sec:no-arity}

The function \isa{arity} can be surprisingly hard
to reason about, particularly when we try to encode higher-order
operators.  Once we have established the basic properties of
$\BL$, we can prove its equivalence to a new definition that does not
involve arities.  Here is another form of definable powerset:
\begin{defsession}
\ \ "DPow'(A)\ ==\ \isacharbraceleft X\ \isasymin \ Pow(A).\isanewline
\ \ \ \ \ \ \ \ \ \ \ \ \ \ \ \ \isasymexists env\ \isasymin \ list(A).\ \isasymexists p\ \isasymin \ formula.\isanewline
\ \ \ \ \ \ \ \ \ \ \ \ \ \ \ \ \ \ \ \ X\ =\ \isacharbraceleft x\isasymin A.\ sats(A,\ p,\ Cons(x,env))\isacharbraceright \isacharbraceright
"
\end{defsession}
This version omits the constraint \isa{arity(p)\ \isasymle \
succ(length(env))} but is otherwise identical to \isa{DPow}.
The point is that if the environment is too short, attempted variable
lookups will yield zero; recall the properties of \isa{nth} from
Sect.\ts\ref{sec:sats}.  If the set \isa{A} is transitive, then it
contains zero as an element.  So the too-short environment can be
padded to the right with zeroes.
\begin{isabelle}\footnotesize
"Transset(A)\ \isasymLongrightarrow\ DPow(A)\ =\ DPow'(A)"
\end{isabelle}
Each \isa{Lset(i)} is a transitive set, so they can be expressed
using \isa{DPow'} rather than \isa{DPow}: 
\begin{isabelle}\footnotesize
"Lset(i)\ =\
transrec(i,\ \%x\ f.\ \isasymUnion y\isasymin x.\ DPow'\ (f\ `\ y))"
\end{isabelle}
The equation above, proved by transfinite induction, lets us
relativize
\isa{Lset} without having to formalize the functions \isa{arity} and
\isa{length}.  That eliminates a lot of work.

The following lemma is helpful for proving instances of separation. 
The first, quantified, premise asks
for an equivalence between the real formula~\isa{P} and the internalized
formula~\isa{p}.  Often we can derive~\isa{p} from~\isa{P}
automatically by supplying a set of suitable inference rules.
\begin{isabelle}\footnotesize
"\isasymlbrakk \isasymforall x\isasymin Lset(i).\ P(x)\ \isasymlongleftrightarrow\ sats(Lset(i),\ p,\ Cons(x,env));\isanewline
\ \ env\ \isasymin \ list(Lset(i));\ \ p\ \isasymin \
formula\isasymrbrakk \isanewline
\ \isasymLongrightarrow\ \isacharbraceleft x\isasymin Lset(i).\ P(x)\isacharbraceright \ \isasymin \
DPow(Lset(i))"
\end{isabelle}
Also, the lemma makes no reference to \isa{arity}, thanks to the
equivalence between \isa{DPow'} and \isa{DPow}.

\subsection{The Zermelo-Fraenkel Axioms Hold in $\BL$}

Following Kunen VI \S2, it is possible to prove that $\BL$ satisfies
the Zermelo-Fraenkel axioms.  Separation is the most difficult case
and is considered later.

\subsubsection{Basic Properties of $\BL$}

We begin with simple closure properties.  Many of
them involve exhibiting an element of \isa{formula} describing the
required set.  We typically begin by starting in
\isa{Lset(i)} and proving that the required set belongs to
\isa{Lset(succ(i))}. 

$\BL$ is closed under unions:
\begin{isabelle}\footnotesize
"X\ \isasymin \ Lset(i)\ \isasymLongrightarrow\ Union(X)\ \isasymin \
Lset(succ(i))"\isanewline
"L(X)\ \isasymLongrightarrow\ L(Union(X))"
\end{isabelle}

$\BL$ is closed under unordered pairs.  More work is necessary because
the sets \isa{a} and \isa{b} may be introduced at different ordinals:
\begin{isabelle}\footnotesize
"a\ \isasymin\ Lset(i)\ \isasymLongrightarrow\ \isacharbraceleft a\isacharbraceright \ \isasymin\ Lset(succ(i))"\isanewline
"\isasymlbrakk a\ \isasymin\ Lset(i);\ \ b\ \isasymin\ Lset(i)\isasymrbrakk \ \isasymLongrightarrow\
\isacharbraceleft a,b\isacharbraceright \ \isasymin\ Lset(succ(i))"\isanewline
"\isasymlbrakk a\ \isasymin\ Lset(i);\ \ b\ \isasymin\ Lset(i);\ \ Limit(i)\isasymrbrakk \
\isasymLongrightarrow\ \isacharbraceleft a,b\isacharbraceright \ \isasymin\ Lset(i)"\isanewline
"\isasymlbrakk L(a);\
L(b)\isasymrbrakk \ \isasymLongrightarrow\ L(\isacharbraceleft a,\ b\isacharbraceright )"
\end{isabelle}

Also, $L_\alpha$ is closed
under ordered pairs provided $\alpha$ is a limit ordinal.  This result
is needed in order to apply the reflection theorem  to $\BL$.
Specifically, it is needed because my version
of the reflection theorem~\cite{paulson-reflection} uses ordered pairs
to cope with the possibility of a formula having any number of free
variables.
\begin{isabelle}\footnotesize
"\isasymlbrakk a\ \isasymin\ Lset(i);\ \ b\ \isasymin\ Lset(i);\ \ Ord(i)\isasymrbrakk \isanewline
\ \isasymLongrightarrow\ \isasymlangle a,b\isasymrangle\ \isasymin\ Lset(succ(succ(i)))"\isanewline "\isasymlbrakk a\ \isasymin\ Lset(i);\ \ b\ \isasymin\ Lset(i);\ \ Limit(i)\isasymrbrakk \ \isasymLongrightarrow\ \isasymlangle a,b\isasymrangle\ \isasymin\ Lset(i)"
\end{isabelle}

\subsubsection{A Rank Function for $\BL$}
\label{sec:lrank}

Some proofs require the $\BL$-rank operator.  Kunen
(VI 1.7) defines $\rho(x)$ to denote the least $\alpha$ such that
$x\in L_{\alpha+1}$:
\begin{defsession}
"lrank(x)\ ==\ \isasymmu i.\ x\ \isasymin \
Lset(succ(i))"
\end{defsession}
Here is one consequence of this definition:
\begin{isabelle}\footnotesize
"Ord(i)\ \isasymLongrightarrow\ x\ \isasymin\ Lset(i)\ \isasymlongleftrightarrow\ L(x)\ \&\ lrank(x)\ \isasymlangle \ i"
\end{isabelle}
A more important result, whose proof involves \isa{lrank}, states
that every set of constructible sets is included in some \isa{Lset}:
\begin{isabelle}\footnotesize
"(\isasymforall x\isasymin A.\ L(x))\ \isasymLongrightarrow\ \isasymexists i.\ Ord(i)\ \&\ A\ \isasymsubseteq \
Lset(i)"
\end{isabelle}
This theorem is useful in proving that $\BL$ satisfies
the separation axiom.  However, note that $A\subseteq\BL$ does not
imply $A\in\BL$, not even if $A$ is a set of natural numbers.

The \isa{lrank} operator is useful for proving that \isa{L}
satisfies the powerset axiom:
\begin{isabelle}\footnotesize
"L(X)\ \isasymLongrightarrow\ L(\isacharbraceleft y\
\isasymin \ Pow(X).\ L(y)\isacharbraceright)"
\end{isabelle}
Note that the powerset of \isa{X} in \isa{L} comprises all subsets of
\isa{X} that belong to~\isa{L}\@.  It is potentially a superset of 
\isa{DPow(X)}.

The \isa{lrank} operator also assists in the proof that \isa{L}
satisfies the replacement axiom.  The idea is to use 
replacement on the ranks of the members of \isa{L}:
\begin{isabelle}\footnotesize
"\isasymlbrakk L(X);\ univalent(L,X,Q)\isasymrbrakk \isanewline
\ \isasymLongrightarrow\ \isasymexists Y.\ L(Y)\ \&\ Replace(X,\ \%x\ y.\ Q(x,y)\ \&\ L(y))\ \isasymsubseteq \
Y"
\end{isabelle}
The proof of replacement is schematic, and therefore independent of the
formula~\isa{Q}\@.  But it is the weak form of replacement.  It
concludes that the range of \isa{Q} (viewed as a class function) is
included in some constructible set~\isa{Y}\@.  Strong replacement,
which is the version we really want, asserts that the range itself is
constructible.  Each instance of strong replacement requires proving
an instance of the axiom of separation.

\subsubsection{Instantiating the Locale \isa{M\_trivial}}

Now we are ready to show that \isa{L} satisfies all the properties
we assumed of the class~\isa{M}, which we used to develop the general
theory of absoluteness.  The class \isa{L} is transitive:
\begin{isabelle}\footnotesize
"\isasymlbrakk y\isasymin x;\ L(x)\isasymrbrakk \ \isasymLongrightarrow\ L(y)"
\end{isabelle}
The facts about \isa{L} proved above can be summarized using the
relativized forms of the ZF axioms:
\begin{isabelle}\footnotesize
"Union\_ax(L)"\isanewline
"upair\_ax(L)"\isanewline
"power\_ax(L)"\isanewline
"replacement(L,P)"
\end{isabelle}
We do not need \isa{L} to satisfy the foundation axiom.  However, this
fact is a trivial consequence of the foundation axiom:
\begin{isabelle}\footnotesize
"foundation\_ax(L)"
\end{isabelle}
The theorems above are all we need to 
prove \isa{"PROP\ M\_trivial(L)"}.  This theorem makes all the results
proved in locale \isa{"M\_trivial"} available as
theorems about~\isa{L}\@.  In particular, the absoluteness and closure
results listed in Sect.\ts\ref{sec:M_trivial} above apply
to~\isa{L}\@.

\section{Comprehension in $\BL$} \label{sec:comprehension}

It remains to show that $\BL$ satisfies the axiom of separation.  The
proof requires the reflection theorem.
As described elsewhere~\cite{paulson-reflection}, my
formalization of that theorem applies to any
class $\BM = \union_{\alpha\in\BON}\,M_\alpha$, where the family
$\{M_\alpha\}_{\alpha\in\BON}$ is increasing
and continuous.  An additional condition is that if $\alpha$ is a
limit ordinal then $M_\alpha$ must be closed under ordered pairing.
Isabelle's locale mechanism captures these requirements, and we
can now instantiate the locale with the class $\BL =
\union_{\alpha\in\BON}\,L_\alpha$.  However, making it ready for
practical use requires additional work.

\subsection{The Reflection Relation}

The reflection theorem states
that if $\phi(x_1,\ldots,x_n)$ is a formula in $n$ variables 
then there exists a closed and unbounded class $\BC$ such that for
all $\alpha\in\BC$ and $x_1$,\ldots, $x_n\in M_\alpha$
we have
\[ \phi^\BM(x_1,\ldots,x_n) \iff \phi^{M_\alpha}(x_1,\ldots,x_n). \]
In fact, we only need the weaker conclusion that $\BC$ is unbounded,
which enables us to find a suitable $\alpha>\beta$ given any
ordinal~$\beta$.  

Applying the reflection theorem yields an Isabelle formula describing
the class~$\BC$.  These formulae may be interesting in the case of
small examples~\cite{paulson-reflection}, but in typical applications
they are huge.  The trivial proofs, which merely refer to other
instances of reflection, take minutes of computer time; the resulting
theorems amount to pages of text.  The obvious solution is to express
the reflection theorem using an existential quantifier, but classes
cannot be quantified over: they are formulae.  

Fortunately, Isabelle
makes a distinction between the object-logic (here first-order logic)
and the metalogic (a fragment of higher-order
logic)~\cite{paulson-found}.  I was able to formalize a
metaexistential quantifier.  It lies outside of first-order logic ---
in particular, Isabelle will reject any attempt to use it in
comprehensions.  However, it can be used in top-level assertions,
which is all we need.  We can now define the reflection relation
between two formulae \isa{P} and~\isa{Q}\@:
\begin{defsession}
\ \ "REFLECTS[P,Q]\ ==\isanewline
\ \ \ \ \ (??C.\ Closed\_Unbounded(C)\ \&\isanewline
\ \ \ \ \ \ \ \ \ \ \ \ (\isasymforall a.\ C(a)\ \isasymlongrightarrow\ (\isasymforall
x\
\isasymin \ Lset(a).\ P(x)\ \isasymlongleftrightarrow\ Q(a,x))))"
\end{defsession}
It relates the formulae  just if there exists a
class \isa{C} satisfying the conclusion of the reflection
theorem~\cite{paulson-reflection}.  That is, \isa{C} is a closed,
unbounded class of ordinals~$\alpha$ such that \isa{P} and \isa{Q}
agree on $L_\alpha$.  The existential quantifier, \isa{??C}, hides
the prohibitively large formula describing this class.
The following lemma illustrates the use of the reflection relation. 
Note that the quantification over classes has disappeared.
\begin{isabelle}\footnotesize
"\isasymlbrakk REFLECTS[P,Q];\ Ord(i)\isasymrbrakk \isanewline
\ \isasymLongrightarrow\ \isasymexists j.\ i<j\ \&\ (\isasymforall x\ \isasymin \ Lset(j).\ P(x)\ \isasymlongleftrightarrow\
Q(j,x))"
\end{isabelle}
If \isa{REFLECTS[P,Q]} and
\isa{i} is an ordinal then there exists a larger ordinal~\isa{j}
for which \isa{P} and
\isa{Q} agree.  Our choice of \isa{i} can make \isa{j} arbitrarily
large.

The general form of the reflection theorem uses the
relativization operator, which cannot be expressed in Isabelle/ZF\@. 
However, given a specific formula
$\phi$, we can generate an instance of the reflection theorem relating
$\phi^\BL$ and $\phi^{L_\alpha}$.  Here is the base case, where
normally \isa{P} should have the form $x\in y$ or $x=y$:
\begin{isabelle}\footnotesize
"REFLECTS[P,\ \isasymlambda a\ x.\ P(x)]"
\end{isabelle}
Reflection relationships can be formed over the propositional
connectives, here negation, conjunction and biconditionals:
\begin{isabelle}\footnotesize
"REFLECTS[P,Q]\ \isasymLongrightarrow\ REFLECTS[\isasymlambda x.\ \isachartilde P(x),\
\isasymlambda a\ x.\ \isachartilde Q(a,x)]"\isanewline
\isanewline
"\isasymlbrakk REFLECTS[P,Q];\ REFLECTS[P',Q']\isasymrbrakk \isanewline
\ \isasymLongrightarrow\ REFLECTS[\isasymlambda x.\ P(x)\ \isasymand \ P'(x),\ \isasymlambda a\ x.\ Q(a,x)\ \isasymand \ Q'(a,x)]"\isanewline
\isanewline
"\isasymlbrakk REFLECTS[P,Q];\ REFLECTS[P',Q']\isasymrbrakk \isanewline
\ \isasymLongrightarrow\ REFLECTS[\isasymlambda x.\ P(x)\ \isasymlongleftrightarrow\ P'(x),\ \isasymlambda a\ x.\ Q(a,x)\ \isasymlongleftrightarrow\
Q'(a,x)]"
\end{isabelle}
Reflection relationships can be formed over the quantifiers:
\begin{isabelle}\footnotesize
"REFLECTS[\ \isasymlambda x.\ P(fst(x),snd(x)),\
\isasymlambda a\ x.\ Q(a,fst(x),snd(x))]\isanewline
\ \isasymLongrightarrow\ REFLECTS[\isasymlambda x.\ \isasymexists z[L].\ P(x,z),\ \isasymlambda a\ x.\ \isasymexists z\isasymin Lset(a).\
Q(a,x,z)]"
\end{isabelle}
In the conclusion, a quantification over $\BL$ is related
to one over $L_\alpha$, as suggested by the general form of
the reflection theorem.  The premise uses the projection operators for
ordered pairs to introduce the new variable,~\isa{z}; syntactically,
\isa{\isasymlambda x.\ P(fst(x),snd(x))} is a unary formula.

\subsection{Internalized Formulae for Some Set-Theoretic Concepts}
\label{sec:union_fm}

Every operator or concept that is used in an
instance of the axiom of separation must be internalized.  If the
defining formula is complicated, then writing the corresponding
element of
\isa{formula} requires a manual (and error-prone) translation into de
Bruijn notation.  The Isabelle/ZF development of constructibility theory
contains about 100 such encodings.  A typical example resembles
that shown in Sect.\ts\ref{sec:subset_fm} above for
\isa{subset\_fm}.  First to be internalized are elementary
concepts such as the empty set, unordered and ordered pairs, unions,
intersections, domain and range.  

The union predicate was defined in Sect.\ts\ref{sec:union_predicate} as
\[ \forall z\,.\; z\in Y \bimp z\in A\disj z\in B.\]
In the corresponding formula, the variables \isa{x}, \isa{y} and
\isa{z} range over de Bruijn indices.
\begin{defsession}
"union\_fm(x,y,z)\ ==\isanewline
\ \ \ Forall(Iff(Member(0,succ(z)),\isanewline
\ \ \ \ \ \ \ \ \ \ \ \ \ \
Or(Member(0,succ(x)),\ Member(0,succ(y)))))"
\end{defsession}

As for \isa{subset\_fm} above, we can prove that \isa{union\_fm}
yields an element of the set \isa{formula}.  
The
theorem about satisfaction now takes the following form:
\begin{isabelle}\footnotesize
"\isasymlbrakk x\ \isasymin \ nat;\ y\ \isasymin \ nat;\ z\ \isasymin \
nat;\ env\ \isasymin \ list(A)\isasymrbrakk \isanewline
\ \isasymLongrightarrow\ sats(A,\ union\_fm(x,y,z),\ env)\ \isasymlongleftrightarrow\isanewline
\ \ \ \ \ union(**A,\ nth(x,env),\ nth(y,env),\ nth(z,env))"
\end{isabelle}
Here, \isa{**A} is Isabelle syntax for the class given by
the set~\isa{A}, that is, $\{x\mid x\in A\}$.  The theorem above simply 
expresses the equivalence between the
relational formula \isa{union} and \isa{union\_fm}, which is its
translation into an element of set \isa{formula}.  Such equivalences
are usually trivial: they simply relate two syntaxes for formulae.
They do not express the equivalence between \isa{union\_fm}
and~$\cup$, which would be an instance of absoluteness. 

After internalizing a predicate such as \isa{union}, it makes sense to
prove its instance of the reflection theorem too, since both
results will be needed when proving instances of
separation.
\begin{isabelle}\footnotesize
"REFLECTS[\isasymlambda x.\ union(L,f(x),g(x),h(x)),\isanewline
\ \ \ \ \ \ \ \ \ \ \isasymlambda i\ x.\
union(**Lset(i),f(x),g(x),h(x))]"
\end{isabelle}
Most reflection proofs are trivial two-line scripts:
\begin{enumerate}
\item Unfold the concept's definition (here \isa{union}).
\item Repeatedly apply existing reflection theorems.
\end{enumerate}
Each predicate is internalized similarly.  Parts of the declarations
and proofs can be copied from those of another predicate.  However, 
getting the definition right requires careful attention to the
original first-order definition.

\subsection{Higher-Order Syntax}
\label{sec:higher-order}

Higher-order syntax is ubiquitous in naive set theory.In the union
$\union_{x\in A}\,B(x)$, the higher-order variable $B$ represents an
indexed family of sets.  In the function abstraction $\lambda_{x\in
A}\,b(x)$, the higher-order variable $b$ represents the function's
body.  Isabelle/ZF additionally uses higher-order syntax to express
many forms of recursion, and so forth.  Although this syntax is
indispensable, it is also illegitimate: formal set theory has no
non-trivial terms, let alone higher-order ones.  We must formalize the
conventions governing higher-order syntax into the language of set
theory.

Converting a higher-order operator such as \isa{\isasymlambda
x\isasymin A.\ b(x)} into relational form yields a higher-order
predicate.  Among its arguments is a predicate \isa{is\_b} 
that expresses the function body,~\isa{b}, in relational form.  If
\isa{is\_b} is purely relational, then so is the definiens of 
\isa{is\_lambda}.
\begin{defsession}
"is\_lambda(M,\ A,\ is\_b,\ z)\ ==\isanewline
\ \ \ \isasymforall p[M].\ p\ \isasymin \ z\ \isasymlongleftrightarrow\isanewline
\ \ \ \ (\isasymexists u[M].\ \isasymexists v[M].\ u\isasymin A\ \&\ pair(M,u,v,p)\ \&\
is\_b(u,v))"
\end{defsession}
This definition states that \isa{z} is a $\lambda$-abstraction
provided its elements are ordered pairs that satisfy~\isa{is\_b}
and whose first component belongs to~\isa{A}\@.

The following predicate expresses that \isa{is\_f} represents the
relational version of~\isa{f} for arguments ranging
over~\isa{A}\@:
\begin{defsession}
"Relation1(M,A,is\_f,f)\ ==\isanewline
\ \ \ \ \isasymforall x[M].\ \isasymforall y[M].\ x\isasymin A\ \isasymlongrightarrow\
is\_f(x,y)\ \isasymlongleftrightarrow\ y\ =\ f(x)"
\end{defsession}
This abbreviation, and similarly \isa{Relation2}, etc.,
are useful for expressing absoluteness results.  If \isa{is\_b} is
the relational equivalent of~\isa{b}, and if the class~\isa{M}
contains each \isa{b(m)} for \isa{m\isasymin A}, then 
\isa{is\_lambda(M,A,is\_b,z)} is the relational version of 
\isa{\isasymlambda x\isasymin A.\ b(x)}.  And thus
$\lambda$-abstraction is absolute:
\begin{isabelle}\footnotesize
"\isasymlbrakk Relation1(M,A,is\_b,b);\ M(A);\ \isasymforall m[M].\
m\isasymin A\ \isasymlongrightarrow\ M(b(m));\ M(z)\isasymrbrakk \isanewline
\ \isasymLongrightarrow \ is\_lambda(M,A,is\_b,z)\ \isasymlongleftrightarrow\ (z\
=\
\isasymlambda x\isasymin A.\ b(x))"
\end{isabelle}
Showing that \isa{M} is closed under $\lambda$-abstraction requires
a separate instance of strong replacement for each~\isa{b}.
\begin{isabelle}\footnotesize
"\isasymlbrakk strong\_replacement(M,\ \isasymlambda x\ y.\
x\isasymin A\ \&\ y\ =\ \isasymlangle x,\ b(x)\isasymrangle
);\isanewline
\ \ M(A);\ \ \isasymforall m[M].\ m\isasymin A\ \isasymlongrightarrow\
M(b(m))\isasymrbrakk\isanewline
\ \isasymLongrightarrow \ M(\isasymlambda
x\isasymin A.\ b(x))"
\end{isabelle}

Internalizing \isa{is\_lambda} is not completely straightforward. 
The predicate argument, \isa{is\_b}, becomes a variable ranging over
the set
\isa{formula}.  
\begin{defsession}
"lambda\_fm(p,A,z)\ ==\ \isanewline
\ \ \ Forall(Iff(Member(0,succ(z)),\isanewline
\ \ \ \ \ \ \ \ \ \ \ Exists(Exists(And(Member(1,A\#+3),\isanewline
\ \ \ \ \ \ \ \ \ \ \ \ \ \ \ \ \ \ \ \ \ \ \ \ \ \ \ \ \ 
And(pair\_fm(1,0,2),\ p))))))"
\end{defsession}
Given a formula and two de Bruijn indices, \isa{lambda\_fm} yields
another formula:
\begin{isabelle}\footnotesize
"\isasymlbrakk p\ \isasymin \ formula;\ x\ \isasymin \ nat;\ y\
\isasymin \ nat\isasymrbrakk \ \isasymLongrightarrow \
lambda\_fm(p,x,y)\ \isasymin \ formula"
\end{isabelle}
But there is no binding mechanism for expressing predicates that take
arguments or refer to local
variables.  The formula \isa{p} must refer to its first argument
using the de Bruijn index~1 and to its second using the 
index~0 (both to be increased in the usual way if \isa{p} contains
quantifiers).  If we are lucky, then we can arrange matters such that
the actual arguments have the right indices, and otherwise
we can force the indices to agree by introducing quantifiers and
equalities: in the internalization of 
$\forall x.\,\forall y.\,x=a\conj y=b\imp p$, the variable with de
Bruijn index~1 will refer to~$a$ and similarly the index~0
will refer to~$b$.
If \isa{p} contains free references to other
variables, their de Bruijn indices must be increased by~3 because
\isa{p} is inserted into a context enclosed by three quantifiers. 

The satisfaction theorem for \isa{is\_lambda} formalizes the remarks
above:
\begin{isabelle}\footnotesize
\isacommand{lemma}\ sats\_lambda\_fm:\isanewline
\ \isakeyword{assumes}\ is\_b\_iff\_sats\ \isasymin\ \isanewline
\ \ \ "!!a0\ a1\ a2.\ \isanewline
\ \ \ \ \isasymlbrakk a0\isasymin A;\ a1\isasymin A;\ a2\isasymin
A\isasymrbrakk \ \isanewline
\ \ \ \ \isasymLongrightarrow \ is\_b(a1,a0)\
\isasymlongleftrightarrow\ sats(A,\ p,\
Cons(a0,Cons(a1,Cons(a2,env))))"\isanewline
\ \isakeyword{shows}\ \isanewline
\ \ \ "\isasymlbrakk x\ \isasymin \ nat;\ y\ \isasymin \ nat;\ env\ \isasymin \ list(A)\isasymrbrakk \isanewline
\ \ \ \ \isasymLongrightarrow \ sats(A,\ lambda\_fm(p,x,y),\ env)\ \isasymlongleftrightarrow\ \isanewline
\ \ \ \ \ \ \ \ is\_lambda(**A,\ nth(x,env),\ is\_b,\
nth(y,env))"
\end{isabelle}
The \isakeyword{assumes}-\isakeyword{shows} syntax eases the use of the
complicated assumption, which states that \isa{is\_b} agrees
with~\isa{p} for the fixed environment~\isa{env} extended with three
additional elements of~\isa{A}\@.  I have not been able to simplify
the form of this theorem while retaining its generality.

It gets more complicated when one higher-order operator refers to
another. One such operator has a quantifier nesting depth of~12.  When
an operator uses its higher-order argument more than once, we must
ensure that the two contexts are similar, adding quantifiers if
necessary to make the nesting depths agree.

Instances of the reflection theorem for higher-order operators must
take into account the possibility of the higher-order argument's
referring to local variables.  Although
\isa{is\_lambda} expects \isa{is\_b} to have only two arguments,
below we formalize it with three arguments (plus its class
argument).  The extra argument is bound by the \isa{REFLECTS}
operator, allowing direct reference to elements of \isa{L} or
\isa{Lset(i)}. 
\begin{isabelle}\footnotesize
\isacommand{theorem}\ is\_lambda\_reflection:\isanewline
\ \isakeyword{assumes}\ is\_b\_reflection:\isanewline
\ \ \ "!!f\ g\ h.\ REFLECTS[\isasymlambda x.\ is\_b(L,\ f(x),\
g(x),\ h(x)),\ \isanewline
\ \ \ \ \ \ \ \ \ \ \ \ \ \ \ \ \ \ \ \ \isasymlambda i\ x.\
is\_b(**Lset(i),\ f(x),\ g(x),\ h(x))]"\isanewline
\ \isakeyword{shows}\ "REFLECTS[\isasymlambda x.\ is\_lambda(L,\
A(x),\ is\_b(L,x),\ f(x)),\ \isanewline
\ \ \ \ \ \ \ \ \isasymlambda i\ x.\ is\_lambda(**Lset(i),\
A(x),\ is\_b(**Lset(i),x),\ f(x))]"
\end{isabelle}

The arity of a higher-order function naturally depends upon that its
function argument.  I found the properties so unintuitive and their
proofs so vexing that I undertook the work described in
Sect.\ts{sec:no-arity}, which eliminates the need for theorems
concerning arities.

\subsection{Proving Instances of Separation}

The set comprehension $\{x\in A\mid\phi(x)\}$ comes from the separation
axiom scheme instantiated to the formula~$\phi$.  The axiom of
replacement yields a set that may be bigger than we want, again 
requiring an appeal to separation.  Because I have not formalized the
metatheory, the Isabelle/ZF development cannot express the proof that
the separation scheme holds for~$\BL$.  Each instance has to be proved
individually.  Fortunately, the proof scripts are nearly identical. 
Given $\phi$, the first step is to prove instance of the reflection
theorem for that formula.  The next step is to run a proof script
corresponding to the sketch in Kunen \cite[p.\ts169]{kunen80}.
The formula~$\phi$ will of course be expressed using the relational
language, using predicates such as \isa{union}.  Executing the proof
script will automatically generate an internalized formula, with
\isa{union\_fm} in the corresponding place.

The lemmas outlined on the preceding pages suffice to
prove many instances of separation.  Consider the instance that
justifies the existence of the intersection \isa{Inter(A)}.  We must
first prove the corresponding instance of the reflection theorem:
\begin{isabelle}\footnotesize
"REFLECTS[\isasymlambda x.\ \isasymforall y[L].\ y\isasymin A\ \isasymlongrightarrow\ x\
\isasymin \ y,\isanewline
\ \ \ \ \ \ \ \ \ \ \isasymlambda i\ x.\ \isasymforall y\isasymin Lset(i).\ y\isasymin A\ \isasymlongrightarrow\ x\ \isasymin \
y]"
\end{isabelle}
Such instances are written manually.  A text editor can replace
quantification over~$\BL$ by quantification over~$L_\alpha$ in the
second formula.  The proof, almost always, is a one-line appeal to
previous reflection theorems.

The statement of each instance of separation comes from the
corresponding locale assumption.  The locale refers to an arbitrary
class~$\BM$, so we must replace $\BM$ by~$\BL$.  The proof scripts are
typically three lines long and follow a regular pattern.  Note that
any parameters used in the separation formula (here $A$) must be
elements of~$\BL$.
\begin{isabelle}\footnotesize
"L(A)\ \isasymLongrightarrow\ separation(L,\ \isasymlambda x.\ \isasymforall y[L].\ y\isasymin A\ \isasymlongrightarrow\ x\isasymin
y)"
\end{isabelle}

The following instance of separation justifies relational
composition.  I leave the corresponding instance of reflection to your
imagination.
\begin{isabelle}\footnotesize
"\isasymlbrakk L(r);\ L(s)\isasymrbrakk \isanewline
\ \isasymLongrightarrow\ separation(L,\ \isasymlambda xz.\ \isasymexists x[L].\ \isasymexists y[L].\ \isasymexists z[L].\ \isasymexists xy[L].\ \isasymexists yz[L].\isanewline
\ \ \ \ \ \ \ \ \ \ \ \ \ pair(L,x,z,xz)\ \&\ pair(L,x,y,xy)\ \&\ pair(L,y,z,yz)\ \&\isanewline
\ \ \ \ \ \ \ \ \ \ \ \ \ xy\isasymin s\ \&\ yz\isasymin
r)"
\end{isabelle}
After proving ten or so instances of separation, we arrive at a
cryptic theorem:
\begin{isabelle}\footnotesize
"PROP\ M\_basic(L)"
\end{isabelle}
This asserts that \isa{L} satisfies the conditions of the locale
\isa{M\_basic}, namely all the instances of separation needed
to derive well-founded recursion.  The absoluteness and closure
results proved in that locale (described in Sect.\ts\ref{sec:M_basic})
--- now become applicable to~\isa{L}\@.

\subsection{Automatic Internalization of Formulae}

Isabelle's ability to translate formulae written in the relational
language into members of \isa{formula} simplifies the proofs of
separation.  Here is an example, from the proof of the instance shown
above (about relational composition).

The first proof step applies a lemma for proving instances of
separations.  It yields a subgoal that has the assumptions
\isa{r\
\isasymin \ Lset(j)} and 
\isa{s\ \isasymin \ Lset(j)}, where \isa{j} is arbitrary.  We have to
prove that the comprehension belongs to the next level of the
constructible hierarchy, namely \isa{DPow(Lset(j))}:
\begin{isabelle}\footnotesize
\isacharbraceleft xz\ \isasymin \ Lset(j)\ .\
\isasymexists x\isasymin Lset(j).\ \isasymexists y\isasymin Lset(j).\
\ldots\isacharbraceright \ \isasymin DPow(Lset(j))
\end{isabelle}

The second proof step applies a lemma for proving membership in
\isa{DPow(Lset(j))}.  It yields three subgoals
(Fig.\ts\ref{fig:synthesis}).  The first is to show the
equivalence between the real formula
\begin{isabelle}\footnotesize
(\isasymexists xa\isasymin Lset(j).\ \isasymexists y\isasymin
Lset(j).\ \ldots)
\end{isabelle}
and \isa{sats(Lset(j),\ ?p3(j),\ [x,r,s])}.  This is the satisfaction
relation applied to \isa{?p3(j)}, a ``logical
variable'' that can be replaced by any expression, possibly involving
the bound variable~\isa{j}.  The third subgoal in
Fig.\ts\ref{fig:synthesis}, namely \isa{?p3(j)\ \isasymin \ formula},
checks that the chosen expression is an internalized formula.  The
second subgoal verifies that the environment, \isa{[r,s]}, is
well-typed --- namely, that it belongs to
\isa{list(Lset(j))}.

The third proof step is this:
\begin{isabelle}\footnotesize
apply (rule sep\_rules | simp)+
\end{isabelle}
It applies some theorem of \isa{sep\_rules}, then simplifies, then
repeats if possible.  This finishes the proof.  All
separation proofs have this form, save only that sometimes
\isa{sep\_rules} needs to be augmented with additional theorems.

\begin{figure}
\begin{isabelle}\footnotesize
\ 1.\ \isasymAnd j\ x.\ \isasymlbrakk L(r);\ L(s);\ r\ \isasymin \ Lset(j);\ s\ \isasymin \ Lset(j);\ x\ \isasymin \ Lset(j)\isasymrbrakk \isanewline
\isaindent{\ 1.\ \isasymAnd j\ x.\ }\isasymLongrightarrow \ (\isasymexists xa\isasymin Lset(j).\isanewline
\isaindent{\ 1.\ \isasymAnd j\ x.\ \isasymLongrightarrow \ (\ \ \ }\isasymexists y\isasymin Lset(j).\isanewline
\isaindent{\ 1.\ \isasymAnd j\ x.\ \isasymLongrightarrow \ (\ \ \ \ \ \ }\isasymexists z\isasymin Lset(j).\isanewline
\isaindent{\ 1.\ \isasymAnd j\ x.\ \isasymLongrightarrow \ (\ \ \ \ \ \ \ \ \ }pair(**Lset(j),\ xa,\ z,\ x)\ \isasymand \isanewline
\isaindent{\ 1.\ \isasymAnd j\ x.\ \isasymLongrightarrow \ (\ \ \ \ \ \ \ \ \ }(\isasymexists xy\isasymin Lset(j).\isanewline
\isaindent{\ 1.\ \isasymAnd j\ x.\ \isasymLongrightarrow \ (\ \ \ \ \ \ \ \ \ (\ \ \ }pair(**Lset(j),\ xa,\ y,\ xy)\ \isasymand \isanewline
\isaindent{\ 1.\ \isasymAnd j\ x.\ \isasymLongrightarrow \ (\ \ \ \ \ \ \ \ \ (\ \ \ }(\isasymexists yz\isasymin Lset(j).\isanewline
\isaindent{\ 1.\ \isasymAnd j\ x.\ \isasymLongrightarrow \ (\ \ \ \ \ \ \ \ \ (\ \ \ (\ \ \ }pair(**Lset(j),\ y,\ z,\ yz)\ \isasymand \isanewline
\isaindent{\ 1.\ \isasymAnd j\ x.\ \isasymLongrightarrow \ (\ \ \ \ \ \ \ \ \ (\ \ \ (\ \ \ }xy\ \isasymin \ s\ \isasymand \ yz\ \isasymin \ r)))\ \isasymlongleftrightarrow \isanewline
\isaindent{\ 1.\ \isasymAnd j\ x.\ \isasymLongrightarrow \ }sats(Lset(j),\ ?p3(j),\ [x,\ r,\ s])\isanewline
\ 2.\ \isasymAnd j.\ \isasymlbrakk L(r);\ L(s);\ r\ \isasymin \ Lset(j);\ s\ \isasymin \ Lset(j)\isasymrbrakk \isanewline
\isaindent{\ 2.\ \isasymAnd j.\ }\isasymLongrightarrow \ [r,\ s]\ \isasymin \ list(Lset(j))\isanewline
\ 3.\ \isasymAnd j.\ \isasymlbrakk L(r);\ L(s);\ r\ \isasymin \ Lset(j);\ s\ \isasymin \ Lset(j)\isasymrbrakk \isanewline
\isaindent{\ 3.\ \isasymAnd j.\ }\isasymLongrightarrow \ ?p3(j)\ \isasymin \ formula%
\end{isabelle}
\caption{Subgoals ready for automatic synthesis of a formula}
\label{fig:synthesis}
\end{figure}

Formula synthesis works in a way familiar to all Prolog programmers. 
Essentially, the theorems in \isa{sep\_rules} comprise a Prolog program
for generating internalized formulae.  Most of the ``program clauses''
relate real formulae to internal ones and are derived from the basic
properties of the satisfaction relation.  For example, this one
relates the real conjunction \isa{P\&Q} with the term \isa{And(p,q)}. 
The first two subgoals concern the synthesis of \isa{p} and~\isa{q}.
The third subgoal expresses a type constraint on~\isa{env}.
\begin{isabelle}\footnotesize
"\isasymlbrakk P\ \isasymlongleftrightarrow\ sats(A,p,env);\ Q\ \isasymlongleftrightarrow\ sats(A,q,env);\ env\
\isasymin \ list(A)\isasymrbrakk \isanewline
\ \isasymLongrightarrow\ (P\ \&\ Q)\ \isasymlongleftrightarrow\ sats(A,\ And(p,q),\ env)"
\end{isabelle}
This ``program clause'' relates the real quantification
\isa{\isasymforall x\isasymin A.\ P(x)} with the term
\isa{Forall(p)}.  The first subgoal concerns the synthesis of~\isa{p}
in an environment augmented with an arbitrary \isa{x\isasymin
A}:
\begin{isabelle}\footnotesize
"\isasymlbrakk !!x.\ x\isasymin A\ \isasymLongrightarrow\ P(x)\ \isasymlongleftrightarrow\ sats(A,\ p,\
Cons(x,\ env));\ env\ \isasymin \ list(A)\isasymrbrakk \isanewline
\ \isasymLongrightarrow\ (\isasymforall x\isasymin A.\ P(x))\ \isasymlongleftrightarrow\ sats(A,\ Forall(p),\
env)"
\end{isabelle}
The environment, which initially contains the parameters of the 
separation formula, gets longer with each nested quantifier.  Each
higher-order operator can add several elements to the environment; as
mentioned above in Sect.\ts\ref{sec:higher-order}.

A base case of synthesis relates the formula
\isa{x\isasymin y} with the term \isa{Member(i,j)}.  
The first two subgoals concern the synthesis of the de Bruijn
indices \isa{i} and~\isa{j}:
\begin{isabelle}\footnotesize
"\isasymlbrakk nth(i,env)\ =\ x;\ nth(j,env)\ =\ y;\ env\ \isasymin \
list(A)\isasymrbrakk \isanewline
\ \isasymLongrightarrow\ (x\isasymin y)\ \isasymlongleftrightarrow\ sats(A,\ Member(i,j),\ env)"
\end{isabelle}

Other base cases concern predicates of the relational language.  This
theorem, which relates the formula
\isa{union(**A,x,y,z)} with the term \isa{union\_fm(i,j,k)}, is just a
reworking of a theorem shown in Sect.\ts\ref{sec:union_fm} above.
\begin{isabelle}\footnotesize
"\isasymlbrakk nth(i,env)\ =\ x;\ nth(j,env)\ =\ y;\ nth(k,env)\ =\ z;\isanewline
\ \ i\ \isasymin \ nat;\ j\ \isasymin \ nat;\ k\ \isasymin \ nat;\
env\ \isasymin \ list(A)\isasymrbrakk \isanewline
\ \isasymLongrightarrow\ union(**A,\ x,\ y,\ z)\ \isasymlongleftrightarrow\ sats(A,\ union\_fm(i,j,k),\
env)"
\end{isabelle}
Given the subgoal \isa{nth(?i,env)\ =\ x}, Isabelle can
synthesize~\isa{?i}.  This de Bruijn index is determined by \isa{x},
which comes from the original formula, and
\isa{env}, which is given in advance.  If \isa{x} matches the head of
the environment, then \isa{?i} should be zero:
\begin{isabelle}\footnotesize
"nth(0, Cons(a, l)) = a"
\end{isabelle}
And if it does not match, then we should discard the head and attempt
to synthesize a de Bruijn index using the tail:
\begin{isabelle}\footnotesize
"\isasymlbrakk nth(n,l)\ =\ x;\ n\ \isasymin \ nat\isasymrbrakk \ \isasymLongrightarrow\ nth(succ(n),\
Cons(a,l))\ =\ x"
\end{isabelle}

The automatic synthesis of internalized formulae saves much work in
proofs of separation.  In principle, we could rewrite
every relational formula into its primitive constituents of membership
and equality, removing the need for
\isa{union\_fm} and 100 similar constants.  But if too few
internalized primitives have been defined, formula synthesis takes
many minutes.

\section{Absoluteness of Recursive Datatypes}
\label{sec:datatypes}

The Isabelle/ZF proofs discussed up to now include the construction of
the class~$\BL$ and the proof that it is a model of the
Zermelo-Fraenkel axioms.  The next step is to show that $\BL$ satisfies
$\BV=\BL$.  That fact follows by the absoluteness of constructibility,
which follows by the absoluteness of satisfaction.  Consulting the
definition of \isa{satisfies} reveals that we must still prove the
absoluteness of lists, formulae, the function
\isa{nth}, and several other notions.  

Isabelle/ZF defines the sets \isa{list(A)} and \isa{formula}
automatically from their user-supplied
descriptions~\cite{paulson-fixedpt-milner}.  These fixedpoint
definitions have advantages, but their use of the powerset operator is
an obstacle to proving absoluteness.  For a start, \isa{Pow(D)} must be
eliminated from this definition:
\begin{defsession}
"lfp(D,h)\ ==\ Inter(\isacharbraceleft X\ \isasymin\ Pow(D). h(X)
\isasymsubseteq\ X\isacharbraceright)"
\end{defsession}

We proceed by formalizing standard concepts from domain theory
\cite[pp.\ts51--56]{davey&priestley}.  A set is \emph{directed} if it
is non-empty and closed under least upper bounds.  A function is
\emph{continuous} if it preserves the unions of directed sets.
\begin{defsession}
"directed(A)\ ==\ A\isasymnoteq 0\ \&\ (\isasymforall x\isasymin
A.\
\isasymforall y\isasymin A.\ x\isasymunion y\ \isasymin
\ A)"\isanewline 
"contin(h)\ \ \ ==\ (\isasymforall A.\ directed(A)\
\isasymlongrightarrow\ h(\isasymUnion A)\ =\ (\isasymUnion X\isasymin A.\ h(X)))"
\end{defsession}
We can prove that the least
fixedpoint of a monotonic, continuous function~\isa{h} can be expressed
as the union of the finite iterations of~\isa{h}.
\begin{isabelle}\footnotesize
"\isasymlbrakk bnd\_mono(D,h);\ contin(h)\isasymrbrakk \ \isasymLongrightarrow\ lfp(D,h)\ =\ (\isasymUnion n\isasymin nat.\
h\isacharcircum n(0))"
\end{isabelle}
This equation not only eliminates~\isa{Pow(D)}, but every
occurrence of~\isa{D}, which is the ``bounding set''
\cite[\S2.2]{paulson-set-II} and is itself typically defined using
powersets.

In order to apply this equation, we must prove that standard
datatype constructions preserve continuity.  
The case bases are that the constant function and the identity
function are continuous:
\begin{isabelle}\footnotesize
"contin(\isasymlambda X.\ A)"\isanewline
"contin(\isasymlambda X.\ X)"
\end{isabelle}
Sums and products preserve continuity:
\begin{isabelle}\footnotesize
"\isasymlbrakk contin(F);\ contin(G)\isasymrbrakk \ \isasymLongrightarrow\ contin(\isasymlambda X.\ F(X)\ +\ G(X))"\isanewline
"\isasymlbrakk contin(F);\ contin(G)\isasymrbrakk \ \isasymLongrightarrow\ contin(\isasymlambda X.\ F(X)\ \isasymtimes\
G(X))"
\end{isabelle}
These four lemmas cover all finitely-branching datatypes, including
lists and formulae.

\subsection{Absoluteness for Function Iteration}

In the equation above for least fixed points, the
term \isa{h\isacharcircum n(0)} abbreviates \isa{iterates(h,n,0)}. 
Isabelle/ZF defines 
\isa{iterates(F,n,x)} by the obvious primitive recursion on
\isa{n\isasymin nat}.  Absoluteness of datatype definitions will
follow from the absoluteness of \isa{iterates}.  

Recall that a well-founded function definition consists of a
relation~$r$ and function body~$H$; recall equation~(\ref{eqn:recfun})
of Sect.\ts\ref{sec:wfrec-abs}.  Relativizing such a function
definition, requires relativizing $H$ by an Isabelle/ZF relation,
say~\isa{MH}\@.  So to relativize \isa{iterates}, we 
declare \isa{is\_iterates} in terms of another predicate
\isa{iterates\_MH}, representing the body of the recursion.
\begin{defsession}
"iterates\_MH(M,isF,v,n,g,z)\ ==\isanewline
\ \ \ is\_nat\_case(M,\ v,\isanewline
\ \ \ \ \ \ \ \ \ \ \ \ \ \ \isasymlambda m\ u.\ \isasymexists
gm[M].\ fun\_apply(M,g,m,gm)\ \&\ isF(gm,u),\isanewline
\ \ \ \ \ \ \ \ \ \ \ \ \ \ n,\ z)"
\isanewline
"is\_iterates(M,isF,v,n,Z)\ ==\isanewline
\ \ \isasymexists sn[M].\ \isasymexists msn[M].\ successor(M,n,sn)\ \&\ membership(M,sn,msn)\ \&\isanewline
\ \ \ \ \ \ \ \ \ \ \ \ \ \ \ \ \ \ \ is\_wfrec(M,\ iterates\_MH(M,isF,v),\ msn,\ n,\
Z)"
\end{defsession}
Incidentally, \isa{is\_nat\_case(M,a,isb,n,z)} expresses case analysis
on the natural number~\isa{n}.  Note that we again work in the general
setting of a class~\isa{M} satisfying certain conditions.  Later, we
shall prove that \isa{L} meets those conditions.

The absoluteness theorem for well-founded recursion requires an
instance of strong replacement for each function being defined.  But
\isa{iterates} is a higher-order function, so technically 
\isa{iterates(F,n,x)} involves a separate instance of well-founded
recursion for each~\isa{F}\@.  The function
\isa{iterates\_replacement} can express each required instance of
replacement; its argument \isa{isF} is the relational form
of~\isa{F}\@.
\begin{defsession}
"iterates\_replacement(M,isF,v)\ ==\isanewline
\ \ \ \isasymforall n[M].\ n\isasymin nat\ \isasymlongrightarrow\isanewline
\ \ \ \ \ \ wfrec\_replacement(M,\ iterates\_MH(M,isF,v),\
Memrel(succ(n)))"
\end{defsession}

Assuming such an instance of replacement, and given
that \isa{isF} is the relational version
of~\isa{F}, the absoluteness of \isa{iterates} is a corollary of
the general theorem about well-founded recursion.
\begin{isabelle}\footnotesize
"\isasymlbrakk iterates\_replacement(M,isF,v);\
relation1(M,isF,F);\isanewline
\ \ n\ \isasymin \ nat;\ M(v);\ M(z);\ \isasymforall x[M].\
M(F(x))\isasymrbrakk \isanewline
\ \isasymLongrightarrow\ is\_iterates(M,isF,v,n,z)\ \isasymlongleftrightarrow\ z\ =\ iterates(F,n,v)"
\end{isabelle}
We similarly find that $\BM$ is closed under function iteration.
\begin{isabelle}\footnotesize
"\isasymlbrakk iterates\_replacement(M,isF,v);\ relation1(M,isF,F);\isanewline
\ \ n\ \isasymin \ nat;\ M(v);\ \isasymforall x[M].\ M(F(x))\isasymrbrakk \isanewline
\ \isasymLongrightarrow\ M(iterates(F,n,v))"
\end{isabelle}

\subsection{Absoluteness for Lists and Formulae}
\label{sec:datatype}

The formal treatment of continuity and \isa{iterates} enables us to
prove that lists and formulae are absolute.  

The definition of lists
generated by the Isabelle/ZF datatype~\cite{paulson-fixedpt-milner} is
too complicated to relativize easily.  Instead, we 
prove its equivalence to a more abstract (and familiar) definition.  
\begin{isabelle}\footnotesize
"list(A)\ =\ lfp(univ(A),\ \isasymlambda X.\ \isacharbraceleft
0\isacharbraceright \ +\ A*X)"
\end{isabelle}
The function given to \isa{lfp} continuous by construction, which
lets us replace the the least fixed point by iteration and eliminate
the non-absolute set \isa{univ(A)}:
\begin{isabelle}\footnotesize
"contin(\isasymlambda X.\ \isacharbraceleft 0\isacharbraceright \ +\
A*X)"\isanewline
"list(A)\ =\ (\isasymUnion n\isasymin nat.\ (\isasymlambda X.\
\isacharbraceleft 0\isacharbraceright \ +\ A*X)\isacharcircum n\
(0))"
\end{isabelle}
Now the absoluteness of \isa{list(A)} is obvious.  But each
element of this equation must be formalized in order to prove 
absoluteness.  We begin by introducing an abbreviation for
finite iterations of \isa{\isasymlambda X.\
\isacharbraceleft 0\isacharbraceright \ +\ A*X} --- that is, for
finite stages of the list construction.
\begin{defsession}
"list\_N(A,n)\ ==\ (\isasymlambda X.\ \isacharbraceleft
0\isacharbraceright \ +\ A*X)\isacharcircum n\ (0)"
\end{defsession}
Next, we relativize the function \isa{\isasymlambda X.\
\isacharbraceleft 0\isacharbraceright \ +\ A*X}.  The predicate
\isa{number1} recognizes the number~1, which equals the set~$\{0\}$.
\begin{defsession}
"is\_list\_functor(M,A,X,Z)\ ==\isanewline
\ \ \ \ \ \ \isasymexists n1[M].\ \isasymexists AX[M].\isanewline
\ \ \ \ \ \ \ number1(M,n1)\ \&\ cartprod(M,A,X,AX)\ \&\
is\_sum(M,n1,AX,Z)"
\end{defsession}
Next, we relativize the function \isa{list\_N}, the finite iterations:
\begin{defsession}
"is\_list\_N(M,A,n,Z)\ ==\isanewline
\ \ \ \ \ \ \isasymexists zero[M].\ empty(M,zero)\ \&\isanewline
\ \ \ \ \ \ \ \ \ \ \ \ \ \ \ \ is\_iterates(M,\ is\_list\_functor(M,A),\ zero,\ n,\
Z)"
\end{defsession}
We relativize membership in \isa{list(A)} as membership in 
\isa{list\_N(A,n)} for some~\isa{n}.  The predicate
\isa{finite\_ordinal} recognizes the natural numbers.
\begin{defsession}
"mem\_list(M,A,l)\ ==\isanewline
\ \ \ \ \isasymexists n[M].\ \isasymexists listn[M].\isanewline
\ \ \ \ \ finite\_ordinal(M,n)\ \&\ is\_list\_N(M,A,n,listn)\ \&\ l\
\isasymin \ listn"
\end{defsession}
Finally, we can relativize the set of lists itself:
\begin{defsession}
"is\_list(M,A,Z)\ ==\ \isasymforall l[M].\ l\ \isasymin \ Z\ \isasymlongleftrightarrow\
mem\_list(M,A,l)"
\end{defsession}

After proving absoluteness of \isa{list\_N(A,n)}, we obtain the
absoluteness of \isa{list(A)} and prove that \isa{M} is closed under
list formation. 
\begin{isabelle}\footnotesize
"M(A)\ \isasymLongrightarrow\ M(list(A))"\isanewline
"\isasymlbrakk M(A);\ M(Z)\isasymrbrakk \ \isasymLongrightarrow\ is\_list(M,A,Z)\ \isasymlongleftrightarrow\ Z\ =\
list(A)"
\end{isabelle}

Formulae are proved absolute in just the same way.  We
express the set
\isa{formula} as an abstract least fixed point of a suitable function,
prove that function to be continuous, and eliminate the \isa{lfp}
operator:
\begin{isabelle}\footnotesize
"formula\ =\ lfp(univ(0),\ \isasymlambda X.\ ((nat*nat)\ +\ (nat*nat))\ +\ (X*X\ +\
X))"\isanewline
"contin(\isasymlambda X.\ ((nat*nat)\ +\ (nat*nat))\ +\ (X*X\ +\
X))"\isanewline "formula\ =\isanewline
\ \ \ \ (\isasymUnion n\isasymin nat.\ (\isasymlambda X.\ ((nat*nat)\ +\ (nat*nat))\
+\ (X*X\ +\ X))\ \isacharcircum \ n\ (0))"
\end{isabelle}
Proceeding as for lists, we define the predicates
\isa{is\_formula\_functor},
\isa{is\_formula\_N},
\isa{mem\_formula} and finally
\isa{is\_formula}.  We obtain the desired theorems:
\begin{isabelle}\footnotesize
"M(formula)"\isanewline
"M(Z)\ \isasymLongrightarrow\ is\_formula(M,Z)\ \isasymlongleftrightarrow\ Z\ =\ formula"
\end{isabelle}

\subsection{Recursion over Lists and Formulae}

We have already seen (Sect.\ts\ref{sec:wfrec}) that functions defined
by well-founded recursion are absolute.  For mathematicians, that is
enough to justify the absoluteness of functions defined recursively on
lists or formulae.  Proof tool users, however, must work through the
details for each instance.  Usually automation makes it easy to apply
general results to particular circumstances.  However, the Isabelle/ZF
translation of recursive function definitions is rather complicated.%
\footnote{See \S\S3.4 and~4.3.1 of Paulson~\cite{paulson-set-II}.}
There are good reasons for this complexity, such as support for
a form of polymorphism.  However, it makes the absoluteness proofs more
difficult: the complications have to be taken apart and relativized one
by one.

At least there is no need to treat recursion over lists.  Defining
the class $\BL$ involves only one list function, namely \isa{nth}. 
Given a natural number~$n$ and a list~$l$, this function returns the
$n^{\textrm{th}}$ element of~$l$, counting from~0.  Obviously this
amounts to taking the tail of the list $n$ times and returning the
head of the result.  The recursion in \isa{nth} is an instance of
\isa{iterates}.

Isabelle/ZF defines the head and tail functions \isa{hd} and
\isa{tl}.  The absoluteness proofs use
modified versions called \isa{hd'} and
\isa{tl'}, which extend \isa{hd} and
\isa{tl} to return 0 if their argument is ill-formed (the details are
unimportant).  Relativization is simpler when a function's behaviour
is fully specified.  Now we can prove an equivalence for \isa{nth}:
\begin{isabelle}\footnotesize
"\isasymlbrakk xs\ \isasymin \ list(A);\ n\
\isasymin \ nat\isasymrbrakk \ \isasymLongrightarrow\ nth(n,xs)\ =\ hd'\
(tl'\ \isacharcircum\ n\ (xs))"
\end{isabelle}
Its relational equivalent, \isa{is\_nth}, has an obvious definition
in terms of the relational equivalents of \isa{iterates},
\isa{tl} and \isa{hd}:
\begin{defsession}
"is\_nth(M,n,l,Z)\ ==\isanewline
\ \ \ \ \isasymexists X[M].\ is\_iterates(M,\ is\_tl(M),\ l,\ n,\ X)\
\&\ is\_hd(M,X,Z)"
\end{defsession}
Absoluteness is proved with no effort:
\begin{isabelle}\footnotesize
"\isasymlbrakk M(A);\ n\ \isasymin \ nat;\ l\ \isasymin \ list(A);\ M(Z)\isasymrbrakk \isanewline
\ \isasymLongrightarrow\ is\_nth(M,n,l,Z)\ \isasymlongleftrightarrow\ Z\ =\ nth(n,l)"
\end{isabelle}

Recursion over lists is absolute in general.  Proving this claim would
require much work, and is unnecessary for proving that
$\BV=\BL$ is absolute.  The function \isa{satisfies} involves
recursion over the datatype of formulae, and its absoluteness proof
consists of several stages.  Isabelle/ZF expresses recursion on
datatypes in terms of
$\in$-recursion, which is recursion on a set's rank
\cite[\S3.4]{paulson-set-II}.  Absoluteness for $\in$-recursion will
follow from that of well-founded recursion once we have established
the absoluteness of $\in$-closure.  Then we shall be in a position to
consider recursion over formulae.

Five instances of strong replacement are necessary for the proofs
sketched above.  There are two each for the absoluteness of
\isa{list(A)} and
\isa{formula}, and one for the absoluteness of \isa{nth(n,l)}.
The locale \isa{M\_datatypes} encapsulates these additional
constraints on the class~\isa{M}\@.  It is one of several locales used
to keep track of instances of separation and replacement in this
development.

\subsection{Absoluteness for $\in$-Closure}

If $A$ is a set, then its $\in$-closure is the smallest transitive set
that includes~$A$.  Formally, the
$\in$-closure of~$A$ is $\union_{n\in\omega} \union^n(A)$.  Here 
$\union^n(A)$ denotes the $n$-fold union of~$A$, defined by
$\union^0(A)=A$ and
$\union^{m+1}(A)=\union(\union^m(A))$.  This is just another
instance of \isa{iterates}, as we can prove:
\begin{isabelle}\footnotesize
"eclose(A)\ =\ (\isasymUnion n\isasymin nat.\ Union\isacharcircum n\
(A))"
\end{isabelle}
Relativization proceeds as it did for lists.  The details are omitted,
but they culminate in the definition of a relational version of
\isa{eclose(A)}:
\begin{defsession}
"is\_eclose(M,A,Z)\ ==\ \isasymforall u[M].\ u\ \isasymin \ Z\
\isasymlongleftrightarrow\ mem\_eclose(M,A,u)"
\end{defsession}
The standard membership and absoluteness results follow:
\begin{isabelle}\footnotesize
"M(A)\ \isasymLongrightarrow\ M(eclose(A))"\isanewline
"\isasymlbrakk M(A);\ M(Z)\isasymrbrakk \ \isasymLongrightarrow\ is\_eclose(M,A,Z)\ \isasymlongleftrightarrow\ Z\ =\ eclose(A)"
\end{isabelle}

\subsection{Absoluteness for \isa{transrec}}

The Isabelle/ZF operator \isa{transrec} expresses $\in$-recursion,
which includes transfinite recursion as a special case:
\[ \isa{transrec}(a,H) = H(a, \lambda_{x\in a}.\isa{transrec}(x,H)). \]
Its definition is a straightforward combination of the operators
\isa{eclose}, \isa{wfrec} (which expresses well-founded recursion),
and \isa{Memrel} (which encodes the membership relation as a set). 
Thus the definition of the relational version, \isa{is\_transrec}, is
also straightforward.  Our previous
results lead directly to a proof of absoluteness:
\begin{isabelle}\footnotesize
"\isasymlbrakk transrec\_replacement(M,MH,i);\ \ relativize2(M,MH,H);\isanewline
\ \ Ord(i);\ \ M(i);\ \ M(z);\isanewline
\ \ \isasymforall x[M].\ \isasymforall g[M].\ function(g)\ \isasymlongrightarrow\ M(H(x,g))\isasymrbrakk \isanewline
\ \isasymLongrightarrow\ is\_transrec(M,MH,i,z)\ \isasymlongleftrightarrow\ z\ =\ transrec(i,H)"
\end{isabelle}
We similarly find that \isa{M} is closed under $\in$-recursion:
\begin{isabelle}\footnotesize
"\isasymlbrakk transrec\_replacement(M,MH,i);\ \
relativize2(M,MH,H);\isanewline
\ \ Ord(i);\ \ M(i);\isanewline
\ \ \isasymforall x[M].\ \isasymforall g[M].\ function(g)\ \isasymlongrightarrow\ M(H(x,g))\isasymrbrakk \isanewline
\ \isasymLongrightarrow\ M(transrec(i,H))"
\end{isabelle}
In these theorems, \isa{transrec\_replacement} abbreviates a specific
use of \isa{wfrec\_replacement}, which justifies this particular
recursive definition (recall Sect.\ts\ref{sec:wfrec_abs}).

\subsection{Recursion over Formulae}

The Isabelle/ZF treatment of recursive functions on datatypes involves
non-absolute concepts, namely the cumulative hierarchy
$\{V_\alpha\}_{\alpha\in\BON}$ and the rank function
\cite[\S3.6]{paulson-set-II}.  For proving
absoluteness, I proved an equation stating that recursion over
formulae could be expressed differently.  The new formulation refers to
the \emph{depth} of a formula, defined by  
\begin{defsession}
"depth(Member(x,y))\ =\ 0"\isanewline
"depth(Equal(x,y))\ \ =\ 0"\isanewline
"depth(Nand(p,q))\ \ \ =\ succ(depth(p)\ \isasymunion \
depth(q))"\isanewline 
"depth(Forall(p))\ \ \ =\ succ(depth(p))"
\end{defsession}
Introducing \isa{depth} seems to be a step backwards, since it
requires relativizing another recursive function on formulae.  But we
can express the depth of a formula in terms of
\isa{is\_formula\_N}, which we need anyway 
(Sect.\ts\ref{sec:datatype});
\isa{is\_formula\_N(M,n,F)} holds just if \isa{F} is the set of
formulae generated by \isa{n} unfoldings of the datatype
definition --- which is all formulae of depth
less than~\isa{n}.  A formula \isa{p} has depth~\isa{n}
if it satisfies \isa{is\_formula\_N(M,succ(n),F)} and not
\isa{is\_formula\_N(M,n,F)}:
\begin{defsession}
"is\_depth(M,p,n)\ ==\isanewline
\ \ \ \isasymexists sn[M].\ \isasymexists formula\_n[M].\ \isasymexists formula\_sn[M].\isanewline
\ \ \ \ is\_formula\_N(M,n,formula\_n)\ \&\ p\ \isasymnotin \ formula\_n\ \&\isanewline
\ \ \ \ successor(M,n,sn)\ \&\isanewline
\ \ \ \ is\_formula\_N(M,sn,formula\_sn)\ \&\ p\ \isasymin \
formula\_sn"
\end{defsession}
Working from this definition, we find that the depth of a formula is
absolute:
\begin{isabelle}\footnotesize
"\isasymlbrakk p\ \isasymin \ formula;\ n\ \isasymin \ nat\isasymrbrakk \ \isasymLongrightarrow\
is\_depth(M,p,n)\ \isasymlongleftrightarrow\ n\ =\ depth(p)"
\end{isabelle}

For relativization, I modified the standard
Isabelle/ZF treatment of recursion over formulae, replacing the set
$V_\alpha$ by \isa{formula} and the rank of a set by the depth of a
formula.  If \isa{f} is a recursive function on formulae, then the
evaluation of \isa{f(p)} begins by determining the depth of~\isa{p},
say~$n$.  Then the recursion equation for~\isa{f} is unfolded $n+1$
times, using transfinite recursion.  The resulting nonrecursive
function is finally applied to~\isa{p}.  This approach unfortunately
needs an explicit
$\lambda$-abstraction over formulae and another instance
of the replacement axiom.  With the benefit of hindsight, I might
have saved much work by seeking simpler ways of expressing
recursion over formulae, such as by well-founded
recursion on the subformula relation.

The recursive definition of a function~\isa{f} is specified by four
parameters \isa{a}, \isa{b}, \isa{c} and~\isa{d}, corresponding to the
four desired recursion equations:
\begin{isabelle}\footnotesize
f(Member(x,y)) = a(x,y)\isanewline
f(Equal(x,y)) \ = b(x,y)\isanewline
f(Nand(p,q)) \ \ = c(p,f(p),q,f(q))\isanewline
f(Forall(p)) \ \ = d(p,f(p))
\end{isabelle}
Given the datatype definition of \isa{formula}, Isabelle/ZF
automatically defines the operator
\isa{formula\_rec} for expressing recursive functions.
The term \isa{formula\_rec(a,b,c,d,p)} denotes the value of the
function~\isa{f} above applied to the
argument~\isa{p}.  More concisely, \isa{formula\_rec(a,b,c,d)}
denotes the the function~\isa{f} itself.  The details of the
definitions are illustrated elsewhere, using the example of lists 
\cite[\S4.3]{paulson-set-II}.

In order to express the recursion theorem, it helps to have first
defined an abbreviation for its case analysis on formulae.
\begin{defsession}
"formula\_rec\_case(a,b,c,d,h)\ ==\isanewline
\ formula\_case\ (a,\ b,\isanewline
\ \ \ \ \ \ \ \ \ \isasymlambda u\ v.\ c(u,\ v,\ h\ `\ succ(depth(u))\
`\ u,\isanewline
\ \ \ \ \ \ \ \ \ \ \ \ \ \ \ \ \ \ \ \ \ \ \ h\ `\ succ(depth(v))\ `\
v),\isanewline
\ \ \ \ \ \ \ \ \ \isasymlambda u.\ d(u,\ h\ `\ succ(depth(u))\ `\
u))"
\end{defsession}
Now we can express recursion on formulae in terms of absolute concepts:
\begin{isabelle}\footnotesize
"p\ \isasymin \ formula\ \isasymLongrightarrow\isanewline
\ formula\_rec(a,b,c,d,p)\ =\isanewline
\ transrec\ (succ(depth(p)),\isanewline
\ \ \ \ \ \ \ \ \ \ \ \isasymlambda x\ h.\ Lambda(formula,\
formula\_rec\_case(a,b,c,d,h)))\ `\ p"
\end{isabelle}
The proof is by structural induction on~\isa{p}.
Note that the argument \isa{h} of \isa{formula\_rec\_case} is a
partially unfolded recursive function taking two curried arguments. 
The second argument is some subformula~\isa{u} and the first is
\isa{succ(depth(u))}.  The intuition behind this theorem may be
obscure, but that is no obstacle to proving absoluteness.  Many
routine details must be taken care of, including relativization and
absoluteness for the formula constructors \isa{Member}, \isa{Equal},
\isa{Nand} and \isa{Forall} and for the operator \isa{formula\_case}.

Obviously \isa{formula\_rec} is a higher-order function.  Its
absoluteness proof depends upon absoluteness assumptions for the
function arguments \isa{a}, \isa{b},
\isa{c} and~\isa{d}.  Its relational version needs those arguments to
be expressed in relational form as  predicates \isa{is\_a},
\isa{is\_b}, \isa{is\_c} and~\isa{is\_d}. The absoluteness theorem
depends upon 10 assumptions in all: two for each of \isa{is\_a},
\isa{is\_b},
\isa{is\_c} and~\isa{is\_d} and two instances of replacement.  
After many intricate but uninteresting details, we arrive at two
key theorems.  If the class \isa{M} is closed under the parameters
\isa{a}, \isa{b}, \isa{c} and~\isa{d} then it is closed under the
corresponding recursion:
\begin{isabelle}\footnotesize
"p\ \isasymin \ formula\ \isasymLongrightarrow\ M(formula\_rec(a,b,c,d,p))"
\end{isabelle}
Recursion over formulae is absolute:
\begin{isabelle}\footnotesize
"\isasymlbrakk p\ \isasymin \ formula;\ M(z)\isasymrbrakk \isanewline
\ \isasymLongrightarrow\ is\_formula\_rec(M,MH,p,z)\ \isasymlongleftrightarrow\ z\ =\
formula\_rec(a,b,c,d,p)"
\end{isabelle}
In this theorem, \isa{MH} abbreviates the
relativization of the argument of \isa{transrec} shown above:
\begin{defsession}
"MH(u::i,f,z)\ ==\isanewline
\ \ \isasymforall fml[M].\ is\_formula(M,fml)\ \isasymlongrightarrow\isanewline
\ \ \ \ \ \ \ is\_lambda\isanewline
\ \ \ (M,\ fml,\ is\_formula\_case\ (M,\ is\_a,\ is\_b,\ is\_c(f),\ is\_d(f)),\
z)"
\end{defsession}

\section{Absoluteness for $\BL$} \label{sec:absoluteness}
\label{sec:absoluteness-l}

In order to prove $\BV=\BL$, we must prove the absoluteness of three
main functions:
\begin{enumerate}
\item \isa{satisfies}, the satisfaction function on formulae
\item \isa{DPow}, the definable powerset function
\item \isa{Lset}, which expresses the levels $L_\alpha$ of the
constructible hierarchy.
\end{enumerate}
Of these functions, \isa{Lset} is defined by transfinite
recursion from \isa{DPow}, which in turn has a straightforward
definition in terms of \isa{satisfies}.  But proving the absoluteness
of
\isa{satisfies} is very complicated.

Absoluteness of \isa{satisfies} is merely an instance
of the absoluteness of recursion over formulae, and is therefore
trivial.  That does not relieve us of the task of formalizing the
details.  The file containing the \isa{satisfies} absoluteness
proof is one of the largest in the entire development.  This file
divides into two roughly equal parts.

The first half contains internalizations and reflection theorems for
operators such as
\isa{depth} and \isa{formula\_case}.  It expresses the four
cases of \isa{satisfies} in both functional and relational form,
and proves absoluteness for each case.  Six instances of strong
replacement are required: one for each case of the recursion
(because each contains a $\lambda$-abstraction), another to justify the
use of \isa{transrec}, and yet another to justify the
$\lambda$-abstraction in \isa{formula\_rec}.  These axioms are
assumed to hold of an arbitrary class model~\isa{M}\@.  They are
used to show that the formalization satisfies the conditions of the
absoluteness theorem for
\isa{formula\_rec} described in the previous section.

The second half of the file is devoted to proving that the six
instances of replacement hold in~\isa{L}\@.  The four cases of the
recursion (in their relational form) must each be internalized.  This
tiresome task involves, as always, translating a definition involving
real formulae into one using internalized formulae.  Then, the six
instances of replacement are justified.  Finally, the pieces are put
together.

\subsection{Proving that \isa{satisfies} is Absolute}

Working in the class~\isa{M}, we assume additional instances of the
replacement axiom and apply them to the definition of
\isa{satisfies}, which is reproduced here:
\begin{defsession}
"satisfies(A,Member(x,y))\ =\isanewline
\ \ \ \ (\isasymlambda env\ \isasymin \ list(A).\ bool\_of\_o\ (nth(x,env)\ \isasymin \ nth(y,env)))"\isanewline
"satisfies(A,Equal(x,y))\ =\isanewline
\ \ \ \ (\isasymlambda env\ \isasymin \ list(A).\ bool\_of\_o\ (nth(x,env)\ =\ nth(y,env)))"\isanewline
"satisfies(A,Nand(p,q))\ =\isanewline
\ \ \ \ (\isasymlambda env\ \isasymin \ list(A).\ not\ ((satisfies(A,p)`env)\ and\isanewline
\ \ \ \ \ \ \ \ \ \ \ \ \ \ \ \ \ \ \ \
\ \ \ \ \ \ (satisfies(A,q)`env)))"\isanewline
"satisfies(A,Forall(p))\ =\isanewline
\ \ \ \ (\isasymlambda env\ \isasymin \ list(A).\ bool\_of\_o\isanewline
\ \ \ \ \ \ \ \ \ \ \ \ \ \ \ \ \ \ \ \ \ \ (\isasymforall
x\isasymin A.\ satisfies(A,p)`(Cons(x,env))\ =\ 1))"
\end{defsession}

Many additional concepts must be internalized.  Consider the
predicate \isa{is\_depth}, which formalizes the depth of a formula:
\begin{defsession}
"depth\_fm(p,n)\ ==\isanewline
\ \ \ Exists(Exists(Exists(\isanewline
\ \ \ \ \ And(formula\_N\_fm(n\#+3,1),\isanewline
\ \ \ \ \ \ \ And(Neg(Member(p\#+3,1)),\isanewline
\ \ \ \ \ \ \ \ And(succ\_fm(n\#+3,2),\isanewline
\ \ \ \ \ \ \ \ \ And(formula\_N\_fm(2,0),\
Member(p\#+3,0))))))))"
\end{defsession}
We prove the usual theorem relating the satisfaction of \isa{depth\_fm}
to the truth of \isa{is\_depth}
\begin{isabelle}\footnotesize
"\isasymlbrakk x\ \isasymin \ nat;\ y\ <\ length(env);\ env\
\isasymin \ list(A)\isasymrbrakk \isanewline
\ \isasymLongrightarrow \ sats(A,\ depth\_fm(x,y),\ env)\ \isasymlongleftrightarrow\isanewline
\ \ \ \ \ is\_depth(**A,\ nth(x,env),\ nth(y,env))"
\end{isabelle}
And we generate yet another instance of the reflection theorem:
\begin{isabelle}\footnotesize
"REFLECTS[\isasymlambda x.\ is\_depth(L,\ f(x),\ g(x)),\isanewline
\ \ \ \ \ \ \ \ \ \ \isasymlambda i\ x.\ is\_depth(**Lset(i),\ f(x),\
g(x))]"
\end{isabelle}

The internalization of \isa{is\_formula\_case} is omitted, but its
definition is 15 lines long and contains 11 quantifiers.  The theorem
statements relating
\isa{is\_formula\_case} to \isa{formula\_case} are also long and
complicated.  And of course they are higher-order, requiring the
methods of Sect.\ts\ref{sec:higher-order}.

In order to relativize \isa{satisfies}, we must first define
constants corresponding to \isa{formula\_rec}'s parameters~\isa{a},
\isa{b}, \isa{c} and~\isa{d}.  Here are the two base cases:
\begin{defsession}
"satisfies\_a(A)\ ==\isanewline
\ \ \isasymlambda x\ y.\ \isasymlambda env\isasymin list(A).\
bool\_of\_o\ (nth(x,env)\ \isasymin \ nth(y,env))"\isanewline
\isanewline
"satisfies\_b(A)\ ==\isanewline
\ \ \isasymlambda x\ y.\ \isasymlambda env\isasymin list(A).\
bool\_of\_o\ (nth(x,env)\ =\ nth(y,env))"
\end{defsession}
In the two recursive cases, the variables \isa{rp} and~\isa{rq} denote
the values returned on the recursive calls for \isa{p} and~\isa{q},
respectively:
\begin{defsession}
"satisfies\_c(A)\ ==\isanewline
\ \ \isasymlambda p\ q\ rp\ rq.\ \isasymlambda
env\isasymin list(A).\ not(rp\ `\ env\ and\ rq\ `\ env)"\isanewline
\isanewline
"satisfies\_d(A)\ ==\isanewline
\ \ \isasymlambda p\ rp.\ \isasymlambda env\isasymin list(A).\ bool\_of\_o\ (\isasymforall x\isasymin A.\ rp\ `\ (Cons(x,env))\ =\
1)"
\end{defsession}

Each of these functions is then re-expressed in relational form.  Here
is the first:
\begin{defsession}
"satisfies\_is\_a(M,A)\ ==\isanewline
\ \isasymlambda x\ y\ zz.\ \isasymforall lA[M].\ is\_list(M,A,lA)\ \isasymlongrightarrow\isanewline
\ \ \ \ \ \ \ is\_lambda(M,\ lA,\isanewline
\ \ \ \ \ \ \ \ \ \ \isasymlambda env\ z.\ is\_bool\_of\_o(M,\isanewline
\ \ \ \ \ \ \ \ \ \ \ \ \isasymexists nx[M].\ \isasymexists
ny[M].\isanewline
\ \ \ \ \ \ \ \ \ \ \ \ \ \ is\_nth(M,x,env,nx)\ \&\
is\_nth(M,y,env,ny)\ \&\ nx\isasymin ny,\ z),\isanewline
\ \ \ \ \ \ \ \ zz)"
\end{defsession}
Once we have done the other three, we can define an instance of
\isa{MH} for
\isa{satisfies}, expressing the body of the recursion as a predicate:
\begin{defsession}
"satisfies\_MH\ ==\isanewline
\ \isasymlambda M\ A\ u\ f\ z.\isanewline
\ \ \ \ \ \ \isasymforall fml[M].\ is\_formula(M,fml)\ \isasymlongrightarrow\isanewline
\ \ \ \ \ \ \ \ \ \ is\_lambda\ (M,\ fml,\isanewline
\ \ \ \ \ \ \ \ \ \ \ \ is\_formula\_case\ (M,\ satisfies\_is\_a(M,A),\isanewline
\ \ \ \ \ \ \ \ \ \ \ \ \ \ \ \ \ \ \ \ \ \ \ \ \ \ \ \ \ \ \
satisfies\_is\_b(M,A),\isanewline
\ \ \ \ \ \ \ \ \ \ \ \ \ \ \ \ \ \ \ \ \ \ \ \ \ \ \ \ \ \ \
satisfies\_is\_c(M,A,f),\isanewline
\ \ \ \ \ \ \ \ \ \ \ \ \ \ \ \ \ \ \ \ \ \ \ \ \ \ \ \ \ \ \
satisfies\_is\_d(M,A,f)),\isanewline
\ \ \ \ \ \ \ \ \ \ \ \ z)"
\end{defsession}
Finally, \isa{satisfies} itself can be relativized:
\begin{defsession}
"is\_satisfies(M,A)\ ==\ is\_formula\_rec\ (M,\
satisfies\_MH(M,A))"
\end{defsession}

This lemma relates the fragments defined above to the original primitive
recursion in \isa{satisfies}.
Induction is not required: the definitions are directly equal!
\begin{isabelle}\footnotesize
"satisfies(A,p)\ =\isanewline
\ \ formula\_rec\ (satisfies\_a(A),\ satisfies\_b(A),\isanewline
\ \ \ \ \ \ \ \ \ \ \ \ \ \ \,satisfies\_c(A),\ satisfies\_d(A),\
p)"
\end{isabelle}

At this point we must assume (by declaring a locale) the six
instances of replacement mentioned above.  That enables us to prove
absoluteness for the parameters \isa{a}, \isa{b}, \isa{c} and
\isa{d} used to define \isa{satisfies}.  For example, the class
\isa{M} is closed under \isa{satisfies\_a}:
\begin{isabelle}\footnotesize
"\isasymlbrakk M(A);\ x\isasymin nat;\ y\isasymin nat\isasymrbrakk \ \isasymLongrightarrow \
M(satisfies\_a(A,x,y))"
\end{isabelle}
This theorem states that 
\isa{satisfies\_is\_a(M,A,x,y,zz)} is the relational equivalent of
\isa{satisfies\_a(A,x,y)} provided \isa{x} and \isa{y} belong
to the set \isa{nat}.  
\begin{isabelle}\footnotesize
"M(A)\ \isasymLongrightarrow\isanewline
\ Relation2(M,\ nat,\ nat,\
satisfies\_is\_a(M,A),\ satisfies\_a(A))"
\end{isabelle}
It can be seen as an absoluteness result subject to typing conditions
on \isa{x} and~\isa{y}.  Proofs are obviously easier if the
absoluteness results are unconditional, but sometimes typing
conditions are difficult to avoid.

Analogous theorems are proved for
\isa{satisfies\_is\_b}, \isa{satisfies\_is\_c} and
\isa{satisfies\_is\_d}.  Thus we use the first four instances of
replacement.  The last two instances, which are specific to
\isa{satisfies}, let us discharge the more general instances of
replacement that are conditions of \isa{formula\_rec}'s absoluteness
theorem.  We ultimately obtain absoluteness for
\isa{satisfies}:
\begin{isabelle}\footnotesize
"\isasymlbrakk M(A);\ M(z);\ p\ \isasymin \ formula\isasymrbrakk \isanewline
\ \isasymLongrightarrow \ is\_satisfies(M,A,p,z)\ \isasymlongleftrightarrow\ z\ =\
satisfies(A,p)"
\end{isabelle}

\subsection{Proving the Instances of Replacement for $\BL$}

Now we must justify those six instances of strong replacement
by proving that they hold in~$\BL$.  Recall that strong
replacement is the conjunction of replacement (which holds
schematically in $\BL$, but may yield too big a set) and an
appropriate instance of separation (Sect.\ts\ref{sec:defining-ZF}).
 
As always, proving instances of separation requires internalizing many
formulae.  Isabelle can do this automatically, but unless it is
given enough internalized formulae to use as building blocks, the
translation  requires much time and space.  I internalized
many concepts manually, declaring their internal counterparts as
constants and proving their correspondence with the original concepts.
Here is the internal equivalent of \isa{satisfies\_is\_a}:
\begin{defsession}
"satisfies\_is\_a\_fm(A,x,y,z)\ ==\isanewline
\ \ Forall(\isanewline
\ \ \ \ Implies(is\_list\_fm(succ(A),0),\isanewline
\ \ \ \ \ \ lambda\_fm(\isanewline
\ \ \ \ \ \ \ \ bool\_of\_o\_fm(Exists(\isanewline
\ \ \ \ \ \ \ \ \ \ \ \ \ \ \ \ \ \ \ \ \ \
Exists(And(nth\_fm(x\#+6,3,1),\isanewline
\ \ \ \ \ \ \ \ \ \ \ \ \ \ \ \ \ \ \ \ \ \ \ \ \ \ \ \ \ \
And(nth\_fm(y\#+6,3,0),\isanewline
\ \ \ \ \ \ \ \ \ \ \ \ \ \ \ \ \ \ \ \ \ \ \ \ \ \ \ \ \ \ \ \ \ \
Member(1,0))))),\ 0),\isanewline
\ \ \ \ \ \ \ \ 0,\ succ(z))))"
\end{defsession}

Obviously, the same task must be done for the other
\isa{satisfies} relations and for the concepts used in their
definitions.  We finally can internalize the body of \isa{satisfies}:
\begin{defsession}
\ "satisfies\_MH\_fm(A,u,f,zz)\ ==\isanewline
\ \ \ Forall(\isanewline
\ \ \ \ \ Implies(is\_formula\_fm(0),\isanewline
\ \ \ \ \ \ \ lambda\_fm(\isanewline
\ \ \ \ \ \ \ \ \ formula\_case\_fm(satisfies\_is\_a\_fm(A\#+7,2,1,0),\isanewline
\ \ \ \ \ \ \ \ \ \ \ \ \ \ \ \ \ \ \ \ \ \ \ \ \ satisfies\_is\_b\_fm(A\#+7,2,1,0),\isanewline
\ \ \ \ \ \ \ \ \ \ \ \ \ \ \ \ \ \ \ \ \ \ \ \ \ satisfies\_is\_c\_fm(A\#+7,f\#+7,2,1,0),\isanewline
\ \ \ \ \ \ \ \ \ \ \ \ \ \ \ \ \ \ \ \ \ \ \ \ \ satisfies\_is\_d\_fm(A\#+6,f\#+6,1,0),\isanewline
\ \ \ \ \ \ \ \ \ \ \ \ \ \ \ \ \ \ \ \ \ \ \ \ \ 1,\ 0),\isanewline
\ \ \ \ \ \ \ \ \ 0,\ succ(zz))))"
\end{defsession}

Now, we can prove the six instances of replacement.  Here is the first
one, for the \isa{Member} case of \isa{satisfies}:
\begin{isabelle}\footnotesize
"\isasymlbrakk L(A);\ x\ \isasymin \ nat;\ y\ \isasymin \ nat\isasymrbrakk \isanewline
\ \isasymLongrightarrow \ strong\_replacement\isanewline
\ \ \ \ \ (L,\ \isasymlambda env\ z.\ \isasymexists bo[L].\ \isasymexists nx[L].\ \isasymexists ny[L].\isanewline
\ \ \ \ \ \ \ \ \ \ env\ \isasymin \ list(A)\ \&\ is\_nth(L,x,env,nx)\ \&\ is\_nth(L,y,env,ny)\
\&\isanewline
\ \ \ \ \ \ \ \ \ \ is\_bool\_of\_o(L,\ nx\ \isasymin \ ny,\ bo)\ \&\isanewline
\ \ \ \ \ \ \ \ \ \ pair(L,\ env,\ bo,\ z))"
\end{isabelle}
The theorem statement may look big, but the proof has only
four commands.  The corresponding instances of the reflection theorem
(not shown) is twice as big, but its proof has only one command.

We proceed to prove the fifth instance of replacement:
\begin{isabelle}\footnotesize
"\isasymlbrakk n\ \isasymin \ nat;\ L(A)\isasymrbrakk \
\isasymLongrightarrow \ transrec\_replacement(L,\ satisfies\_MH(L,A),\
n)"
\end{isabelle}

Finally, we prove the sixth instance of replacement:
\begin{isabelle}\footnotesize
"\isasymlbrakk L(g);\ L(A)\isasymrbrakk \ \isasymLongrightarrow \isanewline
\ strong\_replacement\ (L,\isanewline
\ \ \ \ \isasymlambda x\ y.\ mem\_formula(L,x)\ \&\isanewline
\ \ \ \ \ \ \ \ \ \ (\isasymexists c[L].\ is\_formula\_case(L,\ satisfies\_is\_a(L,A),\isanewline
\ \ \ \ \ \ \ \ \ \ \ \ \ \ \ \ \ \ \ \ \ \ \ \ \ \ \ \ \ \ \ \ \ \ \ \ \
satisfies\_is\_b(L,A),\isanewline
\ \ \ \ \ \ \ \ \ \ \ \ \ \ \ \ \ \ \ \ \ \ \ \ \ \ \ \ \ \ \ \ \ \ \ \ \
satisfies\_is\_c(L,A,g),\isanewline
\ \ \ \ \ \ \ \ \ \ \ \ \ \ \ \ \ \ \ \ \ \ \ \ \ \ \ \ \ \ \ \ \ \ \ \ \
satisfies\_is\_d(L,A,g),\ x,\ c)\ \&\isanewline
\ \ \ \ \ \ \ \ \ \ pair(L,\ x,\ c,\ y)))"
\end{isabelle}

Our reward for this huge effort is that the absoluteness of
\isa{satisfies} now holds for~\isa{L}:
\begin{isabelle}\footnotesize
"\isasymlbrakk L(A);\ L(z);\ p\ \isasymin \ formula\isasymrbrakk \isanewline
\ \isasymLongrightarrow \ is\_satisfies(L,A,p,z)\ \isasymlongleftrightarrow\ z\ =\
satisfies(A,p)"
\end{isabelle}

\subsection{Absoluteness of the Definable Powerset}

Conceptually, the absoluteness of \isa{DPow} is trivial, since it is
just a comprehension involving \isa{satisfies}.  The formal details
require a modest effort.  There are more internalizations, such as
that of
\isa{is\_formula\_rec}.  Note that concepts only have to be
internalized if they appear in an instance of separation, which may
only happen long after the concept is first relativized. 
Unfortunately, \isa{formula\_rec} is a complex higher-order
function; in its relational form, one argument gets enclosed within
11 quantifiers.  Completing this task enables us to internalize 
\isa{is\_satisfies}:
\begin{defsession}
"satisfies\_fm(x)\ ==\
formula\_rec\_fm(satisfies\_MH\_fm(x\#+5\#+6,2,1,0))"
\end{defsession}

Recall that \isa{DPow} is the definable powerset operator.  It has a
variant form, \isa{DPow'}, that does not involve the function
\isa{arity}.  The two operators agree on transitive sets, so in
particular we can use \isa{DPow'} to construct~\isa{L}\@. 
Now we must relativize \isa{DPow'}.  Its definition refers
to the powerset operator, which is not absolute.  It can equivalently
be expressed using a set comprehension, which here represents
an appeal to the replacement axiom:
\begin{isabelle}\footnotesize
"DPow'(A)\ =\ \isacharbraceleft z\ .\ ep\ \isasymin \ list(A)\ \isasymtimes\
formula,\isanewline
\ \ \ \ \ \ \ \ \ \ \ \ \ \ \ \ \ \ \isasymexists env\ \isasymin \ list(A).\ \isasymexists p\ \isasymin \ formula.\isanewline
\ \ \ \ \ \ \ \ \ \ \ \ \ \ \ \ \ \ \ \ \ ep\ =\ \isasymlangle env,p\isasymrangle\ \&\isanewline
\ \ \ \ \ \ \ \ \ \ \ \ \ \ \ \ \ \ \ \ \ z\ =\ \isacharbraceleft x\isasymin A.\ sats(A,\ p,\ Cons(x,env))\isacharbraceright \isacharbraceright
"
\end{isabelle}
Within the comprehension is another comprehension, which
appeals to separation.  The formula \isa{sats(A,\ p,\ Cons(x,env))}
needs to be relativized (as the predicate \isa{is\_DPow\_sats}) and
internalized.  Then, we again extend the list of assumptions about the
class~\isa{M} to include these instances of replacement and
separation.  Using them, we can prove that \isa{M} is closed under
definable powersets:
\begin{isabelle}\footnotesize
"M(A)\ \isasymLongrightarrow\ M(DPow'(A))"
\end{isabelle}
We can also express the equation for \isa{DPow'} shown above in
relational form, defining the predicate \isa{is\_DPow'}, and prove
absoluteness:
\begin{isabelle}\footnotesize
"\isasymlbrakk M(A);\ M(Z)\isasymrbrakk \ \isasymLongrightarrow\ is\_DPow'(M,A,Z)\ \isasymlongleftrightarrow\ Z\ =\
DPow'(A)"
\end{isabelle}
To make these results available for~\isa{L}, we must first prove that
\isa{L} satisfies the new instances of replacement and
separation.  Here is the latter:
\begin{isabelle}\footnotesize
"\isasymlbrakk L(A);\ env\ \isasymin \ list(A);\ p\ \isasymin \ formula\isasymrbrakk \isanewline
\ \isasymLongrightarrow\ separation(L,\ \isasymlambda x.\
is\_DPow\_sats(L,A,env,p,x))"
\end{isabelle}

\subsection{Absoluteness of Constructibility}

The proof that $\BL$ satisfies $\BV=\BL$ nearly finished.  Only the
operator
\isa{Lset}, which denotes the levels of the constructible hierarchy,
remains to be proved absolute.  Recall that it can be expressed using
\isa{DPow'}:
\begin{isabelle}\footnotesize
"Lset(i)\ =\
transrec(i,\ \%x\ f.\ \isasymUnion y\isasymin x.\ DPow'\ (f\ `\ y))"
\end{isabelle}
So now we must internalize the predicate \isa{is\_DPow'}.  First 
we must internalize the operators used in its definition.  Among those
are the predicate \isa{is\_Collect}, which recognizes set
comprehensions.  The equation for \isa{Lset} above involves two further
instances of replacement: one for the use of \isa{transrec} and
another for the indexed union.  Adding them to our list of constraints
on~\isa{M} allows us to prove that that class is closed under the
\isa{Lset} operator:
\begin{isabelle}\footnotesize
"\isasymlbrakk Ord(i);\ \ M(i)\isasymrbrakk \ \isasymLongrightarrow\ M(Lset(i))"
\end{isabelle}
We can also define its relational version:
\begin{defsession}
"is\_Lset(M,a,z)\ ==\isanewline
\ is\_transrec(M,\ \%x\ f\ u.\ u\ =\
(\isasymUnion y\isasymin x.\ DPow'\ (f\ `\ y)),\ a,\ z)"
\end{defsession}
Notice that this definition is not purely relational.  That is all
right because
\isa{is\_Lset} is not used in any instance of separation and thus need
not be internalized.  We can now prove that the
constructible hierarchy is absolute:
\begin{isabelle}\footnotesize
"\isasymlbrakk Ord(i);\ M(i);\ M(z)\isasymrbrakk \
\isasymLongrightarrow\ is\_Lset(M,i,z)\ \isasymlongleftrightarrow\ z\ =\ Lset(i)"
\end{isabelle}
As remarked earlier, results such as this express absoluteness because
the class model~\isa{M} drops out of the right-hand side.  The
left-hand side refers to our
formalization of $L_\alpha$ in~\isa{M}, which by the theorem is
equivalent to $L_\alpha$ itself.
As always, making this result available to~\isa{L} requires
proving the new instances of replacement.  I omit the details, which
contain nothing instructive.

We can finally formalize $\BL^\BM$, the relativization of~$\BL$.  A
set~\isa{x} is constructible (with respect to any class~\isa{M}
satisfying the specified ZF axioms) provided there exists an
ordinal~\isa{i} and a level of the constructible
hierarchy~\isa{Li} such that \isa{x\ \isasymin \ Li}.
\begin{defsession}
"constructible(M,x)\ ==\isanewline
\ \ \ \isasymexists i[M].\ \isasymexists Li[M].\ ordinal(M,i)\ \&\
is\_Lset(M,i,Li)\ \&\ x\ \isasymin \ Li"
\end{defsession}

The following theorem  is a trivial consequence of the absoluteness
results and the definitions of
\isa{constructible} and~\isa{L}\@.
\begin{isabelle}\footnotesize
"L(x)\ \isasymLongrightarrow\ constructible(L,x)"
\end{isabelle}
This theorem expresses our goal, namely that $\BV=\BL$ holds
in~$\BL$ or more formally $(\BV=\BL)^\BL$.  For this statement is
equivalent to $(\forall x.\,\BL(x))^\BL$ and thus to $\forall
x.\,\BL(x)\imp \BL^\BL(x)$.  We can drop the universal quantifier. 
The antecedent of the implication is formalized as
\isa{L(x)} and the consequent as \isa{constructible(L,x)}.
This proof ends the most difficult part of the development.

\section{The Axiom of Choice in $\BL$} \label{sec:ac}
\label{sec:ac-in-l}

The formalization confirms that $\BV=\BL$ is consistent with the axioms
of set theory.  Obviously any consequence of $\BV=\BL$, such as the
axiom of choice, is consistent with those axioms too.  Proving
consequences of $\BV=\BL$ involves working in an entirely
different way, and a much pleasanter one.  Dispensing with the
relational language, relativization, internalization and absoluteness,
we can instead work in native set theory with the additional axiom
$\BV=\BL$.

Assuming $\BV=\BL$, the proof of the axiom of choice is
simple~\cite[p.\ts173]{kunen80}.  It suffices to prove that every set
can be well-ordered.  In fact, we can well-order the whole of~$\BL$. 
The set of internalized formulae is countable, and therefore
well-ordered. The well-ordering of~$\BL$ derives from its cumulative
construction and from the well-ordering of formulae.
For
$x$, $y\in\BL$, say that $x$ precedes $y$ if 
\begin{itemize}
\item $x$ originates earlier than $y$ in the constructible hierarchy ---
that is, there is some $\alpha$ such that $x\in L_\alpha$ and
$y\not\in L_\alpha$.
\item $x$ and $y$ originate at the same level $L_\alpha$, but the
combination of defining formula and parameters for~$x$
lexicographically precedes the corresponding combination for~$y$.  
\end{itemize}
 
Each element of $L_{\alpha+1}$ is a subset of $L_{\alpha}$ that can be
defined by a formula, possibly involving parameters
from~$L_{\alpha}$.  We can assume the induction hypothesis that
$L_{\alpha}$ is well-ordered.  Before we can undertake this
transfinite induction, we must complete several tasks:
\begin{enumerate}
\item exhibiting a well-ordering on lists, for the parameters of a
definable subset
\item exhibiting a well-ordering on formulae
\item combining these to obtain a well-ordering of the definable
powerset
\item show how to extend our well-ordering to the limit case of the
transfinite induction
\end{enumerate}

\subsection{A Well-Ordering for Lists}

First we inductively define a relation on lists: the lexicographic
extension of a relation on the list's elements.  Let \isa{r}
denote a relation over the set~\isa{A}\@.  Then the relation
\isa{rlist(A,r)} is the least set closed under the following rules:
\begin{defsession}
"\isasymlbrakk length(l')\ <\ length(l);\ l'\ \isasymin \
list(A);\ l\ \isasymin \ list(A)\isasymrbrakk \ \isanewline
\ \isasymLongrightarrow \ \isasymlangle l',\ l\isasymrangle\ \isasymin \ rlist(A,r)"\isanewline
\isanewline
"\isasymlbrakk \isasymlangle l',l\isasymrangle\ \isasymin \ rlist(A,r);\ a\
\isasymin \ A\isasymrbrakk \ \isanewline
\ \isasymLongrightarrow \ \isasymlangle Cons(a,l'),\ Cons(a,l)\isasymrangle\ \isasymin \ rlist(A,r)"\isanewline
\isanewline
"\isasymlbrakk length(l')\ =\ length(l);\ \isasymlangle a',a\isasymrangle\
\isasymin \ r;\ \isanewline
\ \ \ \ l'\ \isasymin \ list(A);\ l\ \isasymin \ list(A);\ a'\ \isasymin \ A;\ a\ \isasymin \ A\isasymrbrakk \ \isanewline
\ \isasymLongrightarrow \ \isasymlangle Cons(a',l'),\ Cons(a,l)\isasymrangle\ \isasymin \
rlist(A,r)"
\end{defsession}
Informally, the list \isa{l'}
precedes another list~\isa{l} if 
\begin{enumerate}
\item \isa{l'} is shorter than~\isa{l}, or
\item the lists have the same head and the
tail of \isa{l'} precedes that of~\isa{l}, or
\item the lists have
the same length and the head of \isa{l'} precedes that
of~\isa{l} under the ordering on list elements.
\end{enumerate}
If the element ordering is linear, then so is the list ordering.  This
theorem has a 14-line proof script involving a double structural
induction on lists.
\begin{isabelle}\footnotesize
"linear(A,r)\ \isasymLongrightarrow \ linear(list(A),rlist(A,r))"
\end{isabelle}

If the element ordering is well-founded, then so is the list
ordering.  This
theorem is proved by induction
on the length of the list followed by inductions over the element
ordering and the list ordering.  The proof script is under 20 lines,
but the argument is complicated.
\begin{isabelle}\footnotesize
"well\_ord(A,r)\ \isasymLongrightarrow \ well\_ord(list(A),\
rlist(A,r))"
\end{isabelle}

\subsection{A Well-Ordering on Formulae}
\label{sec:enum}

G\"odel-numbering is the obvious way to well-order the set of
formulae.  An injection from the set of formulae into the set of
natural numbers is easily defined by recursion on the structure of
formulae.  However, it requires an injection from pairs of natural
numbers to natural numbers.  The enumeration function for formulae
takes this injection as its first argument,~\isa{f}:
\begin{defsession}
"enum(f,\ Member(x,y))\ =\ f\ `\ \isasymlangle 0,\ f\ `\ \isasymlangle x,y\isasymrangle\isasymrangle"\isanewline
"enum(f,\ Equal(x,y))\ \ =\ f\ `\ \isasymlangle 1,\ f\ `\ \isasymlangle x,y\isasymrangle\isasymrangle"\isanewline
"enum(f,\ Nand(p,q))\ \ \ =\ f\ `\ \isasymlangle 2,\ f\ `\ \isasymlangle enum(f,p),\
enum(f,q)\isasymrangle\isasymrangle"\isanewline 
"enum(f,\ Forall(p))\ \ \ =\ f\ `\ \isasymlangle succ(2),\
enum(f,p)\isasymrangle"
\end{defsession}

There are several well-known injections from 
$\omega\times\omega$ into~$\omega$, but defining one of them and
proving it to be injective would involve some effort.  Instead we can
appeal to a corollary of $\kappa\otimes\kappa=\kappa$, which is already
available \cite[\S5]{paulson-gr} in Isabelle/ZF:
\begin{isabelle}
\isasymlbrakk well\_ord(A,r);\ InfCard(|A|)\isasymrbrakk \
\isasymLongrightarrow \ A\ \isasymtimes \ A\ \isasymapprox \ A%
\end{isabelle}  
Thus we have $\omega\times\omega\approx\omega$: there is a bijection,
which is also an injection, between $\omega\times\omega$
and~$\omega$.  However, although an injection exists,
we have no means of naming a specific bijection.  Therefore, we conduct
the entire proof of the axiom of choice under the assumption that some
injection exists.  The final theorem is existential, which will allow
the assumption to be discharged.

We declare a locale to express
this new assumption, calling the injection~\isa{fn}. 
Recall that \isa{nat} is Isabelle/ZF's name for the
ordinal~$\omega$:
\begin{defsession}
\isacommand{locale}\ Nat\_Times\_Nat\ =\isanewline
\ \ \isakeyword{fixes}\ fn\isanewline
\ \ \isakeyword{assumes}\ fn\_inj:\ "fn\ \isasymin \ inj(nat*nat,\
nat)"
\end{defsession}

Proving that \isa{enum(fn,p)} defines an injection from formulae
into the naturals requires a straightforward double induction over
formulae:
\begin{isabelle}\footnotesize
"(\isasymlambda p\ \isasymin \ formula.\ enum(fn,p))\ \isasymin
\ inj(formula,\ nat)"
\end{isabelle}
Using the enumeration as a measure function, we find that the set of
formulae is well-ordered:
\begin{isabelle}\footnotesize
"well\_ord(formula,\ measure(formula,\ enum(fn)))"
\end{isabelle}

The functions defined below all have an argument~\isa{f}, which
should range over injections from $\omega\times\omega$ into~$\omega$. 
In proofs, this injection will always be \isa{fn} from locale
\isa{Nat\_Times\_Nat}.  The definiens of a constant definition cannot
refer to~\isa{fn} because it is a variable.

\subsection{Defining the Well-ordering on \isa{DPow(A)}}

The set \isa{DPow(A)} consists of those subsets of~\isa{A} that can
be defined by a formula, possibly using elements of~\isa{A} as
parameters (Sect.\ts\ref{sec:DPow}).  We can define a well-ordering on
\isa{DPow(A)} from one on~\isa{A}\@.  We get a well-ordering
on formulae from their injection into the natural numbers.
To handle the parameters, we define a well-ordering for environments
--- lists over~\isa{A} --- and combine it with the well-ordering of
formulae.  A subset of~\isa{A} might be definable in more than one
way; to make a unique choice, we map environment/formula pairs to
ordinals.  
The well-ordering on environment/formula pairs is the lexicographic
product (given by \isa{rmult}) of the well-orderings on lists
(\isa{rlist}) and formulae (\isa{measure}).  
\begin{defsession}
"env\_form\_r(f,r,A)\ ==\isanewline
\ \ \ rmult(list(A),\ rlist(A,\ r),\isanewline
\ \ \ \ \ \ \ \ \ formula,\ measure(formula,\ enum(f)))"
\end{defsession}
Using existing theorems, it is trivial to prove that this construction
well-orders the set \isa{list(A)\ \isasymtimes\ formula}:
\begin{isabelle}\footnotesize
"well\_ord(A,r)\ \isasymLongrightarrow \ well\_ord(list(A)\ \isasymtimes\
formula,\ env\_form\_r(fn,r,A))"
\end{isabelle}

The order type of the resulting well-ordering yields a map (given by
\isa{ordermap}) from environment/formula pairs into the ordinals. 
For each member of
\isa{DPow(A)}, the minimum such ordinal will determine its place in
the well-ordering.
\begin{defsession}
"env\_form\_map(f,r,A,z)\ ==\isanewline
\ \ ordermap(list(A)\ \isasymtimes\ formula,\ env\_form\_r(f,r,A))\ `\
z"
\end{defsession}

If \isa{r} well-orders \isa{A} and \isa{X} is a definable
subset of~\isa{A}, then let us define \isa{DPow\_ord(f,r,A,X,k)} to
hold if \isa{k} corresponds to some definition of~\isa{X} ---
informally,
\isa{k} defines~\isa{X}:
\begin{defsession}
"DPow\_ord(f,r,A,X,k)\ ==\isanewline
\ \ \ \ \ \ \ \ \isasymexists env\ \isasymin \ list(A).\ \isasymexists p\ \isasymin \ formula.\isanewline
\ \ \ \ \ \ \ \ \ \ arity(p)\ \isasymle \ succ(length(env))\ \&\isanewline
\ \ \ \ \ \ \ \ \ \ X\ =\ \isacharbraceleft x\isasymin A.\ sats(A,\ p,\ Cons(x,env))\isacharbraceright \ \&\isanewline
\ \ \ \ \ \ \ \ \ \ env\_form\_map(f,r,A,\isasymlangle env,p\isasymrangle)\ =\ k"\isanewline
\end{defsession}

Similarly, let us define \isa{DPow\_least(f,r,A,X)} to
be the smallest ordinal defining~\isa{X}\@:
\begin{defsession}
"DPow\_least(f,r,A,X)\ ==\ \isasymmu k.\ DPow\_ord(f,r,A,X,k)"
\end{defsession}
Since \isa{k} determines \isa{env} and \isa{p}, we find that an
ordinal can define at most one element of
\isa{DPow(A)}:
\begin{isabelle}\footnotesize
"\isasymlbrakk DPow\_ord(fn,r,A,X,k);\ DPow\_ord(fn,r,A,Y,k);\
well\_ord(A,r)\isasymrbrakk\isanewline
\ \isasymLongrightarrow \ X=Y"
\end{isabelle}
We also find that every element of \isa{DPow(A)} is defined by some
ordinal, given by \isa{DPow\_least}:
\begin{isabelle}\footnotesize
"\isasymlbrakk X\ \isasymin \ DPow(A);\ well\_ord(A,r)\isasymrbrakk
\isanewline\ \isasymLongrightarrow \ DPow\_ord(fn,\ r,\ A,\ X,\
DPow\_least(fn,r,A,X))"
\end{isabelle}

Now \isa{DPow\_least} can serve as a measure function to define the
well-ordering on \isa{DPow(A)}.
\begin{defsession}
"DPow\_r(f,r,A)\ ==\ measure(DPow(A),\ DPow\_least(f,r,A))"
\end{defsession}
Using general facts about relations defined by measure functions, we
easily find that
\isa{DPow(A)} is well-ordered:
\begin{isabelle}\footnotesize
"well\_ord(A,r)\ \isasymLongrightarrow \ well\_ord(DPow(A),\
DPow\_r(fn,r,A))"
\end{isabelle}

\subsection{Well-Ordering $L_\alpha$ in the Limit Case}

The proof that $L_\alpha$ is well-ordered appeals to transfinite
induction on the ordinal~$\alpha$.  The induction hypothesis is that
$L_\xi$ is well-ordered if $\xi<\alpha$.  In the limit case, $L_\alpha=
\union_{\xi<\alpha}\,L_\xi$.  Recall
(Sect.\ts\ref{sec:lrank}) that $\BL$-rank $\rho(x)$ of~$x$ is the
least
$\alpha$ such that $x\in L_{\alpha+1}$.  If $\alpha$ is a limit
ordinal then we order elements of
$L_\alpha$ first by their $\BL$-ranks; if two elements
have the same $\BL$-rank, say~$\xi$, then we order them using the
existing well-ordering of~$L_{\xi+1}$.

In the Isabelle formalization,
\isa{i} is the limit ordinal and \isa{r(j)} denotes the
well-ordering of \isa{Lset(j)}:
\begin{defsession}
"rlimit(i,r)\ ==\isanewline
\ \ if\ Limit(i)\ then\ \isanewline
\ \ \ \ \isacharbraceleft z\ \isasymin\ Lset(i)\ \isasymtimes\ Lset(i).\isanewline
\ \ \ \ \ \isasymexists x'\ x.\ z\ =\ \isasymlangle x',x\isasymrangle\ \&\isanewline
\ \ \ \ \ \ \ \ \ \ \ \ (lrank(x')\ <\ lrank(x)\ |\isanewline
\ \ \ \ \ \ \ \ \ \ \ \ \ (lrank(x')\ =\ lrank(x)\ \&\ \isasymlangle x',x\isasymrangle\
\isasymin \ r(succ(lrank(x)))))\isacharbraceright \isanewline
\ \ else\ 0"
\end{defsession}
We can prove that the limit
ordering is linear provided the  orderings of previous stages are also
linear:
\begin{isabelle}\footnotesize
"\isasymlbrakk Limit(i);\ \isasymforall j<i.\ linear(Lset(j),\ r(j))
\isasymrbrakk \isanewline
\ \isasymLongrightarrow \ linear(Lset(i),\ rlimit(i,r))"
\end{isabelle}
Under analogous conditions,
the \isa{rlimit(i,r)} is a well-ordering of~\isa{Lset(i)}.  The
proofs are straightforward, and I have omitted many details.
\begin{isabelle}\footnotesize
"\isasymlbrakk Limit(i);\ \isasymforall j<i.\ well\_ord(Lset(j),\ r(j))\isasymrbrakk \isanewline
\ \isasymLongrightarrow \ well\_ord(Lset(i),\ rlimit(i,r))"
\end{isabelle}

\subsection{Transfinite Definition of the Well-Ordering for $\BL$}

The well-ordering on $\BL$ is defined by transfinite recursion.  The
Isabelle definition refers to the cryptic
\isa{transrec} operator, so let us pass directly to the
three immediate consequences of that definition.  For the base case,
the well-ordering is the empty relation:
\begin{defsession}
"L\_r(f,0)\ =\ 0"
\end{defsession}
For the successor case,
the well-ordering is given by
applying \isa{DPow\_r} to the previous level. 
\begin{defsession}
"L\_r(f,\ succ(i))\ =\ DPow\_r(f,\ L\_r(f,i),\ Lset(i))"
\end{defsession}
For the limit case,
the well-ordering is given by \isa{rlimit}. 
\begin{defsession}
"Limit(i)\ \isasymLongrightarrow \ L\_r(f,i)\ =\ rlimit(i,\
L\_r(f))"
\end{defsession}

Thanks to the results proved above, a simple transfinite induction
proves that \isa{L\_r(fn,i)} well-orders the constructible
level~\isa{Lset(i)}.
\begin{isabelle}\footnotesize
"Ord(i)\ \isasymLongrightarrow \ well\_ord(Lset(i),\
L\_r(fn,i))"
\end{isabelle}
Note that this theorem refers to~\isa{fn}, an injection from
$\omega\times\omega$ into~$\omega$.  Recall (Sect.\ts\ref{sec:enum})
that we know such that such functions exist but have not defined a
specific one.  We have been able to prove our theorems by working in a
locale that assumes the existence of~\isa{fn}.  Now, we can eliminate
the assumption.  We use an existential quantifier to hide the
well-ordering in the previous theorem, so that \isa{fn} no longer
appears.  Then, by the mere existence of such an injection, it follows
that every
\isa{Lset(i)} can be well-ordered:
\begin{isabelle}\footnotesize
"Ord(i)\ \isasymLongrightarrow \ \isasymexists r.\
well\_ord(Lset(i),\ r)"
\end{isabelle}

To wrap things up, let us package the axiom $\BV=\BL$ as a locale:
\begin{defsession}
\isacommand{locale}\ V\_equals\_L\ =\isanewline
\ \ \isakeyword{assumes}\ VL:\ "L(x)"
\end{defsession}
The axiom of choice --- in the guise of the well-ordering theorem ---
is a trivial consequence of the previous results.
\begin{isabelle}\footnotesize
\isacommand{theorem}\ (\isakeyword{in}\ V\_equals\_L)\ AC:\
"\isasymexists r.\ well\_ord(x,r)"
\end{isabelle}

\section{Conclusions} \label{sec:conclusions}

What has been accomplished?  I have mechanized the proof of the
relative consistency of the axiom of choice, largely following a
standard textbook presentation.  The formal proof is much longer than
the textbook version because it is complete in all details and uses no
metatheoretical reasoning.  

As noted in Section~\ref{sec:outline}, G\"odel's proof comprises four tasks,
which we can now express more precisely:
\begin{enumerate}
\item defining the class $\BL$ within ZF
\item proving, for every ZF axiom~$\phi$, that $\phi^\BL$ is a ZF theorem
\item proving $(\BV=\BL)^\BL$ in ZF
\item proving that $\mathrm{ZF} + {\BV=\BL}$ implies the axiom of choice
\end{enumerate}
The proof that $\BL$ satisfies $\BV=\BL$ is by far the largest and most difficult part
of the development.  It involves proving $\BL$ to be absolute, which
requires converting every concept used in its definition into
relational form and proving absoluteness.  The sheer number of
concepts is an obstacle, and some of them are hard to express in
relational form, especially those involving recursion.  Most of the
relations have to be re-expressed using an internal datatype
of formulae.

My formalization has two limitations. First,  I
am not able to prove that $\BL$ satisfies the axiom scheme of
comprehension.  Although Isabelle/ZF handles schematic proofs easily,
the proof of comprehension for the formula $\phi$ requires an instance
of the reflection theorem for~$\phi$.  Each instance of
comprehension therefore has a different proof and must be proved
separately.  The
reflection theorem is proved by induction (at the metalevel) on
the structure of~$\phi$; thus, all these proofs are
instances of one algorithm, and they are generated by nearly identical
proof scripts~\cite{paulson-reflection}.  The
inability to prove the comprehension scheme makes the
absoluteness proofs harder: every necessary instance of comprehension
is listed.  Instantiating these proofs to $\BL$ has required
proving that each of those instances held in~$\BL$.  There are about
35 such instances.

My formalization has another limitation.  The proof that $\BL$
satisfies $\BV=\BL$ cannot be combined with the proof that $\BV=\BL$
implies the axiom of choice in order to  conclude that $\BL$
satisfies the axiom of choice. The reason is  that the two instances of
$\BV=\BL$ are formalized differently: one is relativized and the
other is not.  Here I have followed the textbook proofs, which
prove
$\BV=\BL$, declare that the axiom of constructibility can be
assumed, and proceed to derive the consequences of that axiom.

We could remedy both limitations by tackling the whole problem in a
quite different way, by formalizing set theory as a proof system and
working entirely in the metatheory. I leave this as a challenge for the
theorem-proving community.  A by-product of the work is a general theory
of absoluteness for arbitrary class models of~ZF\@.  It could be used
for other formal investigations of inner models.  Future investigators
might also try formalizing the proof that $\BL$ satisfies the
generalized continuum hypothesis and the combinatorial 
principle~$\lozenge$.

\paragraph*{Acknowledgements.}
Krzysztof Gr\c{a}bczewski devoted much effort to an earlier, 
unsuccessful, attempt to formalize this material.  Isabelle work is
supported by the U.K.'s Engineering and Physical Sciences Research
Council, grant GR/M75440.  Markus Wenzel greatly improved Isabelle's
\isa{locale} construct to support these proofs.  Kenneth Kunen gave advice
that helped in my formalization of the reflection theorem.  The referee
made a number of valuable comments on this paper.

\bibliographystyle{plain}\raggedright
\bibliography{string,atp,funprog,general,isabelle,theory,crossref}
\end{document}